\documentclass[usenatbib,onecolumn]{mnras}

\usepackage{newtxtext,newtxmath}

\usepackage[T1]{fontenc}

\DeclareRobustCommand{\VAN}[3]{#2}
\let\VANthebibliography\thebibliography
\def\thebibliography{\DeclareRobustCommand{\VAN}[3]{##3}\VANthebibliography}

\usepackage{graphicx}	
\usepackage{amsmath}	

\usepackage{tensor}
\usepackage{color}
\usepackage{hyperref}
\usepackage{graphicx}
\usepackage{dcolumn}
\usepackage{bm}
\usepackage{hyperref}
\usepackage{footnote}
\usepackage{caption}
\usepackage{subcaption}
\usepackage{physics}
\usepackage{amsmath}
\usepackage{mciteplus}
\usepackage[normalem]{ulem}

\title[Model independent structure growth in $f(R)$]{A model independent approach to the study of structure growth in $f(R)$ gravity}

\author[K. MacDevette et al.]{
Kelly MacDevette,$^{1}$\thanks{E-mail: MCDKEL004@myuct.ac.za}
Jess Worsley,$^{1}$
Peter Dunsby$^{1,2,3}$
and Saikat Chakraborty$^{4,2}$
\\
$^{1}$ Department of Mathematics and Applied Mathematics, University of Cape Town, Rondebosch 7700, Cape Town, South Africa\\
$^{2}$Center for Space Research, North-West University, Potchefstroom 2520, South Africa\\
$^{3}$South African Astronomical Observatory, Observatory 7925, Cape Town, South Africa\\
$^{4}$ The Institute for Fundamental Study ``The Tah Poe Academia Institute", Naresuan University, Phitsanulok 65000, Thailand
}

\pubyear{2024}

\begin{document}
\label{firstpage}

\maketitle

\begin{abstract}
Over the last decade, much attention has been given to the study of modified gravity theories to find a more natural explanation for the late-time acceleration of the Universe. Particular attention has focused on the so-called $f(R)$ dark energy models. 
Instead of focusing on a particular f(R)
model, we present a completely model-independent approach to study the background dynamics and
the growth of matter density perturbations for those f(R) models that mimic the $\Lambda$CDM evolution at the background level. We do this by characterising the dynamics of the gravitational field using a set of dimensionless variables and using cosmography to determine the expansion history. We then illustrate the integrity of this method by fixing the cosmography to be the same as an exact $\Lambda$CDM model, allowing us to test the solution. We compare the exact evolution of the density contrast and growth index with what one obtains from various levels of the quasi-static approximation, without choosing the form of $f(R)$ dark energy. 
\end{abstract}




\newcommand{\kelly}[1]{\textcolor[RGB]{13, 189, 136}{[Kelly: #1]}}
\newcommand{\jess}[1]{\textcolor[RGB]{180, 0, 180}{[Jess: #1]}}





\section{\label{sec:intro} Introduction}
For almost two decades, the cosmological constant has remained the simplest and most compelling explanation for what is driving the present-day accelerated expansion of the universe, assuming of course that the geometry of the universe is well described by the Robertson-Walker metric. While the $\Lambda$CDM model \cite{Blumenthal:1984bp,Ostriker:1995rn} is by far the simplest theoretical description, it is not without its flaws. Dicke was the first to bring up the issue of `fine-tuning' in the cosmological context \cite{DICKE1961} - that a slightly larger value of $\Lambda$ would render large-scale structure impossible. Additionally, the dark energy density having an equal order of magnitude to the matter density at present, despite contrary evolution histories, has been aptly termed the `coincidence problem' \cite{Steinhardt+1998+123+146}. There is also an alarming disagreement between the observed value of the vacuum energy density and the much larger zero-point energy determined from quantum field theory \cite{Adler1995}. Finally, the most recent DESI results appear to support a theory of \textit{evolving} dark energy parameterised by the $w_0w_a$CDM model \cite{tada}. 

In light of these issues, various modifications of General Relativity (GR) have been investigated. One of the simplest non-trivial extensions of GR has an action that is non-linear in the Ricci curvature $R$ and/or other curvature invariants \cite{Stelle:1977ry, DeFelice:2010aj, Sotiriou:2008rp}. An important property of these theories is that the field equations can be constructed in a way which makes comparisons with GR straightforward. This is done by moving all the non-linear corrections onto the right-hand side of the field equations. In this way, we can define an ‘‘effective’’ source term, known as the curvature fluid. One of the properties of the curvature fluid is that it naturally violates the strong energy condition, allowing for a curvature-fluid-driven period of late-time cosmological acceleration \cite{Carroll:2003wy, Capozziello:2007ec}. Unfortunately, this comes at the cost of having to study a much more complicated (fourth-order) set of field equations. This leads to considerable challenges in determining exact and numerical solutions which can be confronted with observations. These difficulties can be addressed somewhat by using dynamical systems methods \cite{Carloni:2004kp, Carloni:2006mr, Carloni:2007br, Abdelwahab:2011dk}. Guided by the Friedmann constraint, a set of dimensionless dynamical variables can be defined, providing a relatively simple method for obtaining exact solutions (via the equilibrium points of the system) and a description of the global dynamics of these models. Unfortunately, the dynamical system obtained is only autonomous for particularly simple forms of $f(R)$ and each theory has to be analysed separately. This is due to a term $\Gamma(R)=d\ln{R}/d\ln{f'}$ remaining in the equations involving $f(R)$ and its first and second derivatives with respect to $R$, which cannot be expressed in terms of the dynamical system variables in general \cite{Carloni:2007br}. This issue can be resolved by introducing additional variables and parameters, but the functional form $f(R)$ still needs to be specified prior to analysing the cosmological dynamics \cite{Carloni_2015}.

Recently it was noticed that this problem can be solved more elegantly by writing $\Gamma(R)$ in terms of the cosmographic parameters \cite{Chakraborty:2021jku}. In this way, it is possible to write a complete set of differential equations for any $f(R)$ theory. Of course, things are not quite that simple. Since the cosmographic series satisfies an infinite hierarchy of equations, a closed system is only obtained when one fixes the expansion history, which determines an algebraic relationship between the cosmographic parameters. Nevertheless, this approach proves to be more general and more closely linked to observations than any other approach thus far. 

In this work, we build on \cite{Chakraborty:2021jku}, focusing on developing a framework for modelling the growth function in a way which is independent of the choice of $f(R)$. All other work in this area either fixes $f(R)$ from the onset or uses some parameterisation to describe the modification to GR (see for example \cite{Carloni:2006mr, Ananda_2009, Tsujikawa:2009ku, Pogosian2008, Bean_2007, gannouji_growth, Gannouji_2008DE, Narikawa:2009lca, Abebe:2013zua}).  We find that our approach is able to capture features that have been observed in various
model-specific methods in a completely general way. Also, by comparing the exact equations with various levels of approximation (semi and full quasi-static (QS)), we are able to determine the shortcomings of the full QS method on large scales, where the perturbations have not spent significant time in the $f(R)$ regime and thus tend to be closer to $\Lambda$CDM.

The outline of the paper is as follows. In Section 2, we give the cosmological equations for $f(R)$ gravity and describe the general model-dependent dynamical systems approach. In Section 3, we give some background on cosmography and describe the cosmographic-based, model-independent dynamical systems formulation in Section 4. In Section 5 we derive the 1+3 covariant equations describing the growth of structure in a way which is independent of the cosmography and the choice of $f(R)$. In Sections 6 and 7, we explain why choosing the $j=1$ cosmographic condition, leads to a background that mimics $\Lambda$CDM in $f(R)$ gravity and in Section 8 we apply this condition to the model-independent background and perturbation equations.



Natural units ($\hbar = c = k_B = 8\pi G = 1$) are used throughout this paper, unless otherwise specified. Latin indices run from 0 to 3. Derivatives with respect to time are indicated by $\dot{X} =  \partial X/\partial t$ and those with respect to the Ricci scalar $R$ are indicated as $X' = \partial X/\partial R$ unless otherwise stated. We use the $(-,+,+,+)$ signature and the symbol $\nabla$ represents the usual covariant derivative. The Riemann tensor is defined as 
\begin{equation}
    \tensor{R}{^a_{b c d}}=\tensor{W}{^a_{b d, c}}-\tensor{W}{^a_{b c, d}}+\tensor{W}{^e_{b d}}\tensor{W}{^a_{c e}}-\tensor{W}{^f_{b c}} \tensor{W}{^a_{d f}},
\end{equation}
where $\tensor{W}{^a_{bd}}$ are the Christoffel symbols defined by 
\begin{equation}
    \tensor{W}{^a_{bd}} = \frac{1}{2}g^{ae}\left(g_{be,d}+ g_{ed,b} - g_{bd,e}\right).
\end{equation}
The Ricci tensor is defined as the contraction of the first and third indices of the Riemann tensor
\begin{equation}
    R_{ab} = g^{cd}R_{cadb}.
\end{equation}
The action for $f(R)$ gravity in natural units is written as
\begin{equation}
    \mathcal{A} = \int \textnormal{d}^4x \sqrt{-g} \left(\frac{1}{2}f(R) + \mathcal{L}_m\right),
\end{equation}
where $R$ is the Ricci scalar, $f(R)$ is a general function (of at least $C^2$) of the Ricci scalar and $\mathcal{L}_m$ is the matter Lagrangian.
\section{Cosmological Equations for $f(R)$ gravity}
\subsection{Background FLRW equations}
In a spatially flat FLRW background, the $f(R)$ equations governing the expansion of the Universe are given by the modified Friedmann and Raychaudhuri equations:
\begin{align}
    3H^2 =& \frac{1}{f'}\bigg[ \rho_m + \frac{1}{2}\left(Rf'-f\right)-3Hf'' \dot{R}\bigg]\;,\label{eq:fried}\\
    3\dot{H} + 3H^2 =& -\frac{1}{2f'}\bigg[\rho_m + 3p_m + f - f'R + 3 H f''\dot{R}
   + 3 f'''\dot{R}^2 + 3f'' \ddot{R}\bigg]\;,\label{eq:raychaud}
\end{align}
where $\rho_m$ represents the energy density of standard matter, a prime represents the derivative with respect to $R$, $H$ is the Hubble parameter defined through the scale factor $a(t)$ as
\begin{equation}
\label{eq:hubbleparamdef}
    H = \frac{\dot{a}}{a}
\end{equation} 
and $R$ is the Ricci scalar:
\begin{equation}
\label{eq:ricciscalardef}
    R = 6\dot{H} + 12H^2.
\end{equation}
The energy conservation equation for standard matter is given by
\begin{equation}
\label{eq:energyconeq}
    \dot{\rho}_m = -3H \rho_m (1+w),
\end{equation}
where $w$ is the equation of state parameter for standard matter. Two requirements for $f$ must be satisfied to maintain the stability of the model. These are: 
\begin{itemize}
    \item $f' > 0$ to prevent a potential ghost degree of freedom, and
    \item $f'' > 0$, at least during the early epoch of matter domination, to prevent an instability in the growth of curvature perturbations (Dolgov-Kawasaki instability).
\end{itemize}
\subsection{Dynamical system analysis}
A general dynamical system for $f(R)$ gravity in flat space can be expressed in terms of the  expansion-normalised dynamical dimensionless variables,
\begin{align}
    x = \frac{\dot{f}'}{f' H}, \quad y = \frac{R}{6H^2}, \quad v = \frac{f}{6f' H^2}, \quad \Omega = \frac{\rho_m}{3 f' H^2}.
    \label{eq:backgroundvariables}
\end{align}
Substituting these into the Friedmann equation \eqref{eq:fried} gives the following constraint, 
\begin{equation}
    - x + y - v + \Omega = 1,
    \label{eq:Friedmann_constraint}
\end{equation}
allowing the dimension of the system to be reduced to 3. 

The variables \eqref{eq:backgroundvariables} are differentiated with respect to the number of e-foldings, $\eta = \ln{a}$, and equations \ref{eq:ricciscalardef} and \ref{eq:energyconeq} are used to produce the full dynamical system:
\begin{subequations}
\label{eq:dynsys_xyzO}
\begin{align}
    \dv{x}{\eta} &= -2x^2 + 2y - (v+2)x - 4v + \Omega(x + 1 - 3w)\;,\\
    \dv{y}{\eta} &= y \left[ x(\Gamma -2) - 2(v-1) + 2\Omega \right]\;, \\
    \dv{v}{\eta} &= xy\Gamma + v( -3x -2v + 2\Omega + 2 )\;, \\
    \dv{\Omega}{\eta} &= \Omega ( -3x -2v + 2\Omega - 3w -1 )\;,
\end{align}
\label{eq:cosmo_params_dyn_sys}
\end{subequations}
where $\Gamma = \Gamma(R)$ is defined as
\begin{equation}
    \Gamma(R) \equiv \dv{\ln{R}}{\ln{f'}} = \frac{f'}{Rf''} = \frac{\Dot{R}f'}{R\Dot{f}'}\,.
    \label{eq:Gamma_def}
\end{equation}

To close the system, it is necessary to express $\Gamma(R)$ in terms of the dynamical variables $x$, $y$, $v$ and $\Omega$. Finding $R(y/z)$ and subsequently $\Gamma(y/z)$ requires the invertibility of the relation $y/z = RF/y$. This has previously been a limitation on which models of $f(R)$ can be studied with this method. Alternative methods have been proposed to tackle this issue, generally involving the introduction of further parameters or variables. In \cite{Carloni_2015} the dynamical system was reformulated to include the variables,
\begin{align}
    A = \left(\frac{H}{m}\right)^2, \quad Q = \frac{3}{2H} \dv{H}{\tau}, \quad J = \frac{1}{4H} \dv[2]{H}{\tau},
\end{align}
where $m$ is a constant having the dimension of mass specific to the $f(R)$ theory under consideration. The variables $Q$ and $J$ can be related to the cosmographic deceleration ($q$) and jerk ($j$) parameters, respectively. Expressing these as a dynamical system using the same method as above yields a 4-dimensional phase space. This alternative introduces two auxiliary quantities, $X$ and $Y$ (see In \cite{Carloni_2015} for their definition), instead of the single auxiliary quantity $\Gamma$, used in the earlier formulation. A closed autonomous dynamical system can be achieved by expressing these auxiliary quantities in terms of the dynamical variables. Unlike the previous method, which depends on the invertibility of a specific relation to find $R$, this alternative approach avoids this limitation by defining $X$ and $Y$ as functions of $H^2$ and $R$. Depending on the form of $f(R)$, the expressions for $X(A,y)$ and $Y(A,y)$ can end up becoming very complex. The benefit is that this method can handle any form of $f(R)$. 

The drawback of these approaches (including those in the previous sections) is that they adopt a top-down approach, \textit{requiring} the functional form of $f(R)$ to be specified beforehand. The cosmological dynamics of a given $f(R)$ can then be analyzed using one of these methods. This means that investigating consequences of $f(R)$ dark energy theories must be done model-by-model, which is both inefficient and time-consuming. Ideally one would like to reconstruct the theory of gravity directly from observations or at the very least constrain it by requiring the background cosmology to be consistent to observational data. 

It is for this reason that a bottom-up approach, where the form of $f(R)$ is totally unspecified to begin with and must satisfy certain cosmological conditions, is developed in the rest of the paper.

\section{Some remarks about cosmography}\label{sec:cosmography}

As pointed out in \cite{Chakraborty:2021jku} the key to closing the dynamical system presented in the last section is to write the function $\Gamma(R)$ in terms of cosmography. The cosmography is determined by a set kinematic (cosmographic) parameters arising from the Taylor expansion of the Hubble parameter \cite{Dunsby:2015ers}. The $0$-th to $5$-th order cosmographic parameters are as follows:
\begin{subequations}\label{eq:CP}
    \begin{align}
        H &\equiv  \frac{\dot a}{a}\,,
        \\
        q &\equiv  -\frac{1}{aH^2}\frac{d^2 a}{dt^2} = -1-\frac{\dot H}{H^2} \,,
        \\
        j &\equiv  \frac{1}{aH^3}\frac{d^3 a}{dt^3} = \frac{\ddot H}{H^3} - 3q - 2 \,,
        \\
        s &\equiv  \frac{1}{aH^4}\frac{d^4 a}{dt^4} = \frac{\dddot H}{H^4} + 4j + 3q(q+4) + 6 \,,
        \\
        l &\equiv \frac{1}{aH^5}\frac{d^5 a}{dt^5} = \frac{H^{(4)}}{H^5} + 5s - 10(j+3q)(q+2) - 24 \,,
    \end{align}
\end{subequations}
and are called the Hubble, deceleration, jerk, snap, and lerk parameters, respectively. Of course, one could go on to construct even higher-order cosmographic parameters. In fact there are an infinite hierarchy of them \footnote{For some comprehensive reviews see \cite{Dunsby:2015ers, Bolotin:2018xtq}.}. 
\begin{subequations}\label{CP_rel}
    \begin{align}
        j &= 2q^{2} + q - \frac{dq}{d\tau}\,,
        \\
        s &= \frac{dj}{d\tau} - j(2 + 3q)\,,
        \\
        l &= \frac{ds}{d\tau} - s(3+4q)\,.
    \end{align}
\end{subequations}
A given cosmic evolution, if believed to be the solution of either GR or some $f(R)$ theory, can always be specified by an algebraic relation between a finite number of cosmographic parameters. Let us elaborate on this point using a simple example from GR. The cosmic evolution corresponding to the standard $\Lambda$CDM model of cosmology is a solution of GR plus a cosmological constant, and satisfies the field equation
\begin{equation}\label{eq:lcdm}
    H^2 + \frac{k}{a^2} = \frac{\rho_0}{3a^3} + \frac{\Lambda}{3}\,.
\end{equation}
The above equation contains $H$ and the quantities $\left(\frac{k}{a^2},\frac{\rho_0}{a^3},\Lambda\right)$. The first and second time derivatives of \eqref{eq:lcdm} give
\begin{subequations}
    \begin{align}
        \dot H - \frac{k}{a^2} &= - \frac{\rho_0}{2a^3}\,,
        \\
        \frac{\ddot H}{H} + \frac{2k}{a^2} &= \frac{3}{2}\frac{\rho_0}{a^3}\,.
    \end{align}
\end{subequations}
The above equations, along with \eqref{eq:lcdm}, can be used to solve $\left(\frac{k}{a^2},\frac{\rho_0}{a^3},\Lambda\right)$ in terms of $(H,\dot{H},\ddot{H})$ from these three equations. If we now, the third time derivative of \eqref{eq:lcdm} and substitute the expressions for $\left(\frac{k}{a^2},\frac{\rho_0}{a^3},\Lambda\right)$, the resulting equation can be expressed entirely in terms of the cosmographic parameters \cite{Dunajski:2008tg}:
\begin{equation}
    s + 2(q+j) + qj = 0.
\end{equation}
For the special case of spatially flat \emph{or} vacuum cosmologies, one need not go up to the third derivative of  \eqref{eq:lcdm} and the snap parameter is therefore unnecessary. This is the case for spatially flat $\Lambda$CDM cosmology which gives rise to the simpler and well-known cosmographic condition $j=1$.

A similar exercise could be performed for cosmic solutions of $f(R)$ theories. If a cosmic evolution $a(t)$ is believed to be a solution of some $f(R)$ theory in the presence of a perfect fluid with a constant equation of state $w$, then it must satisfy the modified Friedmann equation \eqref{eq:fried}, which contains up to the second derivative of the Hubble parameter, $\ddot{H}$, and the quantities $\left(\frac{k}{a^2},\frac{\rho_0}{a^{3(1+w)}}\right)$. Taking a time derivative of this equation gives the modified Raychaudhuri equation \eqref{eq:raychaud}, which contains up to the third derivative of the Hubble parameter $\dddot{H}$, and the quantities $\left(\frac{k}{a^2},\frac{\rho_0}{a^{3(1+w)}}\right)$. These two equations can be used to solve for the two constants $\left(\frac{k}{a^2},\frac{\rho_0}{a^{3(1+w)}}\right)$ in terms of the $(H,\dot{H},\ddot{H},\dddot{H})$. If we take the third derivative of the modified Friedmann equation \eqref{eq:fried} and substitute the expressions for $\left(\frac{k}{a^2},\frac{\rho_0}{a^{3(1+w)}}\right)$. The resulting equation can be expressed purely in terms of the cosmographic parameters $H,\,q,\,j,\,s,\,l$. One does not need to go beyond the lerk parameter, in agreement with the $f(R)$ cosmography analysis presented in \cite{Capozziello:2008qc}. Again, for spatially flat \emph{or} vacuum cosmological solutions of $f(R)$ theories, one need not go up to the third derivative of \eqref{eq:fried} and the lerk parameter is unnecessary. For spatially flat \emph{and} vacuum cosmological solutions, one need not even consider the modified Raychudhuri equation \eqref{eq:raychaud} and the modified Friedmann equation itself represents a cosmographic relation going only up to the jerk parameter \cite{Dunajski:2008tg}. 

\section{A \textit{model-independent} dynamical system formulation}\label{sec:mod_ind_dyn_sys}
Recently \cite{Chakraborty:2021jku} a \emph{model-independent} dynamical system formulation was proposed, which does not require one to specify a-priori an $f(R)$ model, while still allowing one to obtain a closed autonomous system.

In the standard model-\textit{dependent} approach, the dynamical system \eqref{eq:dynsys_xyzO} is not closed due to the presence of the model-dependent factor $\Gamma(R)$ \eqref{eq:Gamma_def}. As described in \cite{Chakraborty:2021jku}, the key to solving this issue was to write $\Gamma(R)$ in terms of the cosmographic parameters \eqref{eq:CP}:
\begin{equation}
     xy\Gamma = \frac{\dot{R}}{6H^3} = j-q-2\;,
\end{equation}
using the the Ricci scalar definition \eqref{eq:ricciscalardef}. 



Differentiating \eqref{eq:backgroundvariables} and \eqref{eq:CP} with respect to the redshift $z$, the resulting \textit{model-independent} dynamical system for a barotropic perfect fluid with equation of state parameter $w$ is 
\begin{subequations}\label{eq:ds}
\begin{align}
        \dv{x}{z} =& \frac{1}{(z+1)}\Big[ 3(1+w)\Omega - (2+q-x)x-2q-2\Big]\;,\\
        \dv{\Omega}{z} =& \frac{\Omega}{(1+z)}\Big[x +1+3w-2q\Big]\;,\\
        \dv{h}{z} =& \frac{h}{(1+z)}\Big[1+q\Big]\;,\\
        \dv{q}{z} =& \frac{1}{(1+z)}\Big[j-q-2q^2\Big]\;,\\
        \dv{j}{z} =& -\frac{1}{(1+z)}\Big[s+j(2+3q)\Big]\;,\\
        \dv{s}{z} =& -\frac{1}{(1+z)}\Big[s(3+4q)+l\Big]\;,\\
        \label{eq:ds_s}
\end{align}
\end{subequations}
where $h\equiv\frac{H}{H_0}$. The constraint  \eqref{eq:Friedmann_constraint} and an additional relation coming from the definition of the Ricci scalar: \begin{equation}
    y = 1-q\;, \label{eq:ricciconstraint}
\end{equation}
has been used to eliminate $y$ and $v$. 

There is a decoupling structure in the above system. The last four equations, which are purely kinematic in nature, by themselves form a closed dynamical system, provided, of course, one specifies a cosmic evolution as $l=l(h,q,j,s)$\footnote{One can compare this decoupled kinematic dynamical system with the hierarchy of inflationary slow-roll parameters \cite{Kinney:2002qn,Liddle:2003py,Spalinski:2007ef}}. The gravitational dynamics enters the system through the $x$ and $\Omega$ equations.

In principle, one could go on and obtain an infinite number of equations since there is an infinite hierarchy of cosmographic parameters, but, as we discussed in Section \ref{sec:cosmography}, any cosmic evolution that is a solution of some $f(R)$ theory can be expressed as an algebraic constraint involving up to the lerk parameter. Therefore, one does not need to go beyond \eqref{eq:ds_s}. When one tries to reconstruct the $f(R)$ functional form starting from a given cosmological solution $a(t)$, one of course believes that there is an underlying $f(R)$ theory of which $a(t)$ is a solution. The dynamical system formulation based on the cosmographic parameters is in some sense an alternative to the reconstruction method \cite{Nojiri:2009kx, Dunsby:2010wg, Carloni:2010ph}. Consequently, it is completely justified to take into consideration only up to \eqref{eq:ds_s}. This particular formulation allows one to study the phase space of all those $f(R)$ theories that can reproduce a cosmic evolution given by the cosmographic condition $l=l(q,j)$, without explicitly reconstructing the underlying form of $f(R)$.

\begin{figure}
    \centering
    \includegraphics[width=0.5\linewidth]{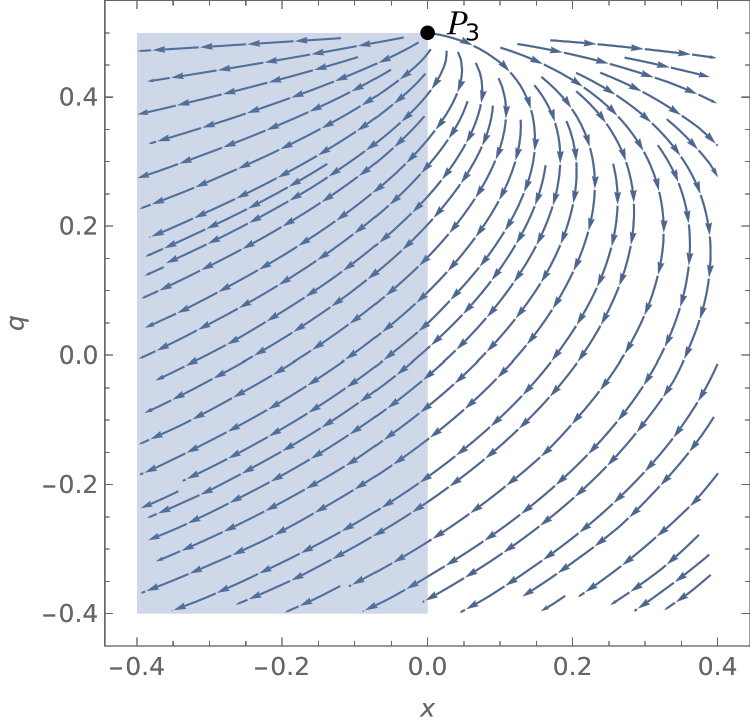}
    \caption{Corrected phase plot from \protect\cite{Chakraborty:2021jku} showing viable phase space at $ x<0$ instead of $x>0$, with $\Omega=1$. Note that as this is a 2 dimensional projection of 3 dimensional system it does not show the evolution with $\Omega$, only a slice of the phase space. An accurate trajectory, taking into account the $\Omega$ evolution, is shown in Fig \ref{fig:parametricplot}.}
    \label{fig:phaseplot_Omega1}
\end{figure}

Additionally, it is noted that using the definition of $\Gamma$ one can write 
\begin{equation}
    \frac{1}{y\Gamma}=\frac{6f''H^{2}}{f'}=\frac{x}{j-q-2} \;.  
\end{equation}
We recall that the absence of ghost and tachyonic instability requires $f'>0,\,f''\geq0$. Assuming that the condition $f'>0$ is met, demanding $f''\geq0$ puts the following constraints on the phase space\footnote{This constraint was stated incorrectly in a previous paper \cite{Chakraborty:2021jku} as $x/(j-q-2) \le 0$. Both the constraint and associated plot (see Figure \ref{fig:phaseplot_Omega1}) have been amended here. }
\begin{equation}\label{v_cond}
    \frac{x}{j-q-2}\geq0\,.    
\end{equation}
It follows that only a portion of the phase space is physically viable. The physically viable region of the phase space, where the theory is free from ghost and tachyonic instability, must \emph{necessarily} satisfy the constraint \eqref{v_cond}. One should be careful, though; this is \emph{not} a sufficient condition. It is certainly possible that a region of the phase space satisfying the condition \eqref{v_cond} is plagued by both ghost and tachyonic instability ($f'<0$ and $f''<0$). To be sure, one must explicitly calculate $f',\,f''$ along a phase trajectory of interest. In a region of the phase space, if the condition \eqref{v_cond} is satisfied and also the dynamical variable $\Omega\equiv\frac{ \rho}{3f'H^2}\geq0$, then one can definitely say that the region is physically viable.

\section{Structure growth}
\subsection{Covariant gauge-invariant density perturbations in $f(R)$ gravity}
In order to analyse the growth of large scale structure we use the 1+3 Covariant and Gauge Invariant approach introduced by Ellis \& Bruni \cite{Ellis:1989jt} and developed for $f(R)$ theories of gravity in \cite{Carloni_PRD77}. Our focus will be on density perturbations, so it will be enough to extract only the scalar parts of the density gradient using a local decomposition \cite{dunsby1992covariant,MBruni_1992} through repeated application of the operator $D_a= \tensor{h}{^b_a} \nabla_b$, where $h_{ab}$ is the projection tensor into the tangent 3-spaces orthogonal to the 4-velocity $u^a$. We thus define the scalar quantities
\begin{align}
    \Delta_m = \frac{a^2}{\rho_m} D^2 \rho_m, \quad Z = 3a^2D^2H, \quad C = a^2 D^2 \tilde{R},\quad
    \mathcal{R} = a^2 D^2 R, \quad \mathfrak{R} = a^2 D^2 \dot{R}\;,
\end{align}
representing the fluctuations in the matter energy density $\rho_m$, the expansion $\Theta = 3H$, the 3-Ricci scalar $\tilde{R}$, the Ricci scalar $R$ and its momentum $\dot{R}$ respectively. Here $D^2 \equiv D_a D^a$.

We then perform a harmonic decomposition using the eigenfunctions of the spatial Laplace-Beltrami operator defined in \cite{Ellis:1989jt,dunsby1992covariant,MBruni_1992}, $D^2Q = -(k^2/a^2) Q^k$, where $k=2\pi a/\lambda$ is the wavenumber and $\dot{Q}=0$, allowing us to expand every first order quantity in the perturbation equations in terms of the harmonics as 
\begin{equation}
    X(t, \textbf{x}) = \sum X^{(k)}(t)Q^{(k)}(\textbf{x})\;.
\end{equation}
$\sum$ stands for a summation over a discrete index or an integration over a continuous one. 

Starting from the equations governing the gravitational dynamics linearised around an FLRW background (see \cite{Carloni_PRD77} for details), one can derive a pair of coupled second order equations describing the evolution of the $k$th mode of density perturbations in $f(R)$ gravity.

These equations can be expressed in terms of derivatives of the function $f(R)$ and time derivatives of the background Ricci scalar \cite{Carloni_PRD77}:

\begin{subequations}
    \begin{align}
    \ddot{\Delta}_m^k& -\left(\left(3w-2\right)H + \frac{f''\dot{R}}{f'}\right)\dot{\Delta}_m^k + \left( w\frac{k^2}{a^2}+\frac{(w-1)\rho_m}{f'}-w\frac{f}{f'} \right)\Delta_m^k \nonumber\\
    =&  
    \left(-\frac{3H(1+w) f''}{f'}\right)\dot{\mathcal{R}}^k+\left(\frac{(1+w)}{2}\left[-1+\frac{f f''}{ f'^2}-\frac{2\rho_m f''}{f'^2}-\frac{2f''}{f'}\frac{k^2}{a^2}+6H\dot{R}\left(\frac{f''}{f'}\right)^2 - \frac{6H f''' \dot{R}}{f'}\right]\right)\mathcal{R}^k,
\end{align}

\begin{align}
\ddot{\mathcal{R}}^k +&\left(3H+\frac{2f''' \dot{R}}{f''}\right)\dot{\mathcal{R}}^k+\left(-\frac{R}{3}+\frac{k^2}{a^2}+\frac{f'}{3f''}+\frac{3H f''' \dot{R}}{f''}+\frac{f^{(4)}\dot{R}^2}{f''}+\frac{f''' \ddot{R}}{f''}\right)\mathcal{R}^k =  \left(\frac{(1-w)\dot{R}}{(1+w)}\right)\dot{\Delta}_m^k \nonumber\\
&- \left(\frac{(3w-1)\rho_m}{3f''} + \frac{2w}{w+1}\left[ \ddot{R}+\dot{R}\left(3H+ \frac{f''' \dot{R}}{f''}\right)\right]\right)\Delta_m^k  .
\end{align}
\label{eq:deppertsys}
\end{subequations}

Since the system \eqref{eq:deppertsys} is very complicated, we employ the quasi-static (QS) approximation, as is typically done in the literature \cite{Silvestri2013,Noller2014}. This approximation consists of two separate premises, the first being that temporal fluctuations of $\mathcal{R}^k$ are suppressed in comparison to the perturbation itself and can thus be neglected, $|\dot{X}| \lesssim H |X| $ for $X=\mathcal{R}^k,\dot{\mathcal{R}}^k$. This assumption alone allows us to simplify the system to one equation for $\Delta_m$:

    \begin{align}
        \ddot{\Delta}&+   \left[\frac{3 (1-w) f'' \dot{R} \left(- f'' \left(6 H f'' \dot{R}+f-2 \rho_m\right)+ f'^2+2 \frac{k^2}{a^2} f' f''\right)}{2 f'^2 \left(f'' \left(3 \frac{k^2}{a^2}- R\right)+ f'+3  \left(f^{(4)} \dot{R}^2+3 f^{(3)} H \dot{R}+f^{(3)} \ddot{R}\right)\right)}-\frac{f'' \dot{R}}{f'}+H (2-3 w)\right]\dot{\Delta}\nonumber\\
        +\Bigg[&\frac{\left(6 w \left(f^{(3)} \dot{R}^2+3 H f'' \dot{R}+f'' \ddot{R}\right)+\rho_m  \left(3 w^2+2 w-1\right)\right) \left( f'' \left(6 H f'' \dot{R}+f-2 \rho_m \right)- f'^2-2 \frac{k^2}{a^2} f' f''\right)}{2 f'^2 \left(f'' \left(3 \frac{k^2}{a^2}- R\right)+ f'+3  \left(f^{(4)} \dot{R}^2+3 f^{(3)} H \dot{R}+f^{(3)} \ddot{R}\right)\right)} +\frac{k^2 w}{a^2}-\frac{w f}{f'}+\frac{\rho_m  (w-1)}{f'}\Bigg] \Delta =0 \nonumber\\
        \label{eq:semiQSgeneralnon}
    \end{align}

The second QS premise involves the sub-horizon assumption $ k \gg aH $, which leads to the approximation $k^2/(aH)^2 \gg f^{(n)}/f'$ for $n = 2,3,4$. This simplifies the system to the recognisable form
\begin{align}
&\ddot{\Delta}_m^k+(2-3w)H\dot{\Delta}_m^k=\left[w\left(\frac{f}{f'}-\frac{k^2}{a^2}\right)+\frac{\rho_m}{2f'}\left(\frac{(3w^2+1) +2(3w^2-w+2)\frac{f''}{f'} \frac{k^2}{a^2} }{1+3\frac{f''}{f'}\frac{k^2}{a^2}   }\right)\right]\Delta_m^k ,
      \label{eq:QSgeneraleqnon}
\end{align}
which mirrors the quasi-static approximation for the density contrast $\delta = \delta \rho_m/\rho_m$ in the Bardeen perturbation formalism \cite{tsuj2007} when $w=0$:
\begin{equation}
    \ddot{\delta}+(2-3w)H\dot{\delta}-\frac{\rho_m}{2f'}\left(\frac{1 +4\frac{f''}{f'} \frac{k^2}{a^2} }{1+3\frac{f''}{f'}\frac{k^2}{a^2}   }\right)\delta =0.
\end{equation}

The validity of this approximation in the synchronous and conformal gauges has been extensively studied \cite{Hojjati2012, Noller2014, Sawicki2015} and performs well in certain model-specific, sub-horizon contexts, particularly for models with background evolution very close to $\Lambda$CDM \cite{de_la_Cruz_Dombriz_2008}. However, concern remains whether neglecting both higher order $f$ derivatives and time derivatives of the curvature perturbation is too aggressive   \cite{Bean_2007} for models deviating from $\Lambda$CDM. For this reason, it is better to consider the QS method both including and excluding the higher order $f$ derivatives. These are referred to as the semi and full quasi-static approximations respectively from here on.

All of these approaches retains explicit dependence on $f(R)$ and are thus only applicable once a particular model of $f(R)$ is chosen. Additionally, to study the perturbations in relation to results obtained from dynamical systems analyses, it is necessary to express all quantities in the coefficients in terms of the chosen dynamical systems variables, including up to fourth order derivatives of $f(R)$ in the case of the full covariant system. This adds a further restriction on the particular models of $f(R)$ which can be studied 
with this approach. 

\subsection{Model-independent formulation}
As in the background system, this model-dependence can be avoided by employing the cosmographic approach. Time derivatives of the Ricci scalar \eqref{eq:ricciscalardef} can be expressed in terms of the cosmographic parameters \eqref{eq:CP}:

\begin{subequations}
\begin{align}
     R &= 6H^2\left(\frac{\dot{H}}{H^2} + 2\right) = 6H^2(1-q),  \\
    \dot{R} &= 6H^3\left(\frac{\ddot{H}}{H^3}+4\frac{\dot{H}}{H^2}\right) = 6H^3\left(j-q-2\right), \\
    \ddot{R} &=  6H^4\left(s+q^2+8q+6\right) ,\\
    \dddot{R} &= 6H^5\left(l-s-2qj-8j-18q^2-48q-24\right).
    \end{align}
    \label{eq:ricciscalarCP}
\end{subequations}

Then, taking time derivatives of increasing order of the Friedmann equation \eqref{eq:fried} yields higher order $R$ derivatives of $f$ which can then be expressed exclusively in terms of the dynamical systems variables and $R$ derivatives and thus the cosmographic parameters using \eqref{eq:ricciscalarCP}. For example, taking the first time derivative of \eqref{eq:fried} allows us to express $f^{(3)}$ as:
    \begin{align}
        f''' = \frac{\dot{f}''}{\dot{R}} &= \frac{f' H^2}{\dot{R}^2}\left(  x -3(1+w)\Omega +2  (q+1) -\frac{\ddot{R} x }{H \dot{R}} \right),\nonumber\\
        &= \frac{f' H^2}{\dot{R}^2}\left(  x -3(1+w)\Omega +2  (q+1) -\frac{(s+q^2+8q+6) x }{(j-q-2)} \right).
    \end{align}

Continuing in this way, we can express all coefficients in \eqref{eq:deppertsys} in terms of the dynamical system variables and the cosmographic parameters, resulting in a pair of second order equations describing the evolution of the $k$th mode of density perturbations in $f(R)$ gravity in a totally model-independent way. In terms of the redshift $z$, these are


\begin{subequations}\label{eq:covperteqsredshift}
    \begin{align}
 h^2 (1+z)^2 &\frac{d^2\Delta_m^k}{dz^2}+ h^2 (1+z) (q+x+3w) \frac{d\Delta_m^k }{dz}+h^2\left(w\left(6(q+x) +\frac{\hat{k}^2}{h^2}(1+z)^2 \right)-3(1+w)\Omega \right)\Delta_m^k =\nonumber\\
 & (1+z)\left(\frac{(1+w)  x}{2(j-q-2)} \right)\frac{d\hat{\mathcal{R}}^k}{dz}-\frac{(1+w)}{6(j-q-2)}\left(\frac{\hat{k}^2 }{h^2 }(1+z)^2x+3 j+3 q (x-1)-6\right)\hat{\mathcal{R}}^k,\label{eq:generaldeltaequation}
\end{align}
\begin{align}
&\frac{d^2\hat{\mathcal{R}}^k}{dz^2}+ \left(\frac{ \left(q ((j+18) x-4 j-6 (1+w)\Omega +12)+2 x (-2 j+s+10)+2 (j-2) (3(1+w) \Omega -2)+q^2 (x+4)\right)}{x (1+z) (j-q-2)}+\frac{1}{(1+z)}\right)\frac{d\hat{\mathcal{R}}^k}{dz} \nonumber\\
&+  \frac{1}{(1+z)^2}\Bigg[\frac{\hat{k}^2 }{h^2}(1+z)^2+\frac{1}{x (j-q-2)^2} \bigg(x \left(11 j^2-j (l+4 s+34)+2 (l+(s+4) (s+12))+ q (j (5 j-22)+l+36 s+206)\right)\nonumber\\
&-6 q(w+1) \Omega  (3 j (w-3)+s-6 w+24)-4 j q(s+14)+12 qs+136q+q^3 (-4 j-6 (w+1) \Omega +22 x+44)+2 q^4 (x+2)\nonumber\\
&+q^2 (x (5 j+4 s+133)+4 (-9 j+s+32)+3 (w+1) \Omega  (2 j+3 (w-7)))+ 3 (w+1) \Omega  (3 (j-2) w-j+2 s+14)(j-2)\nonumber\\
&-4 (s+6)(j-2)\bigg)\Bigg] \hat{\mathcal{R}}^k = \left(\frac{6 h^2 (j-q-2)(4w(1+q+2x)-(1+w)(1-3w) \Omega ) }{(1+z)^2(1+w)x}\right) \Delta_m^k
-\left(\frac{6 h^2   (1-w)(j-q-2)}{(1+z)(1+w)}\right) \frac{d\Delta_m^k }{dz}.
\label{eq:generalRequation}
\end{align}
\end{subequations}

where the dimensionless quantities $\hat{\mathcal{R}}^k = \mathcal{R}^k/H_0^2$, $\hat{k}=k/a_0 H_0$ and $h = \frac{H}{H_0}$ have been used. 





This same approach can be applied to the semi and full quasi-static perturbation equations \eqref{eq:semiQSgeneralnon} and \eqref{eq:QSgeneraleqnon}, allowing us to express these in a model-independent way. In terms of redshift, the corresponding perturbation equations are respectively

    \begin{align}
&(1+z)^2\frac{d^2\Delta_m^k}{dz^2} + (1+z)\left(q+x+3w + \frac{(w-1)(3j-6+3q(x-1)+\frac{\hat{k}^2}{h^2}(1+z)^2 x)}{D}\right)\frac{d\Delta_m^k}{dz} + \left(w \left[\frac{\hat{k}^2 }{h^2}(1+z)^2+6 q+6 x\right]\right.\nonumber\\
&\left.-3(1+w) \Omega-\frac{\Big(4 w (q+2 x+1)-(w+1) (3 w+1) \Omega \Big) \left(\frac{\hat{k}^2  }{h^2}(1+z)^2 x+3 j+3 q (x-1)-6\right)}{xD}\right)\Delta_m^k = 0
\label{eq:semiQSgeneraleq}
\end{align}
where 
\begin{align*}
D =& \frac{\hat{k}^2}{h^2} (1 + z)^2+ \frac{1}{(j-q-2)^2 x}\Bigg[x(11 j^2 - j (34 + l + 4 s) + 
      2 (l + (4 + s) (12 + s)))  + 2 q^4 (2 + x) \\
      &+ 
   q^3 (44 - 4 j + 22 x - 6 (1 + w) \Omega) + 
   q^2 (4 (32 - 9 j + s) + (133 + 5 j + 4 s) x + 
      3 (2 j + 3 (-7 + w)) (1 + w) \Omega) \\
      &+ 
   q (136 + 12 s - 
      4 j (14 + s) + x(206 + j (-22 + 5 j) + l + 36 s)  - 
      6 (24 + s + 3 j (-3 + w) - 6 w) (1 + w) \Omega) \\
      &+ (j-2) (-4 (6 + s) + 
      3 (1 + w) (14 - j + 2 s + 3 (j-2) w) \Omega)\Bigg],
\end{align*}
and 
\begin{align}
    (1+z)^2 \frac{d^2\Delta_m^k}{dz^2}+(1+z)(q+3w) \frac{d\Delta_m^k}{dz}=\left[\Omega\left(\frac{3\frac{(3w^2+4w+1)(j-q-2)}{(1+z)^2} +(3w^2+5w+2)x\frac{\hat{k}^2}{h^2} }{2\frac{(j-q-2)}{(1+z)^2}+x\frac{\hat{k}^2}{h^2} }\right) -w\left(6(x+q)+\frac{\hat{k}^2}{h^2}(1+z)^2\right)\right]\Delta_m^k .
      \label{eq:QSgeneraleq}
\end{align}

In the GR limit $f(R) = R$, we see that system \eqref{eq:deppertsys} (and thus each model-independent  method \eqref{eq:covperteqsredshift} - \eqref{eq:QSgeneraleq} derived therefrom) reduces to 
\begin{subequations}
    \begin{align}
    \ddot{\Delta}_m^k -\left(3w-2\right)H \dot{\Delta}_m^k +\left( w\frac{k^2}{a^2}+\left(\frac{-1+2w-3w^2}{2} \right)\rho_m\right)\Delta_m^k
     = 0
\end{align}
\begin{align}
\mathcal{R}^k =  
 (1-3w)\rho_m\Delta_m^k  ,
\end{align}
\end{subequations}
which gives the well-known equation 
\begin{equation}
    \ddot{\Delta}_m^k +2H \dot{\Delta}_m^k 
    -\frac{1}{2} \rho_m\Delta_m^k
     = 0\;,
\end{equation}
for the dust equation of state $w=0$. Obvious differences between the $f(R)$ perturbation equations and those in GR include their fourth-order nature and, most importantly, the scale-dependence of density perturbations for any equation of state. If detected, this would point toward clear deviation from $\Lambda$CDM.
\subsection{Growth function and growth index for matter perturbations}
Additional methods of quantifying the growth of the matter perturbations may provide another way to discriminate between modified gravity models and GR DE models ($\Lambda$CDM and otherwise). These include the growth function $S$ and growth index $\gamma$, defined by
\begin{equation}
    S \equiv \frac{d\ln \delta}{d\ln a} = \Omega_m^{\enskip\gamma},
\end{equation}
where $\Omega_m $ is the standard matter density parameter,
\begin{equation}\label{eq:densityparam}
    \Omega_m  \equiv \frac{\rho_m}{3H^2} = f'\Omega,
\end{equation}
where we have used $\Omega_m(z=0) = 0.3$. In the covariant formalism, in terms of redshift, this is equivalently 
\begin{equation}\label{eq:growthfunction}
    S \equiv -(1+z)\frac{d\ln \Delta}{dz},
\end{equation}
from which the growth index $\gamma$ can be calculated as 
\begin{equation}\label{eq:growthindexparam}
    \gamma = \frac{\ln(-(1+z)\frac{d\ln \Delta}{dz})}{\ln \Omega_m}.
\end{equation}

When it was introduced, the growth index was taken to be constant at low redshift \cite{Peebles1980,Lightman1990}. While there is clear time dependence, this approximation proves appropriate for models within GR with constant or smoothly varying equations of state which display $|\gamma'_0 \equiv \frac{d\gamma}{dz}(z=0)| \lesssim 0.02$ \cite{POLARSKI2008439}, with $\Lambda$CDM specifically having $\gamma_0 \approx 0.55$ with very little variation for $z<0.5$ \cite{growthrateconstraints}. Deviation from these values, significant time variation on small redshifts and scale-dependence of $\gamma_0$ and $\gamma_0'$ could all provide strong evidence of a modification to GR. 

\section{The cosmographic condition $j=1$}

Before moving on to the main analysis of this paper, we wish to comment on the accuracy and specificity of the $\Lambda$CDM cosmographic condition. 


It is well known that the $\Lambda$CDM model gives rise to a cosmic evolution that corresponds to the cosmographic condition $j=1$ \cite{Dunajski:2008tg}. While model independent estimates of the jerk parameter  relying on cosmic chronometers and supernovae data are consistent with the $\Lambda$CDM condition $j \approx 1$  within $3\sigma$ \cite{Mehrabi:2021cob,Mukherjee:2020ytg}, the reconstructed cosmographic quantities are in good agreement with those of $\Lambda$CDM when some specific parametric deviations from the concordance values are assumed in the analysis and then constrained by the data fitting \cite{Zhai:2013fxa,Amirhashchi:2018vmy,Mukherjee:2016trt,Mukherjee:2016shl}. However, we think it is necessary at this point to clarify the notion of cosmographic (one may also say kinematic) degeneracy with $\Lambda$CDM. 

The redshift evolution of the jerk parameter is (see e.g. \cite[Eq.(4)]{Mehrabi:2021cob}): 
\begin{equation}
j(z)=\frac{H(z) (1+z)^2 \frac{d^2 H(z)}{dz^2}+\left[(1+z) \frac{d H(z)}{dz}-H(z)  \right]^2}{H^2(z)}\,.
\end{equation}
Solving $j(z)=1$ we obtain \cite{Zhai:2013fxa,Amirhashchi:2018vmy,Mukherjee:2016trt,Mukherjee:2016shl} 
\begin{equation}
 \label{HC1C2}
h^{2}(z)  \equiv \frac{H^2(z)}{H_0^2} = {\mathcal C}_1 (1+z)^3 + {\mathcal C}_2,   
\end{equation}
where ${\mathcal C}_1$ and ${\mathcal C}_2$ are two arbitrary constants satisfying ${\mathcal C}_1 + {\mathcal C}_2=1$. In general, for nonzero $\mathcal{C}_1$ and $\mathcal{C}_2$, the family of solutions given by \eqref{HC1C2} specifies a \emph{$\Lambda$CDM-like cosmic evolution history}, in the sense that it has two clearly defined asymptotic limits: an {\it effective} CDM limit at asymptotic past $z\rightarrow\infty$ (or $a\rightarrow0$) where the cosmic evolution goes like $h^2 \sim (1+z)^3$, and an {\it effective} $\Lambda$-limit at asymptotic future $z\rightarrow-1$ (or $a\rightarrow\infty$) where the evolution goes like $h^2 \sim constant$. The particular solution of the above family, specified by ${\mathcal C}_1 = \Omega_{m0}$ and ${\mathcal C}_2 = \Omega_{\Lambda 0}$, corresponds to the particular $\Lambda$CDM cosmic history that our universe is going through. 

It should be appreciated, though, that the condition $j(z)=1$ by itself, although implying a $\Lambda$CDM-like cosmic evolution history, may not necessarily single out the $\Lambda$CDM model. This is best understood in terms of the statefinder parameters $\{r,s\}$ introduced in \cite{Sahni:2002fz}. The first statefinder parameter $r$ is precisely the cosmographic jerk parameter $j$, but this alone does \emph{not} specify a model. Specifying a model requires specification of both the statefinder parameters $\{r,s\}$. If one puts $r=1$ in \cite[Eq.3]{Sahni:2002fz}, one sees that the equation may be satisfied in general by other models as well, including dynamical dark energy models and models with a dark sector interaction \footnote{For a phase space analysis of some simple dark sector interaction models satisfying the cosmographic condition $j=1$, see \cite{Chakraborty:2022evc}. For the cases considered in that paper, it was found that achieving the condition $j=1$, although possible in principle, is still problematic (e.g. the set of acceptable solutions are of measure zero), leading the authors to conclude that within GR and respecting the FLRW symmetry, $\Lambda$CDM is still the natural choice for $j=1$.}. However, whatever the inherent model is, it must give rise to an evolution of the form of \eqref{HC1C2}, having clearly defined CDM-limit and $\Lambda$-limit with non-zero integration constants $\mathcal{C}_{1,2}$, and therefore corresponding to a $\Lambda$CDM-like evolution. 

Also, note that the condition $j=1$ by itself does not provide either any physical interpretation for the two integration constants ${\mathcal C}_1$ and ${\mathcal C}_2$, nor any route of fixing them. A physical meaning of these two arbitrary constants can only be provided with respect to a particular model in mind. For example, in the $\Lambda$CDM model, dark matter and dark energy evolve independently, and their energy densities redshift as $(1+z)^3$ and $(1+z)^0$ respectively. Then and only then could we identify ${\mathcal C}_1$ with $\Omega_{m0}$ and ${\mathcal C}_2$ with $\Omega_{\Lambda 0}$. We could not have concluded the same, for example, had the equation of state of dark energy been dynamical or had there been an interaction in the dark sector. In other words, the functional relationships ${\mathcal C}_1 = {\mathcal C}_1 (\Omega_{\rm DE0},\Omega_{m0})$ and ${\mathcal C}_2 = {\mathcal C}_2 (\Omega_{\rm DE0}, \Omega_{m0})$ are sensitive to the choice of the model \cite{Mehrabi:2021cob} \footnote{For example, if one takes the dark energy to be modeled by a fluid with $w_{\rm DE}\neq-1$ and still impose kinematic degeneracy with $\Lambda$CDM by imposing the condition $j(z)=1$, one obtains $\{\mathcal{C}_1,\mathcal{C}_2\}=\{1+w_{\rm DE}-w_{\rm DE}\Omega_{m0},-w_{\rm DE}\Omega_{\rm DE0}\}$. See \cite[Eq.17]{Chakraborty:2022evc}}. 

In summary, the condition $j=1$ is sufficient for reproducing the kinematics of $\Lambda$CDM (i.e. similar cosmic evolution), but \emph{not} its dynamics (i.e. the model itself). From the astrophysical perspective, this remark means that when testing the cosmic history (\ref{HC1C2}) with respect to cosmographic data, one may very well get ${\mathcal C}_1=0.3$ and ${\mathcal C}_2=0.7$, which however should not be taken naively to represent the abundances of dark matter and dark energy at the present day \cite{Zhai:2013fxa}. 
\section{The condition $j=1$ in the context of $f(R)$ gravity} \label{sec:j=1 LCDM}
In the previous section we have discussed the cosmographic kinematic condition $j=1$ from a generic context. Let us focus now on the model under consideration in this paper, i.e., a late time $f(R)$ model with no dark-sector interaction and see what the condition $j=1$ implies in this case. The cosmological field equations \eqref{eq:fried},\eqref{eq:raychaud} can be rearranged in the following form
\begin{subequations}
    \begin{eqnarray}\label{field_eqs_new}
        3H^{2} = \rho_{tot} = \rho + \rho_{\rm DE}\,,\label{eq:fried_new}\\
        -\left(2\dot{H} + 3H^{2}\right) = P_{tot} = P_{\rm DE}\,,\label{eq:Raychoudhuri_new}
    \end{eqnarray}     
\end{subequations}
where $\rho$ is the energy density of the non-relativistic matter, and we have defined the energy density and pressure of the dark matter as follows
\begin{subequations}
    \begin{align}
         \rho_{DE}&=\frac{1}{2}(Rf'-f)-3H\dot{F}+3H^{2}(1-f')\,,\label{eq:DEed}\\
         P_{DE}&=\ddot{f'}+2H\dot{f'}-\frac{1}{2}(RF-f)-(2\dot{H}+3H^2)(1-f')\,. \label{eq:DEp}
    \end{align}
\end{subequations}
The dark energy equation of state $\omega_{\rm DE} = \frac{P_{\rm DE}}{\rho_{\rm DE}}$ is 
\begin{equation}
    \omega_{\rm DE} = \frac{\ddot{f'}+2H\dot{f'}-\frac{1}{2}(Rf'-f)-(2\dot{H}+3H^2)(1-f')}{\frac{1}{2}(Rf'-f)-3H\dot{f'}+3H^{2}(1-f')}\,.
\end{equation}
Using the field equations \eqref{field_eqs_new} the above can be written as
\begin{equation}\label{DE_eos}
    \omega_{\rm DE} = - \frac{2\dot{H}+3H^{2}}{3H^{2}-\rho} = \frac{H^{2}-\frac{R}{3}}{3H^{2}-\rho}\,.
\end{equation}
In this form, the non-relativistic matter and the dark energy is separately conserved
\begin{subequations}\label{eq:cons}
    \begin{eqnarray}
        && \dot\rho + 3H\rho = 0\,.\\
        && \dot\rho_{\rm DE} + 3H\rho_{\rm DE}(1+\omega_{\rm DE}) = 0\,.
    \end{eqnarray}
\end{subequations}
Dividing the numerator and denominator by $H^2$ and utilizing the definitions of the dynamical variables and the cosmographic parameters, we arrive at
\begin{equation}\label{eq:wDE}
 \omega_{\rm DE} = \frac{2q-1}{3-3\Omega_m}\,,  
\end{equation}
where in the last step we have utilized the relation \eqref{eq:ricciconstraint} and $\Omega_m$ is the usual matter density parameter defined in \eqref{eq:densityparam} which obeys the dynamical equation
\begin{equation}\label{eq:dyn_Omegatilde}
    \frac{d\Omega_m}{d\tau} = -\Omega_m (1-2q)\,.
\end{equation}
Imposing $w_{\rm DE}=-1$ one obtains
\begin{equation}
    2q - 3\Omega_m=-2\,.
\end{equation}
Taking a $\tau$-derivative, utilizing \eqref{eq:dyn_Omegatilde} and the first equation of \eqref{CP_rel} and after performing some rather straightforward steps, one can arrive at the condition $j=1$. This clearly shows that if the curvature part of the total energy density (i.e. $\rho_{\rm DE}$) behaves like a cosmological constant ($w_{\rm DE}=-1$), it clearly gives rise to a cosmology with $j(z)=1$, as expected.

Consider, now, a dynamical dark energy model with $w_{\rm}=-\alpha(\tau)$. Imposing this in \eqref{eq:wDE} and taking a $\tau$-derivative, after some straightforward steps one arrives at 
\begin{equation}
    j-1 = \left(q-\frac{1}{2}\right)\left(3-3\alpha-\frac{d\alpha/d\tau}{\alpha}\right)\,.
\end{equation}
One possibility to get $j=1$, i.e. kinematic degeneracy with $\Lambda$CDM is $q=\frac{1}{2}$, but it cannot be true for the entire evolution history. Another possibility is $\alpha=1$, which takes us back to the case $w_{\rm DE}=-1$. The third possibility is that $\alpha(\tau)$ satisfies the equation
\begin{equation}
    \frac{d\alpha}{d\tau} = 3\alpha(1-\alpha)\,,
\end{equation}
which, upon solving, gives 
\begin{equation}
    \alpha(\tau) = \frac{\alpha_0 e^{3\tau}}{\alpha _0 e^{3\tau} + (1-\alpha_0)}\,,
\end{equation}
with $\alpha_0=\alpha(0)$. This implies a dark energy equation of state that can be expressed as 
\begin{equation}\label{eq:DDE}
    w_{\rm DE}(\tau) = -\frac{w_{\rm DE}(0)e^{3\tau}}{w_{\rm DE}(0)e^{3\tau} - (1+w_{\rm DE}(0))}\,.
\end{equation}
Except for the special case $w_{\rm DE}(0)=-1$, the above corresponds, in general, to a dynamical dark energy, which is still compatible with the cosmographic condition $j(z)=1$. 

One can, in fact, go one step further and calculate from the dark energy conservation equation of \eqref{eq:cons} that
\begin{equation}
     \rho_{\rm DE} \propto \frac{1}{\alpha} = \tilde{\mathcal{C}}_1e^{3\tau}+\tilde{\mathcal{C}}_2 = \tilde{\mathcal{C}}_1(1+z)^3+\tilde{\mathcal{C}}_2\,.
\end{equation}
so that from \eqref{eq:fried_new} one can write 
\begin{equation}
    h^2(z) = \left(\Omega_{m0}+\frac{\tilde{\mathcal{C}}_1}{3H_0^2}\right)(1+z)^3 + \frac{\tilde{\mathcal{C}}_2}{3H_0^2}\,. 
\end{equation}
In other words, even though \eqref{eq:DDE} may represent in general a dynamical dark energy, the cosmic evolution is still $\Lambda$CDM-like, in the sense that there is a clearly defined effective CDM limit at asymptotic past and a clearly defined effective $\Lambda$-limit at asymptotic future. Since we will base our subsequent analysis on the condition $j=1$, such models are also included in our picture. 
\section{Application to $\Lambda$CDM Cosmology}
In this section, we study the background cosmic evolution and growth of matter perturbations for an $f(R)$ theory mimicking a $\Lambda$CDM cosmic history. As discussed in Section \ref{sec:j=1 LCDM}, this can be achieved in the cosmographic formulation by fixing $j=1$. This automatically fixes the snap parameter $s$ which in turn fixes the lerk parameter $l$:
\begin{align}
    s &= \frac{dj}{d\tau} - j(2 + 3q) = -(2+3q),\\
    l &= \frac{ds}{d\tau} - s(3+4q) = 6q^2 + 14q + 9.
\end{align}

Using these expressions, both the background and perturbation equations can be written in terms of the dynamical variables and the deceleration parameter $q$. 
\subsection{Background evolution for $j=1$}
Fixing $j=1$ closes the dynamical system \eqref{eq:ds} which, for a dust equation of state ($w=0$), gives
\begin{subequations}
\begin{align}
        \dv{x}{z} &= \frac{1}{(z+1)}\Big[ 3\Omega - (2+q-x)x-2q-2\Big],\\
        \dv{\Omega}{z} &= \frac{\Omega}{(1+z)}\Big[x +1-2q\Big],\\
        \dv{h}{z} &= \frac{h}{(1+z)}\Big[1+q\Big],\\
        \dv{q}{z} &= \frac{1}{(1+z)}\Big[1-q-2q^2\Big].
\end{align}
\end{subequations}

\begin{figure}
    \centering
    \includegraphics[width=0.5\linewidth]{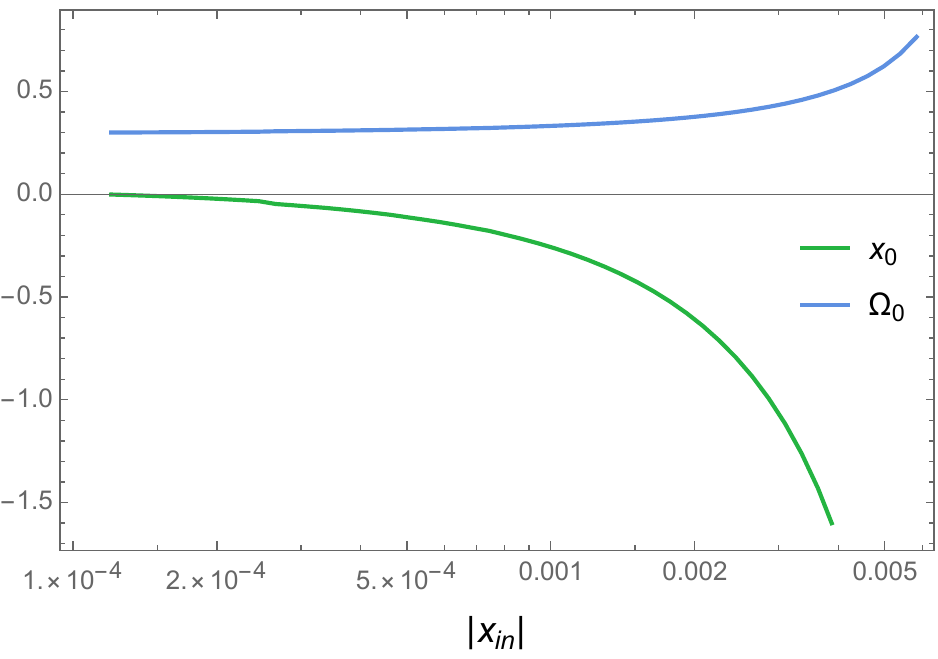}
    \caption{Dependence of the present day values $x_0$, $\Omega_0$ of the variables $x$ and $\Omega$ on the initial condition $|x_{in}|$. }
    \label{fig:todaydependence}
\end{figure}

\begin{figure}
    \centering
    \begin{subfigure}[b]{0.5\linewidth}
        \includegraphics[width=\linewidth]{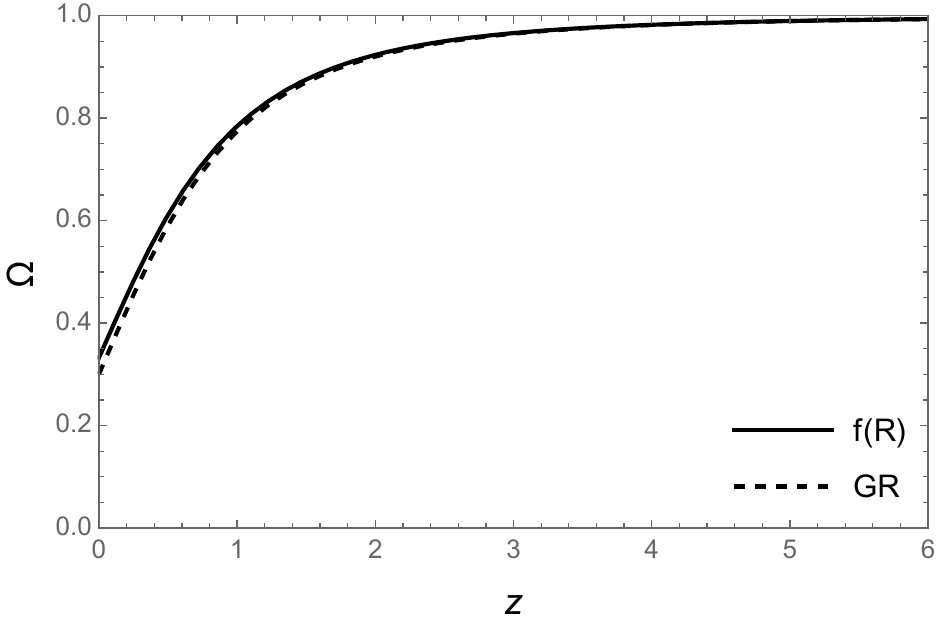}
        \caption{}
        \label{fig:omegabackground}
    \end{subfigure}
    \begin{subfigure}[b]{0.5\linewidth}
        \includegraphics[width=\linewidth]{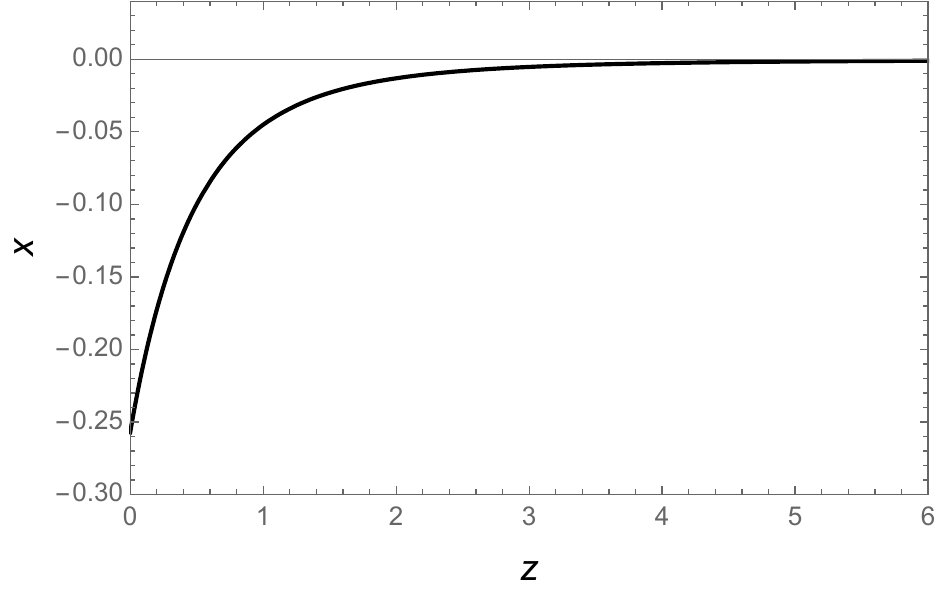}
        \caption{}
        \label{fig:xbackground}
    \end{subfigure}
    \begin{subfigure}[b]{0.5\linewidth}
        \includegraphics[width=\linewidth]{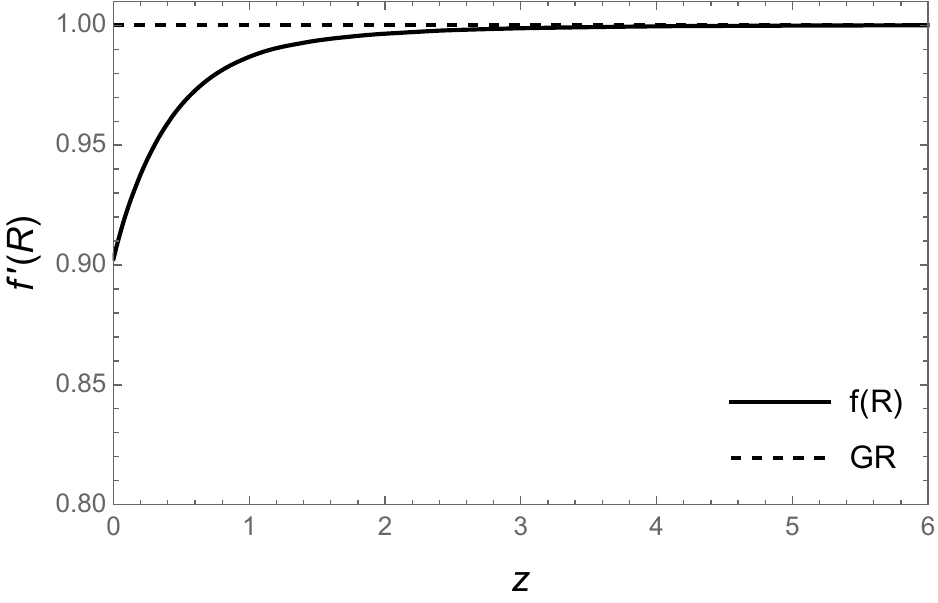}
        \caption{}
         \label{fig:fprime}
    \end{subfigure}
    \caption{(a)-(b) Evolution of the background variables $\Omega$ and $x$ for the trajectory with initial conditions \eqref{eq:init_cond}. (c) Evolution of $f'(R)$, obtained from $f'(R) = \Omega_m/\Omega$.}
    \label{fig:background}
\end{figure}

We set the initial conditions for the background variables deep in the matter-dominated era\footnote{The dynamical system is extremely sensitive to initial conditions, often up to the fourth decimal place. Therefore, setting initial conditions at $z=0$, if not set very precisely, can result in an incorrect evolution that does not feature a matter-dominated era but rather a scalaron-dominated one for $z\gtrsim 2$. Setting initial conditions arbitrarily deep in the matter-dominated era then ensures an evolution that complies with $\Lambda$CDM.}. Initial conditions for $\Omega_{in}$ and $ q_{in}$ are chosen to match with $\Lambda$CDM values, assuming $\Omega^{\Lambda\text{CDM}}_{0} = \Omega_m(z=0) = 0.3$ and $q_0 = q(z=0) = -0.55$. For the $j=1$ case, the viability condition \eqref{v_cond} gives an upper bound on $x$ such that 
\begin{equation}
    \frac{x}{1+q} \leq 0.
\end{equation}
Since $q > -1$, this implies $x \leq 0$ at all redshifts. As $x$ encodes the deviation from GR ($x \rightarrow 0$ being the GR limit), $|x|$ should be small in the matter-dominated era. Selecting exactly how small the initial value of $x_{in}$ should be is a matter of delicate balance. As shown in Fig \ref{fig:todaydependence}, for very small values of $|x_{in}|\sim 10^{-5}$, the variable $x$ evolves to a value $x_0>0$, violating the viability condition \eqref{v_cond}. Additionally, larger values of $|x_{in}|$ lead to larger values of $|x_0|$, as the deviation from GR grows faster. This in turn affects the value of $\Omega_0$, driving it  further from the observed value of the matter abundance parameter $\Omega_{m0}=0.3$. To achieve this balance, we choose a value of $x_{in}$ which ensures $x<0$ throughout its evolution up to $z=0$ and results in a value of $\Omega_0 \approx 0.3$ . 
The selected initial conditions are then
\begin{equation}
    x_{in}= -0.001 ,\quad \Omega_{in} = 0.993243 ,\quad  q_{in}=0.489865,
    \label{eq:init_cond}
\end{equation}
which give the background evolution for the $f(R)$ variables $x$ and $\Omega$ shown in Fig \ref{fig:background}.

\begin{figure}
    \centering
    \includegraphics[width=0.5\linewidth]{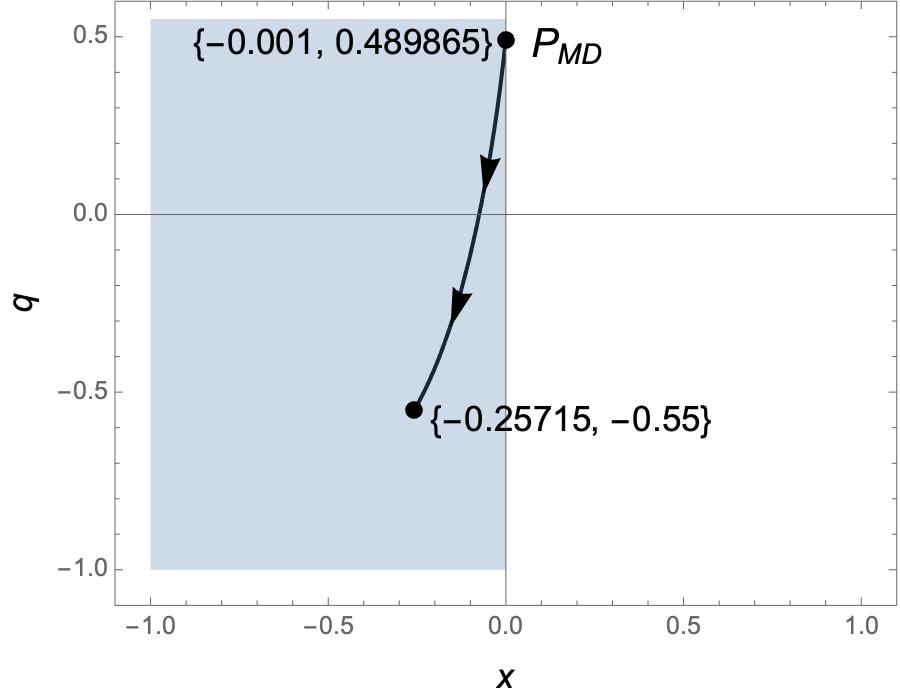}
    \caption{Parametric plot of specific trajectory with the initial conditions \ref{eq:init_cond}, starting in the matter dominated era near the saddle point $P_{MD}$ and evolving to the present values $x_0 = -0.25715$, $q_0 = -0.55$. }
    \label{fig:parametricplot}
\end{figure}

\begin{figure}
    \centering
    \begin{subfigure}[b]{0.5\linewidth}
        \includegraphics[width=\linewidth]{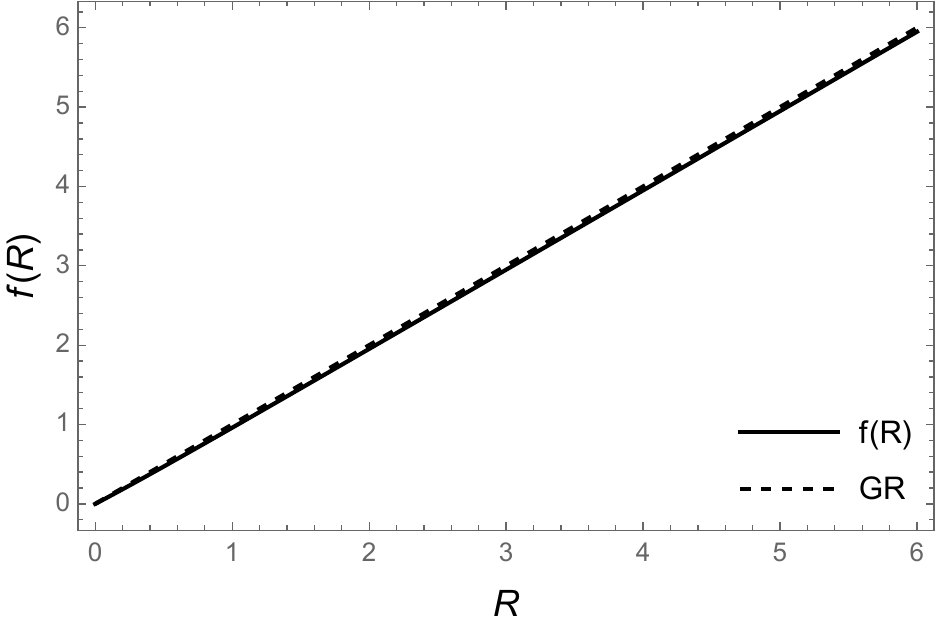}
        \caption{}
        \label{fig:freconstructa}
    \end{subfigure}
    \begin{subfigure}[b]{0.53\linewidth}
        \includegraphics[width=\linewidth]{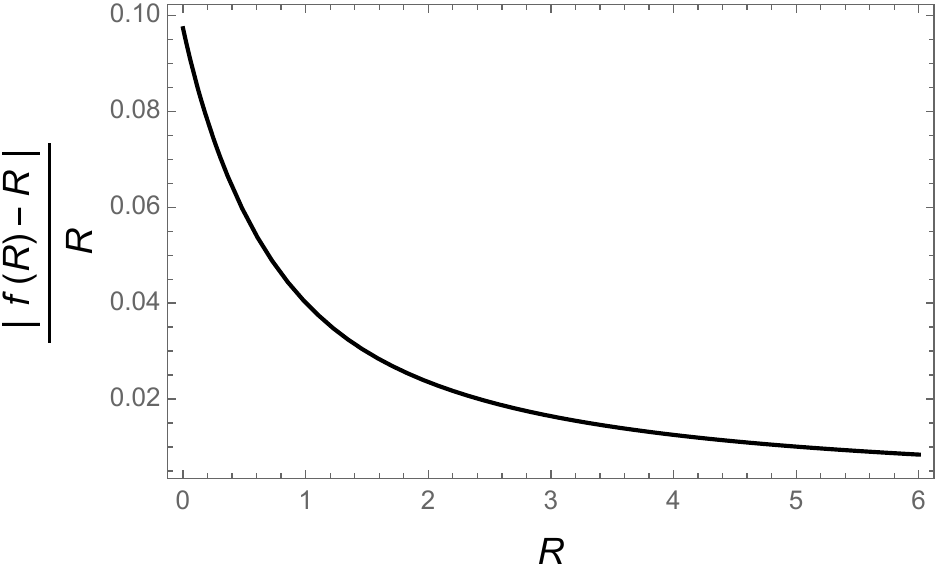}
        \caption{}
        \label{fig:fdiff}
    \end{subfigure}
    \caption{(a) Numerically reconstructed $\Lambda$CDM-mimicking $f(R)$ obtained from integration of $f'(R)$ shown in Fig \ref{fig:fprime}. (b) Percentage deviation of the reconstructed $f(R)$ from GR.}
    \label{fig:freconstruct}
\end{figure}

This trajectory meets all the viability conditions since $x < 0$ for all time and $f' > 0$ as shown in Fig \ref{fig:fprime}. Since condition \eqref{v_cond} is satisfied, this ensures that $f''> 0$.  For completeness, the numerical expression $f'(R) = \frac{\Omega_m}{\Omega}$ in Fig \ref{fig:fprime} is used to numerically reconstruct the function $f(R)$, shown in Fig \ref{fig:freconstructa}, and the percentage difference plotted in Fig \ref{fig:fdiff} indicates the modification to standard GR. 

\subsection{Matter perturbations with $j=1$}
We now turn our attention to the matter perturbations for an $f(R)$ model with a $\Lambda$CDM background and a dust equation of state. 

\begin{figure}
    \centering
    \includegraphics[width=0.5\linewidth]{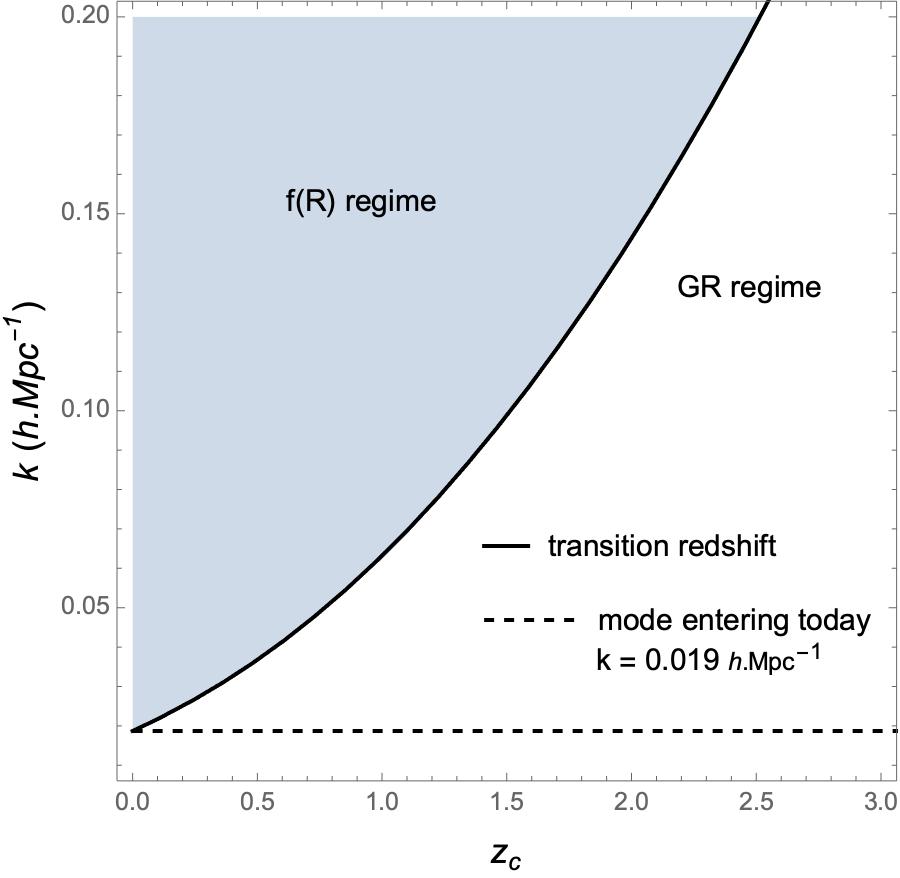}
    \caption{Scale dependence of the transition redshift from GR regime to $f(R)$ regime for matter perturbations. The shaded region represents the $f(R)$ regime where the fifth force is felt and the unshaded region represents the GR regime where this force is suppressed. The dotted line shows the perturbation mode entering the $f(R)$ regime at $z=0$, being $k = 0.019 \,h.$Mpc$^{-1}$. All scales larger than this (smaller $k$) will still be in the GR regime today. }
    \label{fig:transitionredshift}
\end{figure}

An important consideration in modified gravity theories is the transition from the ``GR regime" to the modified regime. In  the case of $f(R)$ models, this is characterised by the parameter 
\begin{equation}
    M^2 \equiv \frac{f'}{3f''} =  \frac{-2(1+q)}{x},
\end{equation}
which corresponds to the mass squared of the scalaron in the regime $M^2 \gg R$ \cite{Tsujikawa:2009ku}. 
In the $f(R)$ regime, $M \ll k/aH$, the scalaron mass is light, leading to a finite range ``fifth force" which is felt by perturbations \cite{Pogosian2008}. In the GR regime $M \gg k/aH$, the scalaron is massive and the fifth force is suppressed, leading to negligible difference with GR. The transition between regimes is time and scale dependent, meaning different scales (i.e. different $k$-modes) pass from the GR regime into the $f(R)$ regime at different times, thus feeling the effect of the deviation from GR differentially. For a perturbation $k$-mode, this transition occurs when $M = \frac{k}{aH}$, which corresponds to the redshift $z_c$ for which
\begin{equation}
   M = \sqrt{\frac{-2(1+q(z_c))}{x(z_c)}}= \frac{k}{aH}.
\end{equation}
The scale-dependence of the transition redshift is shown in Fig \ref{fig:transitionredshift}. For smaller scales (larger $k$), the transition occurs earlier, while larger modes (smaller $k$) have either entered the $f(R)$ regime very close to $z=0$ or have not yet entered at all. In the latter case, these modes are not distinguishable from $\Lambda$CDM as they have till now always been in the GR regime. 

\begin{figure}
    \centering
    \includegraphics[width=0.5\linewidth]{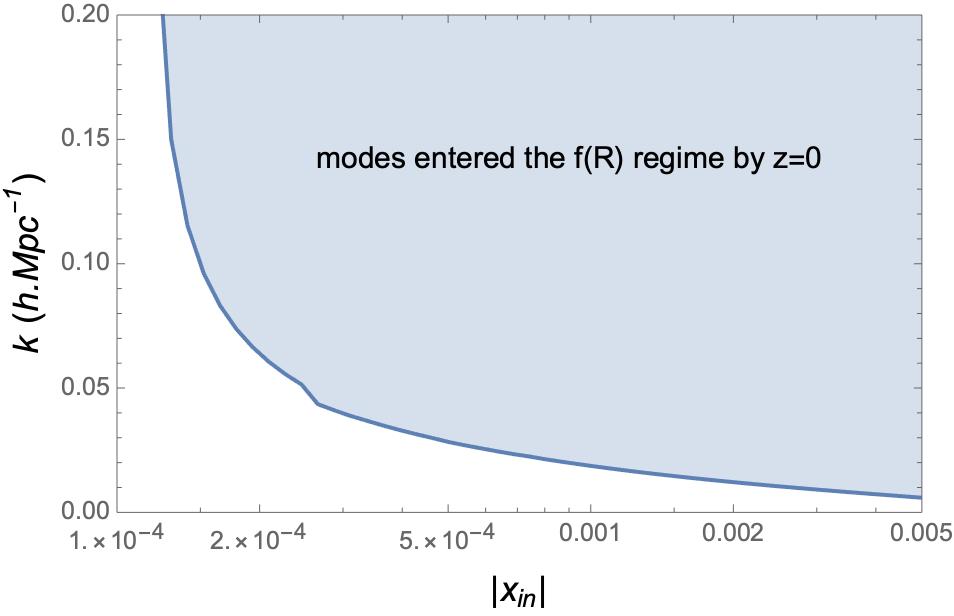}
    \caption{Dependence of the range of perturbations modes that have already crossed from the GR regime to the $f(R)$ regime on the initial value $|x_{in}|$. The shaded region represents modes that have entered the $f(R)$ regime before or at $z=0$. For a background evolution with very little deviation from GR $|x_{in}|\sim 10^{-4}$ in the matter dominated era, all perturbations modes of interest are still in the GR regime.}
    \label{fig:transitiondependence}
\end{figure}

As this transition redshift depends explicitly on the evolution of $x$, varying the initial condition $x_{in}$ will affect which modes have entered the $f(R)$ regime by today. This is shown in Fig \ref{fig:transitiondependence}. For larger values of $|x_{in}|$, say $|x_{in}| \sim 0.003$, all the modes relevant to the galaxy power spectrum, $0.01 \,h.Mpc^{-1} \leq k \leq 0.2\, h. Mpc^{-1}$, have already entered the $f(R)$ regime by today, whereas for very small values of $|x_{in}|$, all of these modes are still inside the GR regime, making them indistinguishable from $\Lambda$CDM. This further motivates the choice of $x_{in} = -0.001$, as most of the relevant perturbation modes have entered the $f(R)$ regime and the $f(R)$ modification is applicable. 

As for the background, we set $j=1$ in the perturbation evolution equations for each of the three approaches: the exact covariant, semi QS and full QS approaches. Respectively, these then simplify to 

\begin{subequations}\label{eq:full_cov}
    \begin{align}
 h^2\frac{d^2\Delta_m^k }{dz^2}+  h^2\frac{(q+x)}{(1+z)} \frac{d\Delta_m^k}{dz}+3  h^2\frac{\Omega}{(1+z)^2}  \Delta_m^k=
-\frac{ x }{2(q+1)(1+z)} \frac{d\hat{\mathcal{R}}^k}{dz}+\left( \frac{\frac{\hat{k}^2 }{h^2 }(1+z)^2 x+3 q (x-1)-3}{6(q+1)(1+z)^2}\right)\hat{\mathcal{R}}^k,
\end{align}
\begin{align}
 \frac{d^2\hat{\mathcal{R}}^k}{dz^2}-\frac{ (q (x+4)+11 x-6 \Omega +4) }{x(1+z)}\frac{d\hat{\mathcal{R}}^k}{dz}+\left(\frac{\hat{k}^2 }{h^2}+\frac{2 q (q (x+2)+6 x-3 \Omega +10)-27 \Omega +16+34x}{x(1+z)^2}\right)\hat{\mathcal{R}}^k& \nonumber\\
=\frac{6 h^2 (q+1)}{(1+z)}\frac{d\Delta_m^k}{dz}-\frac{6 h^2 (q+1) \Omega }{x(1+z)^2} \Delta_m^k&, 
\end{align}
\end{subequations}

for the exact covariant equations,
\begin{align}\label{eq:semi_QS}
 & \frac{d^2\Delta_m^k}{dz^2}+\left(\frac{q+x(1 -\tilde{C})}{(1+z)}\right)\frac{d\Delta_m^k}{dz} 
 =\Omega\left(\frac{3-\tilde{C}}{(1+z)^2}\right)   \Delta_m^k,
\end{align}
with
\begin{align}
      \tilde{C} = \frac{\left(\frac{\hat{k}^2  (z+1)^2}{h^2}x -3(1+q)+3 qx\right) }{   \left(\frac{\hat{k}^2 (z+1)^2}{h^2}x+(34+2q^2+12q)x+4(6q^2+5q+4)-(6q  +27) \Omega \right)},
\end{align}
for the semi QS equation, and 
\begin{equation}\label{eq:full_QS}
    (1+z)^2 \frac{d^2\Delta_m^k}{dz^2}+(1+z)q \frac{d\Delta_m^k}{dz}=\Omega \left(\frac{2 \frac{\hat{k}^2}{h^2} x- 3 \frac{(q+1)}{(1+z)^2} }{\frac{\hat{k}^2}{h^2} x -2  \frac{(q+1)}{(1+z)^2}}\right)\Delta_m^k,
\end{equation}
for the full QS method. 

\begin{figure}
    \centering
    \begin{subfigure}[b]{0.53\linewidth}
    \includegraphics[width=\linewidth]{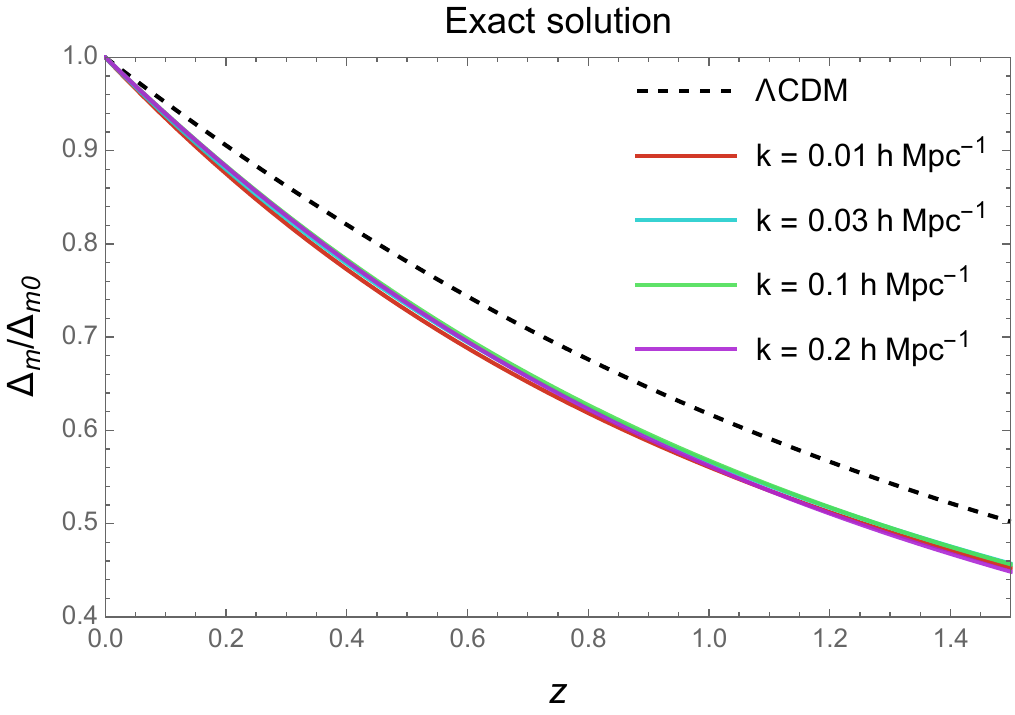}
    \caption{}
    \label{fig:fullpert}
    \end{subfigure}
    \hfill
    \begin{subfigure}[b]{0.53\linewidth}
    \includegraphics[width=\linewidth]{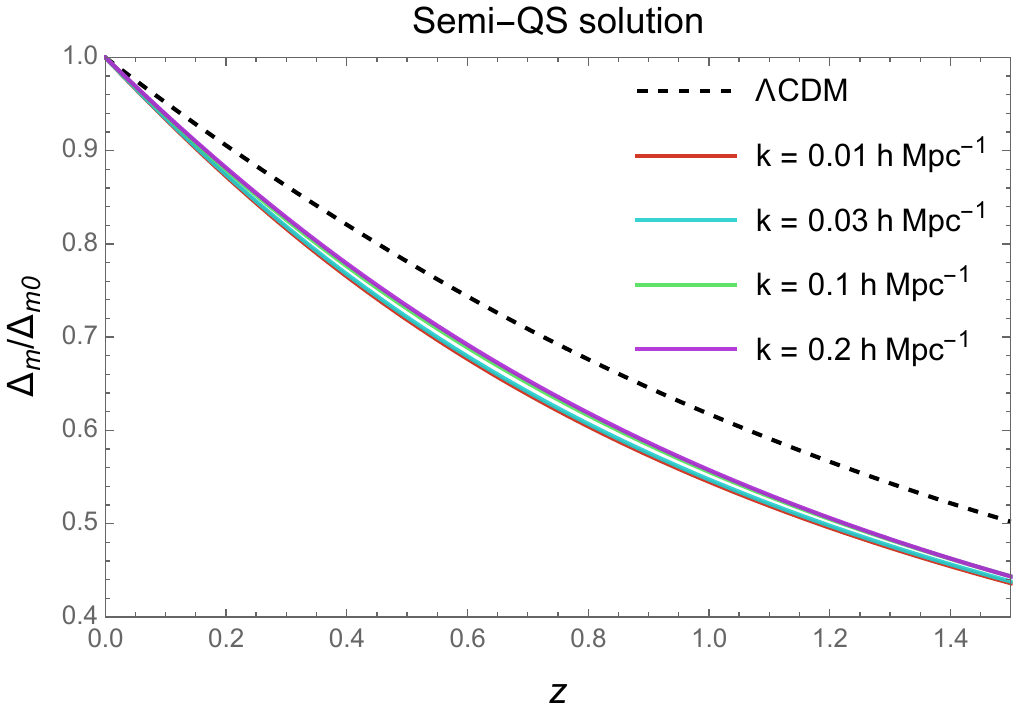}
    \caption{}
    \label{fig:semipert}
    \end{subfigure}
    \hfill
    \begin{subfigure}[b]{0.53\linewidth}
    \includegraphics[width=\linewidth]{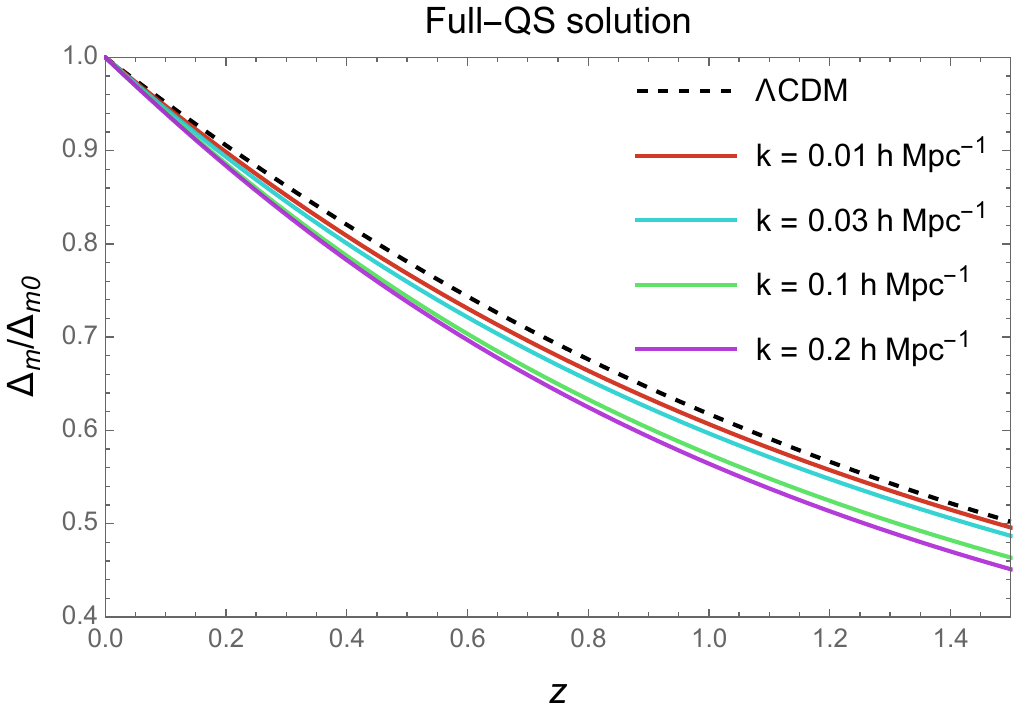}
    \caption{}
    \label{fig:QSpert}
    \end{subfigure}
    \caption{Evolution of subhorizon matter density perturbation $\Delta_m^k$ for different values of $k$ normalised at $z=0$ using (a) Full covariant method (Eqs.\eqref{eq:full_cov}), (b) Semi Quasi-Static approximation (Eq.\eqref{eq:semi_QS}), (c) Full Quasi-Static approximation (Eq.\eqref{eq:full_QS}). The dashed line in each plot represents the evolution of perturbations in a $\Lambda$CDM background with the same set of initial conditions \eqref{eq:initcondpert}.}
    \label{fig:perturbations}
\end{figure}

\begin{figure}
    \centering
    \begin{subfigure}[b]{0.53\linewidth}
    \includegraphics[width=\linewidth]{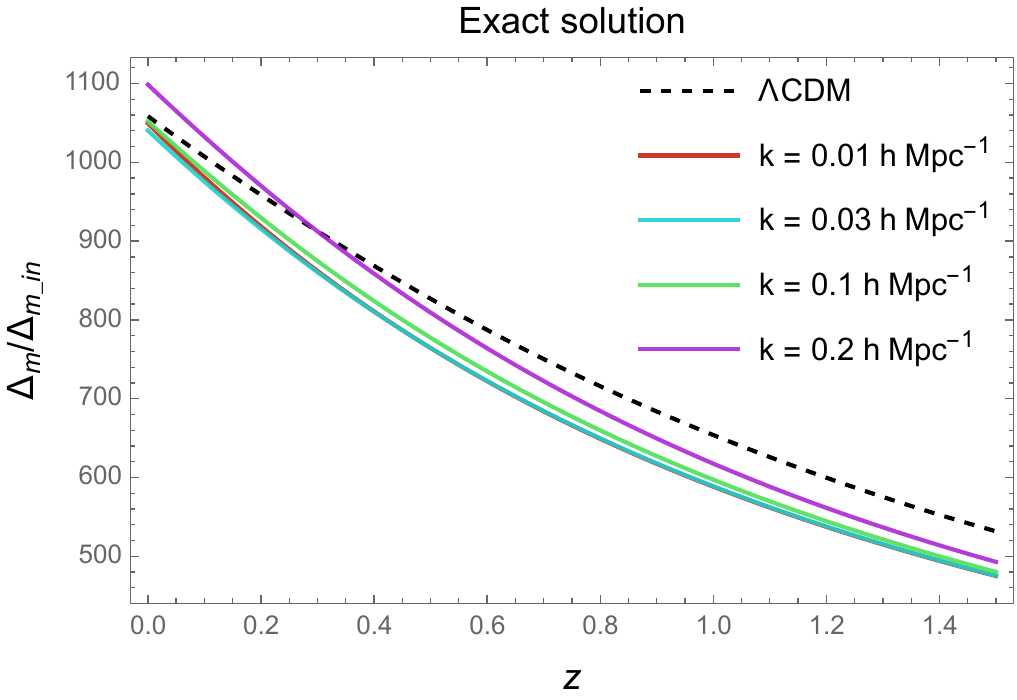}
    \caption{}
    \label{fig:fullpertlast}
    \end{subfigure}
    \hfill
    \begin{subfigure}[b]{0.53\linewidth}
    \includegraphics[width=\linewidth]{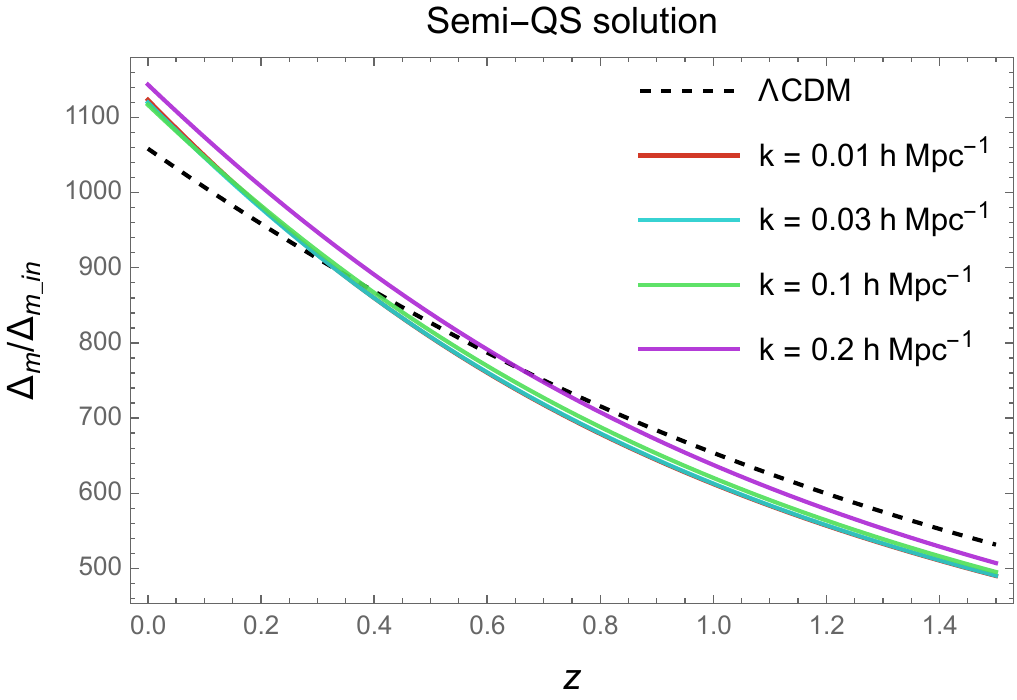}
    \caption{}
    \label{fig:semipertlast}
    \end{subfigure}
    \hfill
    \begin{subfigure}[b]{0.53\linewidth}
    \includegraphics[width=\linewidth]{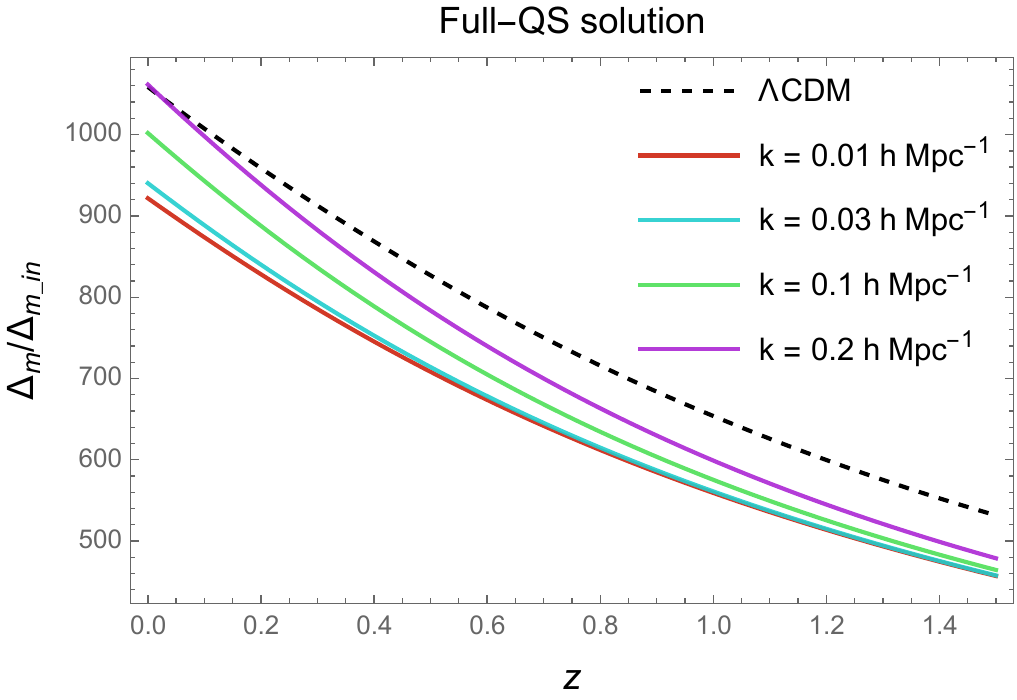}
    \caption{}
    \label{fig:QSpertlast}
    \end{subfigure}
    \caption{Evolution of subhorizon matter density perturbation $\Delta_m^k$ for different values of $k$ normalised at $z=2000$ using (a) Full covariant method (Eqs.\eqref{eq:full_cov}), (b) Semi Quasi-Static approximation (Eq.\eqref{eq:semi_QS}), (c) Full Quasi-Static approximation (Eq.\eqref{eq:full_QS}). The dashed line in each plot represents the evolution of perturbations in a $\Lambda$CDM background with the same set of initial conditions \eqref{eq:initcondpert}.}
    \label{fig:perturbationsnormcmb}
\end{figure}

In line with \cite{Abebe:2013zua}, we set the scale-independent initial conditions in the CMB era at $z_{in}=2000$ to be 
\begin{equation}
    \Delta_m^k|_{in} = \hat{\mathcal{R}}^k|_{in} = 10^{-5}, \quad  
    (\Delta_m^k)'|_{in} = (\hat{\mathcal{R}}^k)'|_{in} = 0.
    \label{eq:initcondpert}
\end{equation}
For the background trajectory specified in Figs.\ref{fig:parametricplot}, the associated matter perturbation evolution for each method are shown in Figs. \ref{fig:perturbations} and \ref{fig:perturbationsnormcmb}. In Fig. \ref{fig:perturbations} $\Delta_m$ is normalized at $z=0$ to facilitate a comparison with similar plots appearing in model-dependent analysis of \cite{gannouji_growth}, which investigates perturbation evolution in Starobinsky's model. However, such a plot does not clearly portray the scale dependence of $\Delta_m$ at low redshift. To show more clearly the scale-dependence of $\Delta_m$ at very low redshifts ($z\approx0$), we have included the corresponding plots where $\Delta_m$ is normalised at $z=2000$ in Fig. \ref{fig:perturbationsnormcmb}\footnote{We believe that normalizing at high redshift is more logical in the sense that at high redshift a viable $f(R)$ theory should asymptotically tend to GR, for which there is no scale-dependence.}.

Little scale dependence is observed across all the relevant modes for the full covariant and semi QS methods at low redshifts. The scale dependence is more pronounced for the full QS method.

\begin{figure}
    \centering
    \begin{subfigure}[b]{0.49\linewidth}
    \includegraphics[width=\linewidth]{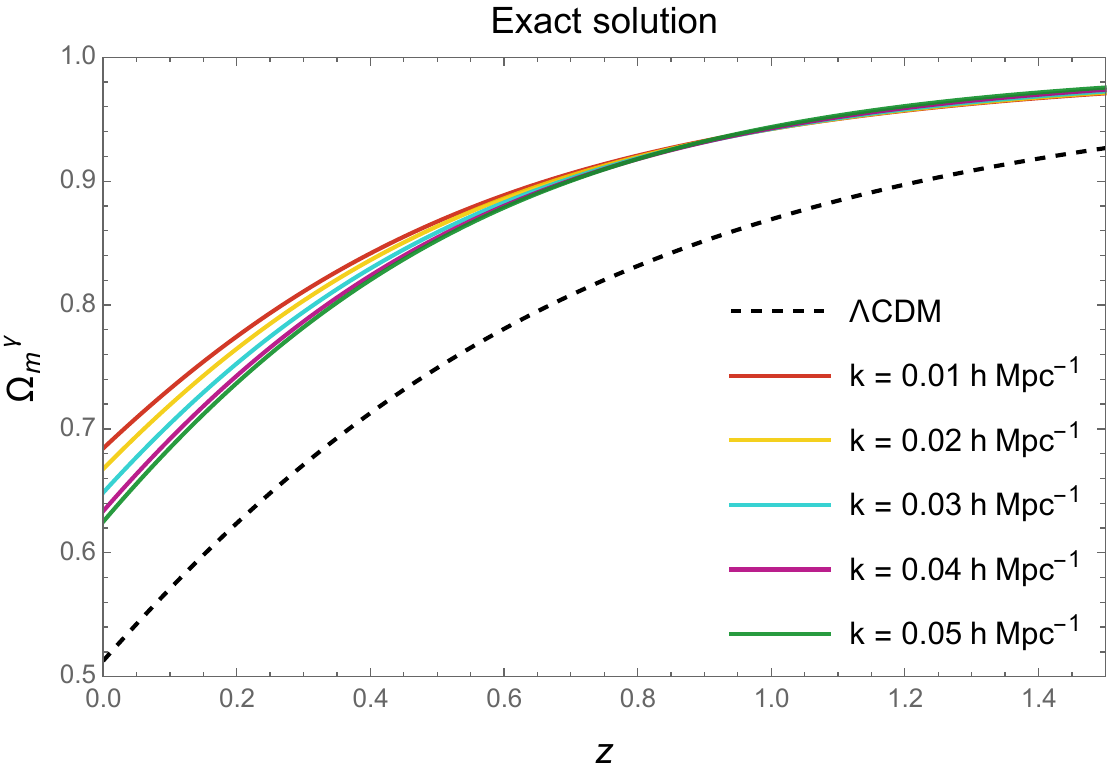}
     \caption{}
        \label{fig:growthfuncfullsmallk}
    \end{subfigure}
    \begin{subfigure}[b]{0.49\linewidth}
    \includegraphics[width=\linewidth]{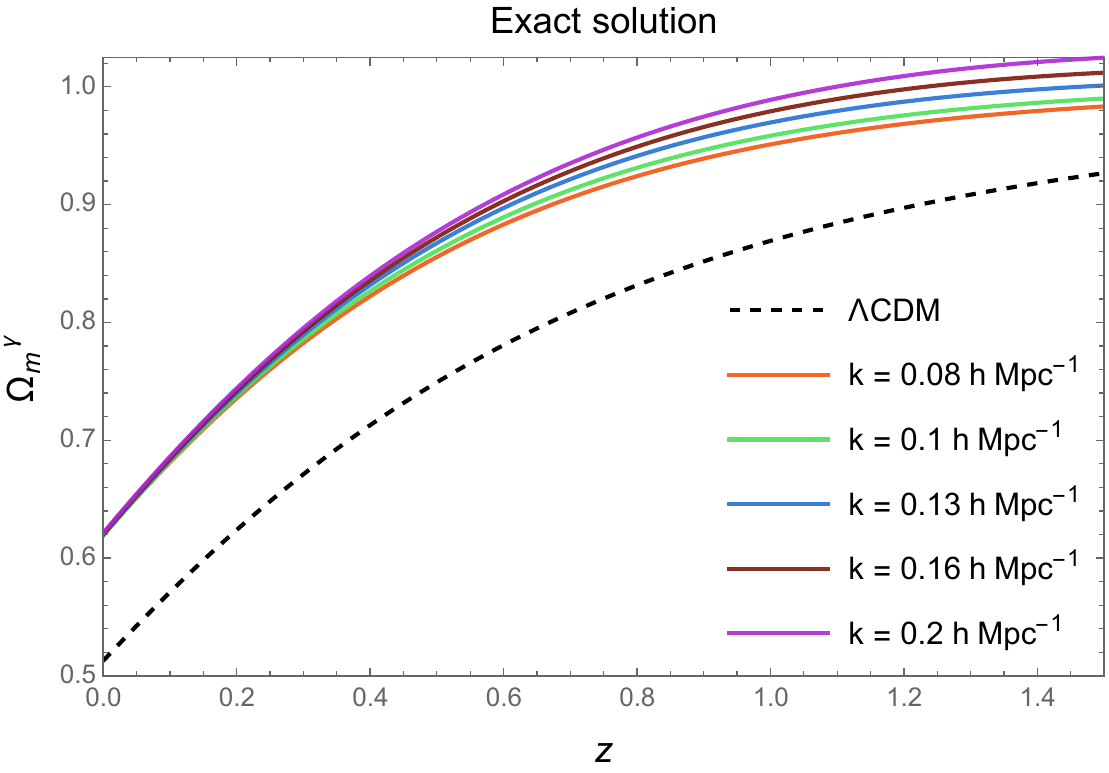}
     \caption{}
        \label{fig:growthfuncfulllargek}
    \end{subfigure}
    \begin{subfigure}[b]{0.49\linewidth}
    \includegraphics[width=\linewidth]{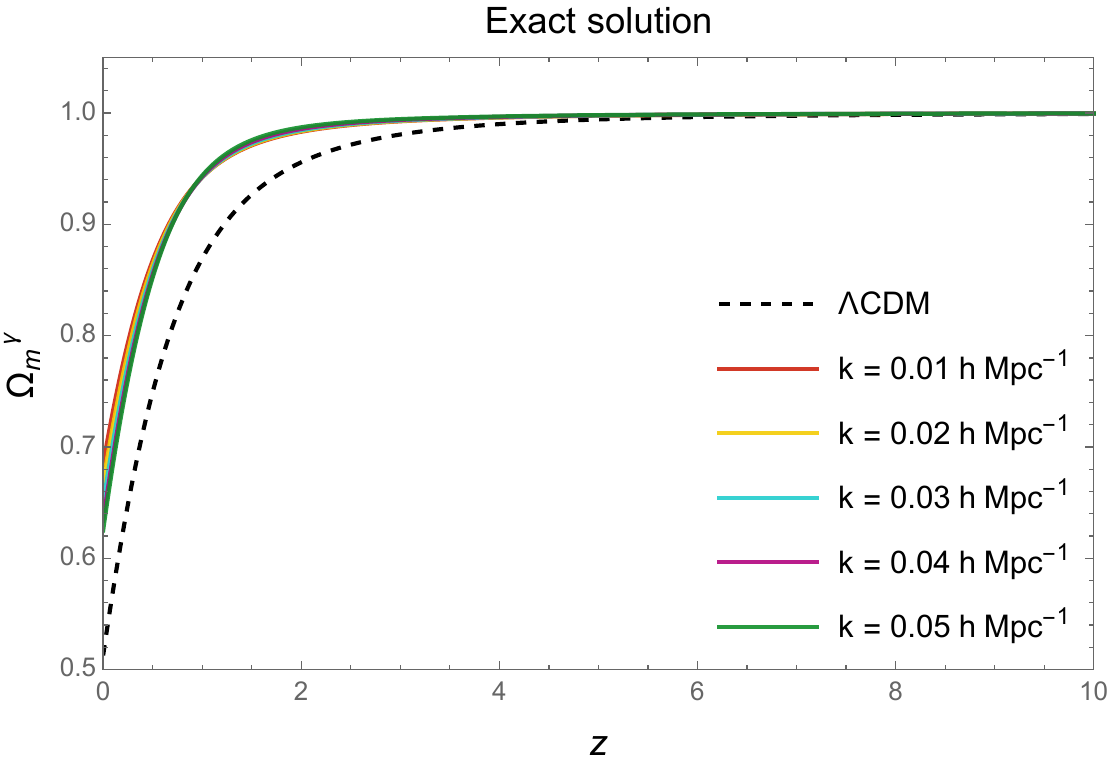}
     \caption{}
        \label{fig:growthfuncfullsmallk2}
    \end{subfigure}
    \begin{subfigure}[b]{0.49\linewidth}
    \includegraphics[width=\linewidth]{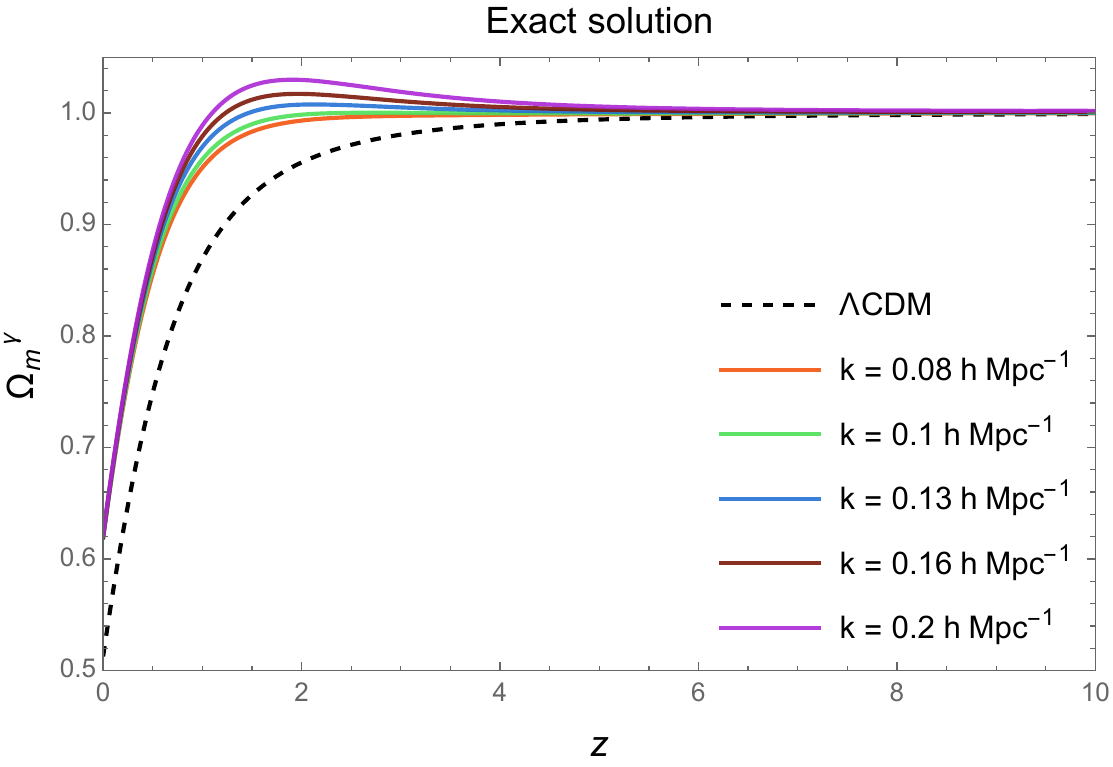}
     \caption{}
        \label{fig:growthfuncfulllargek2}
    \end{subfigure}
    \begin{subfigure}[b]{0.49\linewidth}
    \includegraphics[width=\linewidth]{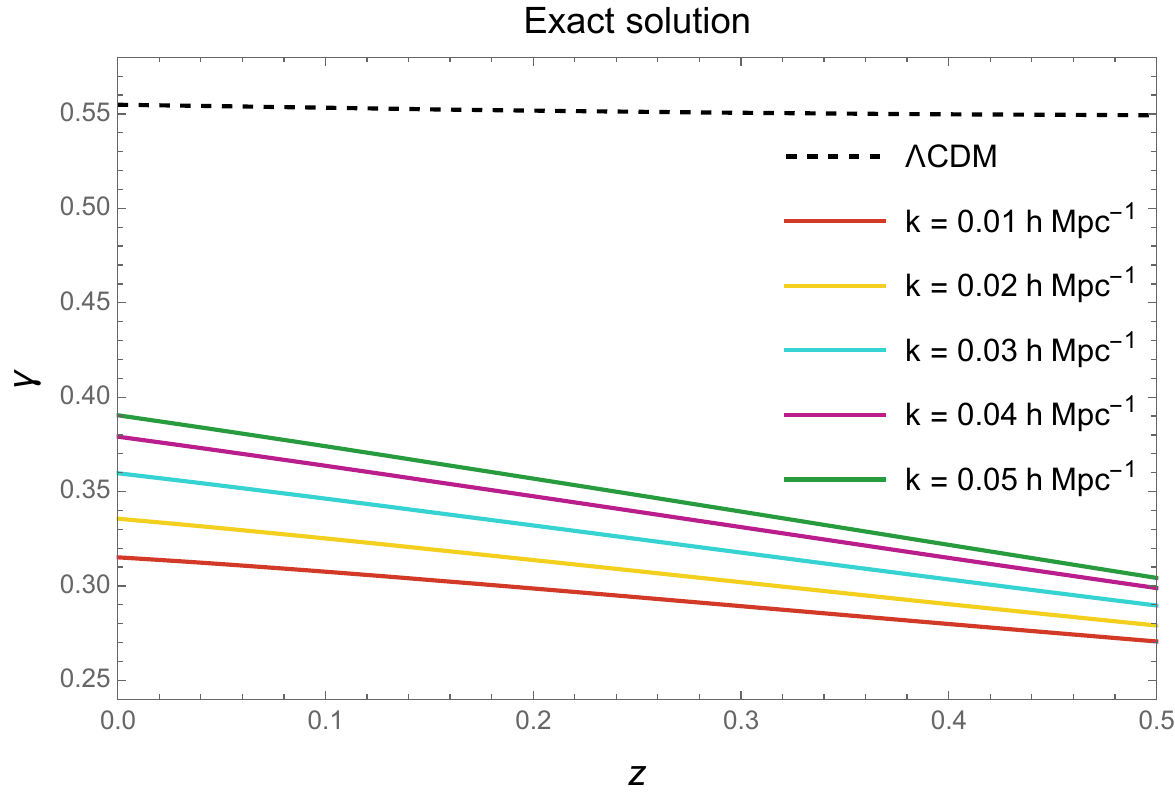}
     \caption{}
        \label{fig:growthindexfullsmallk}
    \end{subfigure}
    \begin{subfigure}[b]{0.49\linewidth}
    \includegraphics[width=\linewidth]{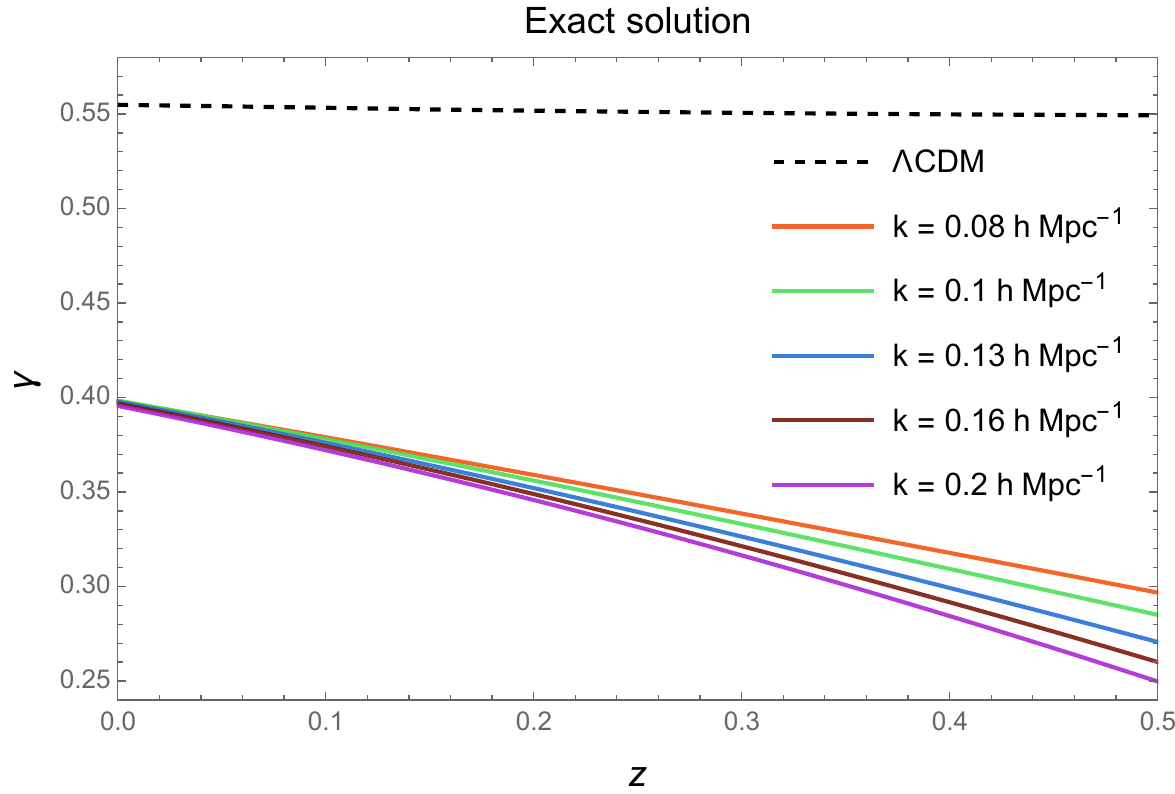}
     \caption{}
        \label{fig:growthindexfulllargek}
    \end{subfigure}
    \caption{(a-d) Growth function $S$ and (e-f) growth index $\gamma$ for the full covariant perturbation method. Left column is large scale, right column is small scale. The difference between the figures (a,b) and the figures (c,d) is that (c,d) show the behaviour for a larger range of $z$, clearly showing that in the far past the scale dependence is suppressed as the growth function for all the modes converge to the $\Lambda$CDM value.}
    \label{fig:growthfull}
\end{figure}

\begin{figure}
    \centering
    \begin{subfigure}[b]{0.49\linewidth}
    \includegraphics[width=\linewidth]{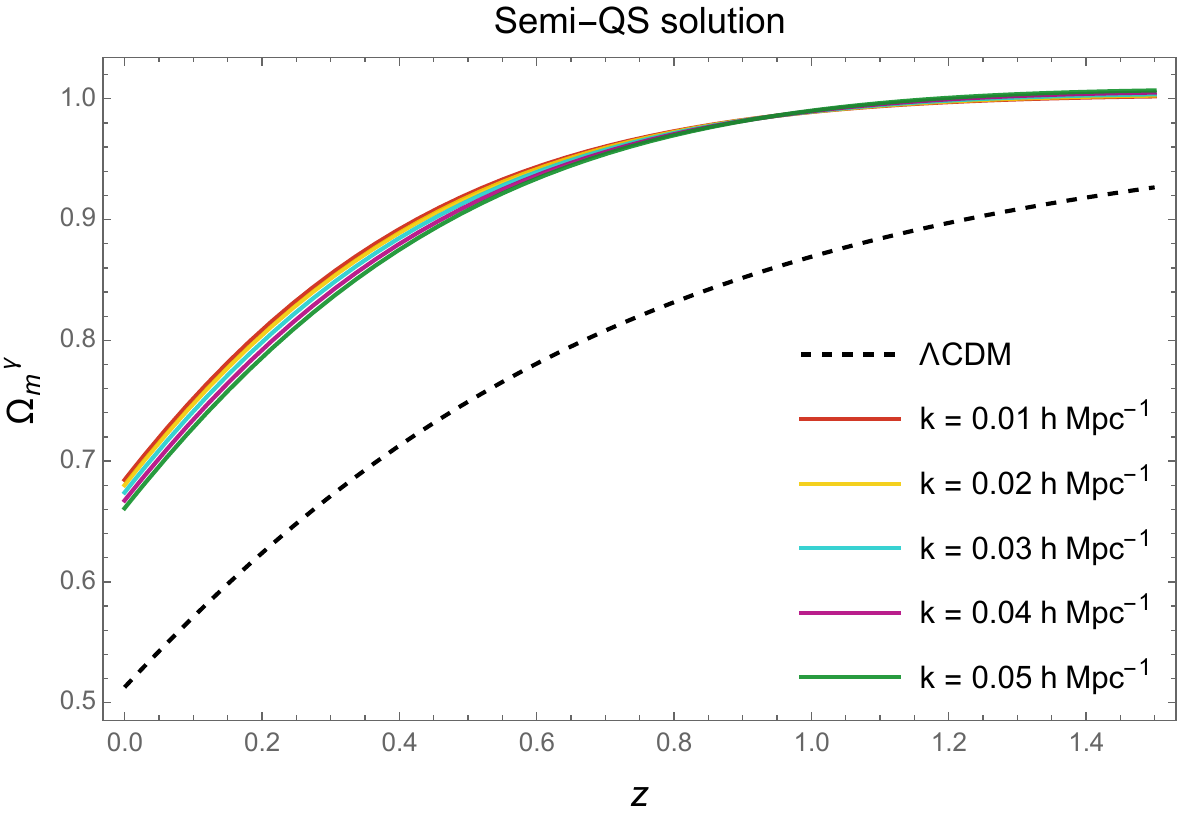}
     \caption{}
        \label{fig:growthfuncsemismallk}
    \end{subfigure}
    \begin{subfigure}[b]{0.49\linewidth}
    \includegraphics[width=\linewidth]{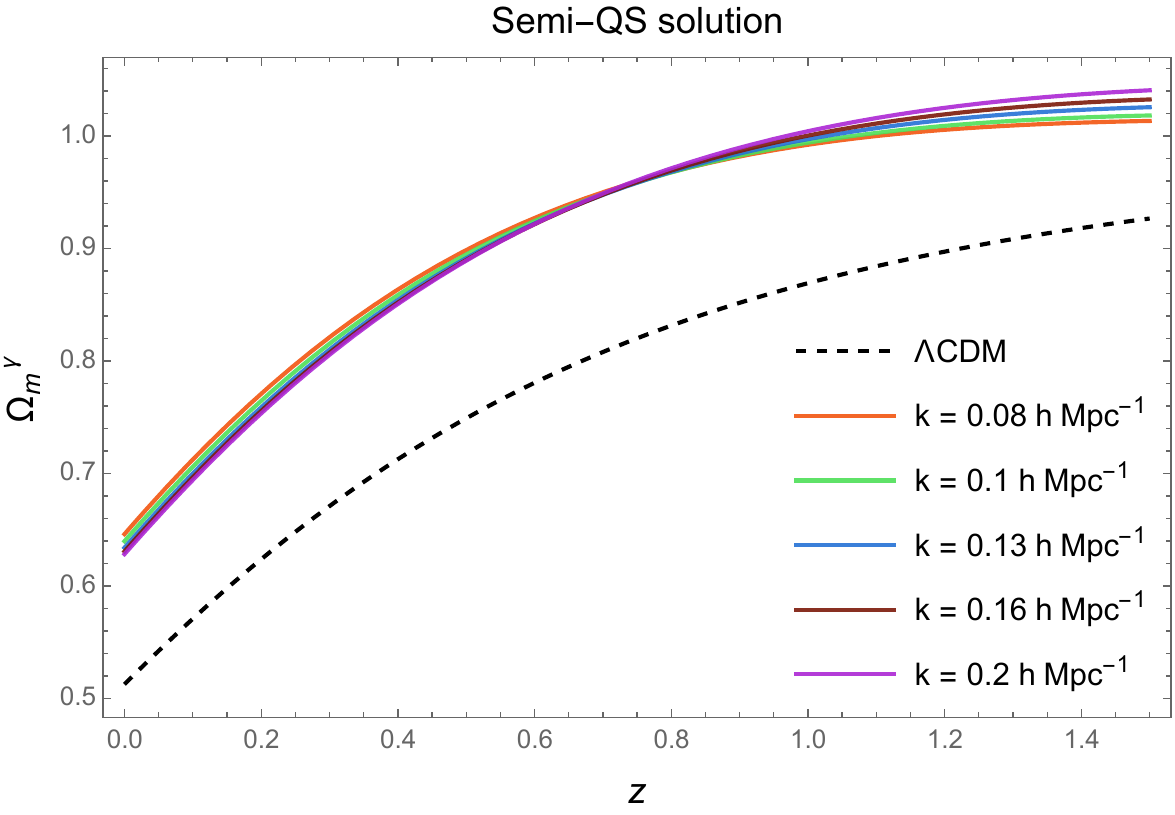}
     \caption{}
        \label{fig:growthfuncsemilargek}
    \end{subfigure}
    \begin{subfigure}[b]{0.49\linewidth}
    \includegraphics[width=\linewidth]{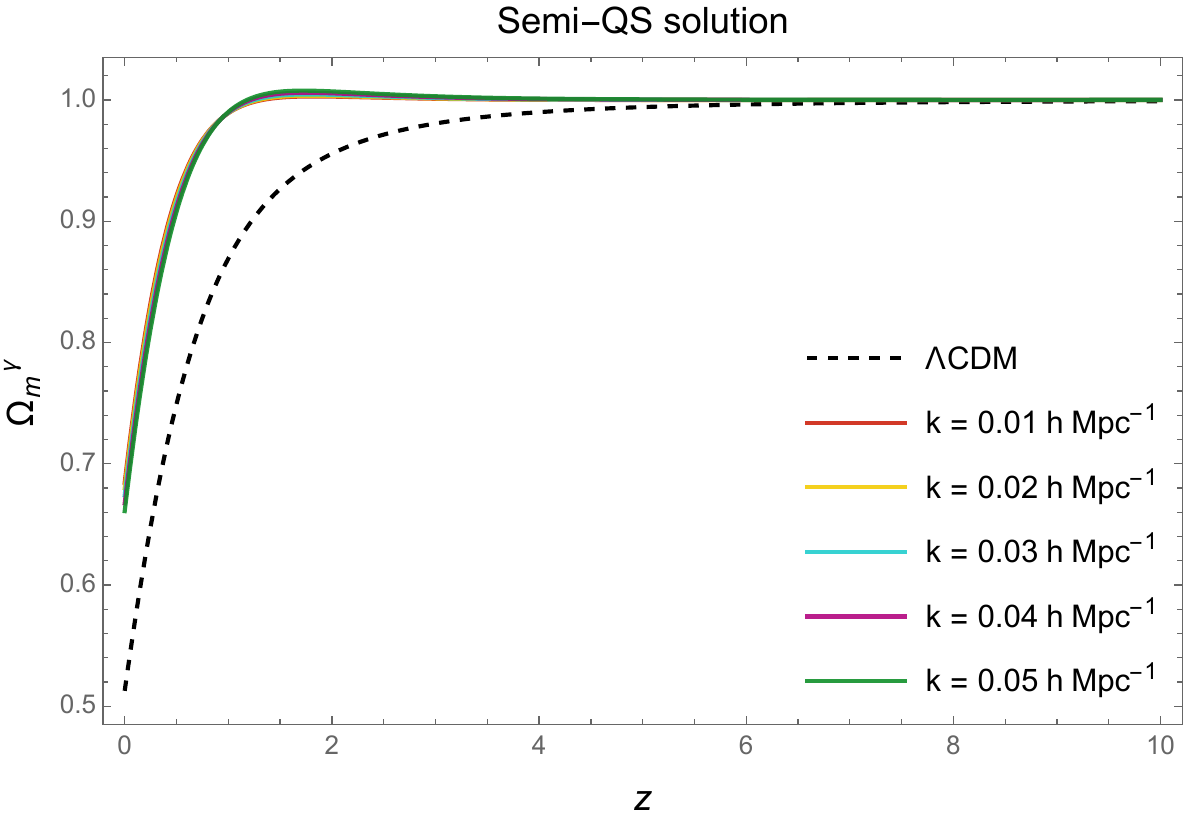}
     \caption{}
        \label{fig:growthfuncsemismallk2}
    \end{subfigure}
    \begin{subfigure}[b]{0.49\linewidth}
    \includegraphics[width=\linewidth]{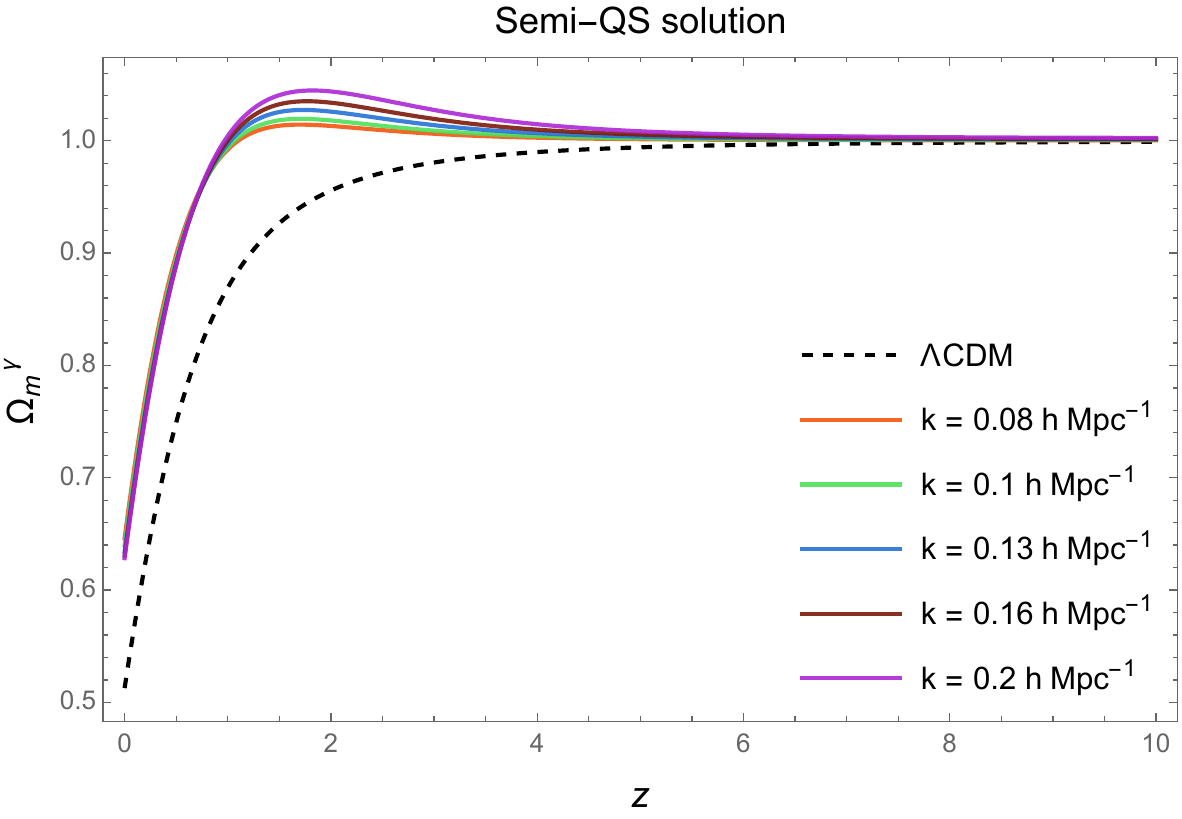}
     \caption{}
        \label{fig:growthfuncsemilargek2}
    \end{subfigure}
    \begin{subfigure}[b]{0.49\linewidth}
    \includegraphics[width=\linewidth]{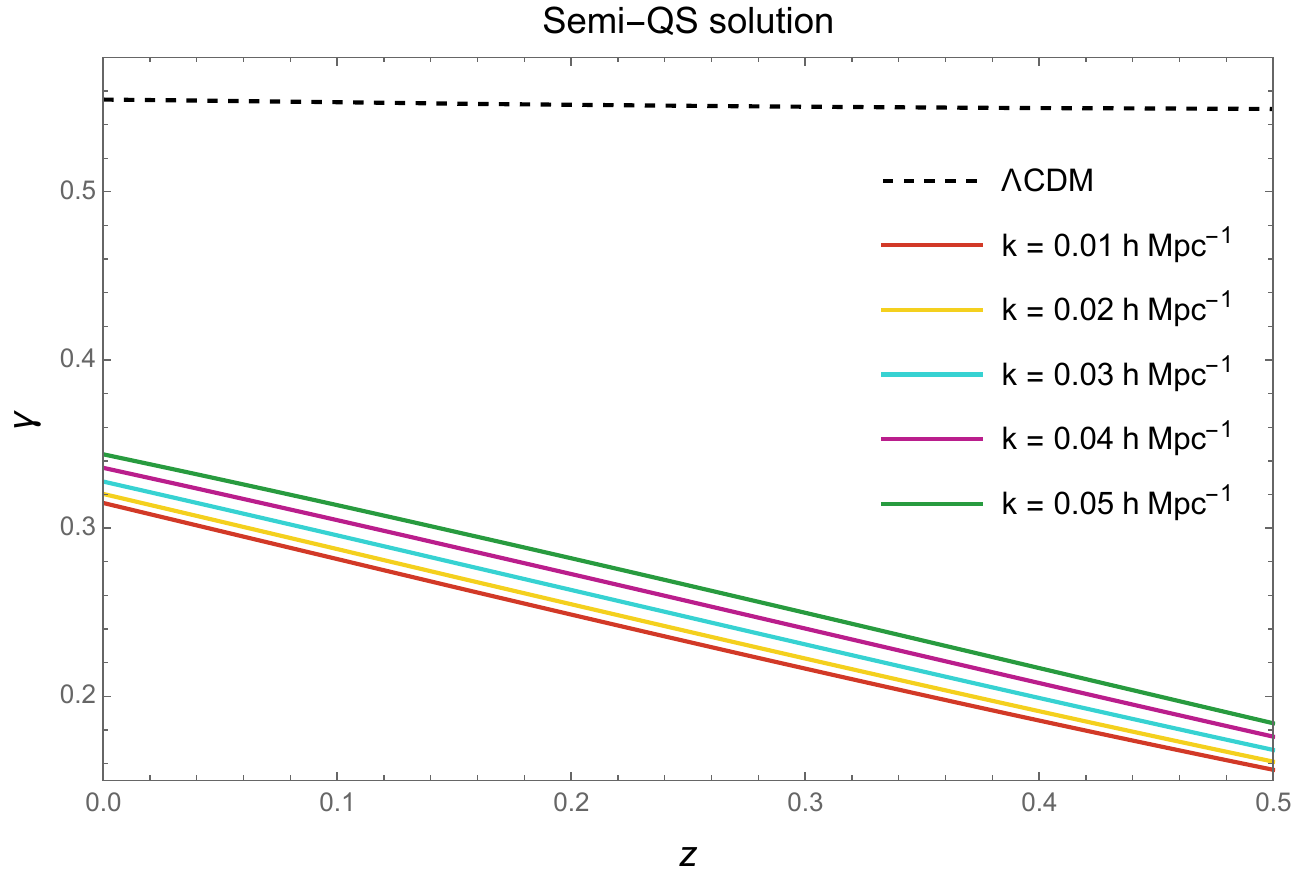}
     \caption{}
        \label{fig:growthindexsemismallk}
    \end{subfigure}
    \begin{subfigure}[b]{0.49\linewidth}
    \includegraphics[width=\linewidth]{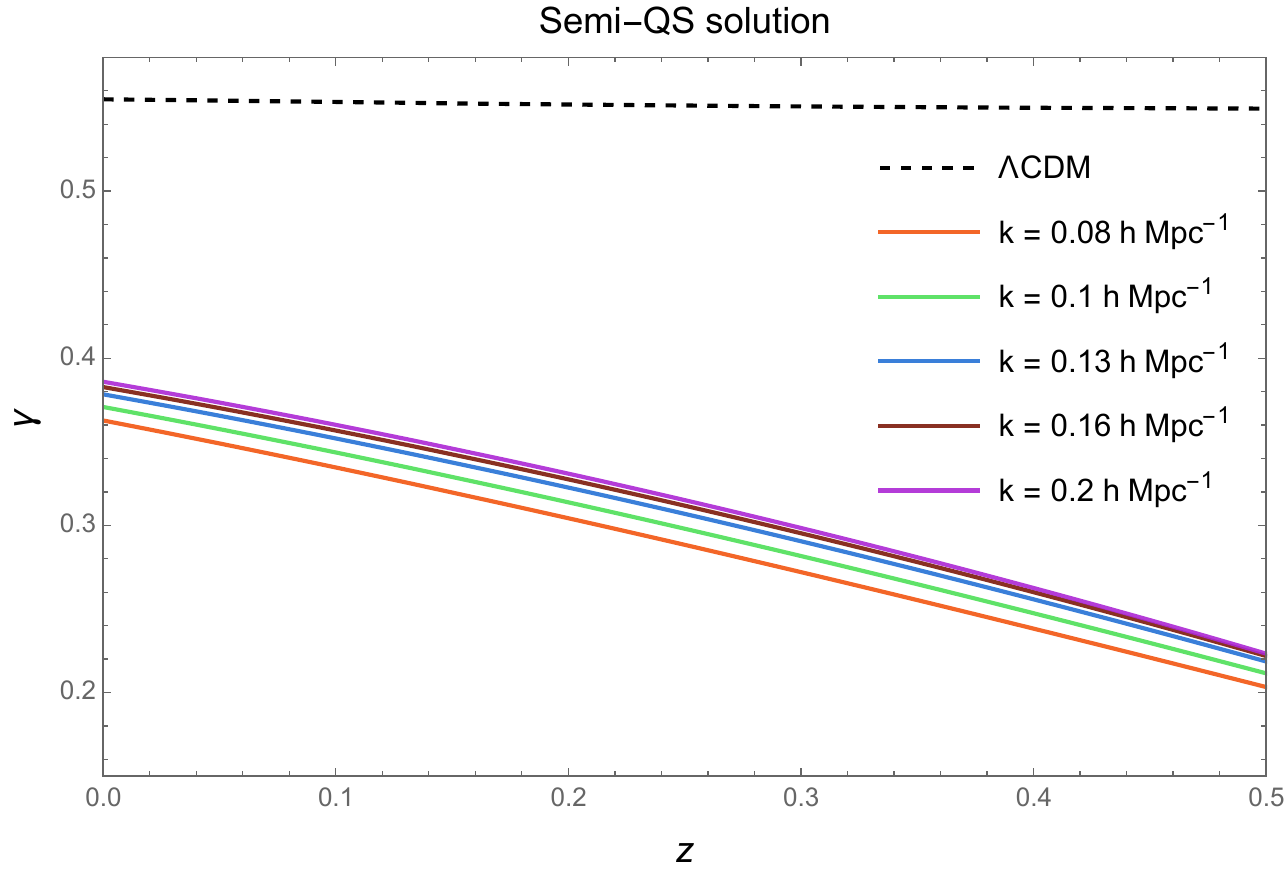}
     \caption{}
        \label{fig:growthindexsemilargek}
    \end{subfigure}
    \caption{(a-d) Growth function $S$ and (e-f) growth index $\gamma$ for the semi quasi-static perturbation method. Left column is large scale, right column is small scale.}
    \label{fig:growthsemi}
\end{figure}

\begin{figure}
    \centering
    \begin{subfigure}[b]{0.49\linewidth}
    \includegraphics[width=\linewidth]{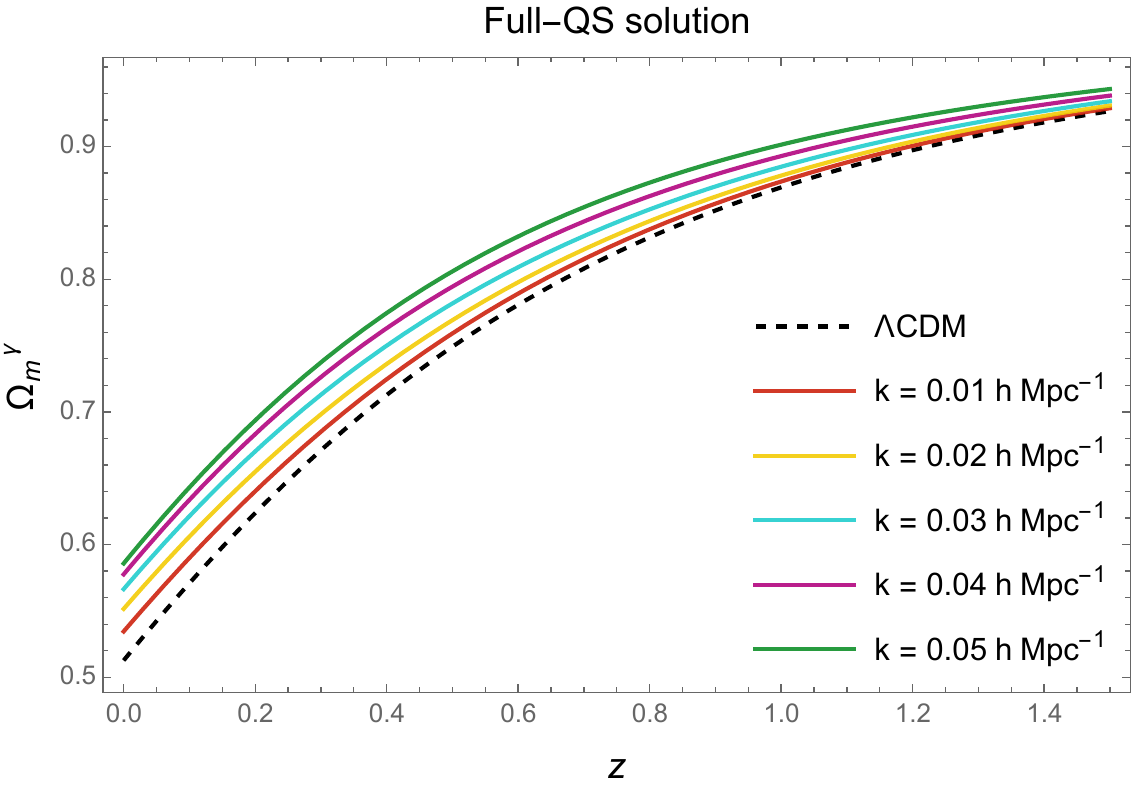}
     \caption{}
        \label{fig:growthfuncQSsmallk}
    \end{subfigure}
    \begin{subfigure}[b]{0.49\linewidth}
    \includegraphics[width=\linewidth]{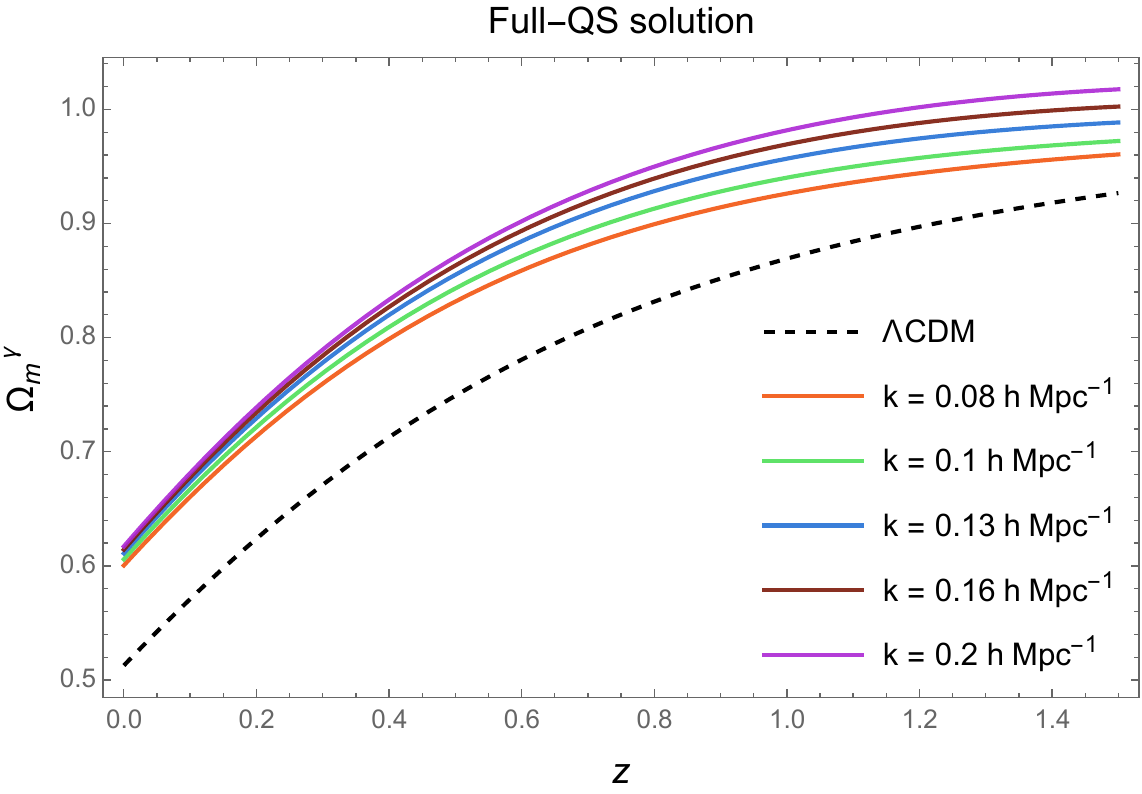}
     \caption{}
        \label{fig:growthfuncQSlargek}
    \end{subfigure}
    \begin{subfigure}[b]{0.49\linewidth}
    \includegraphics[width=\linewidth]{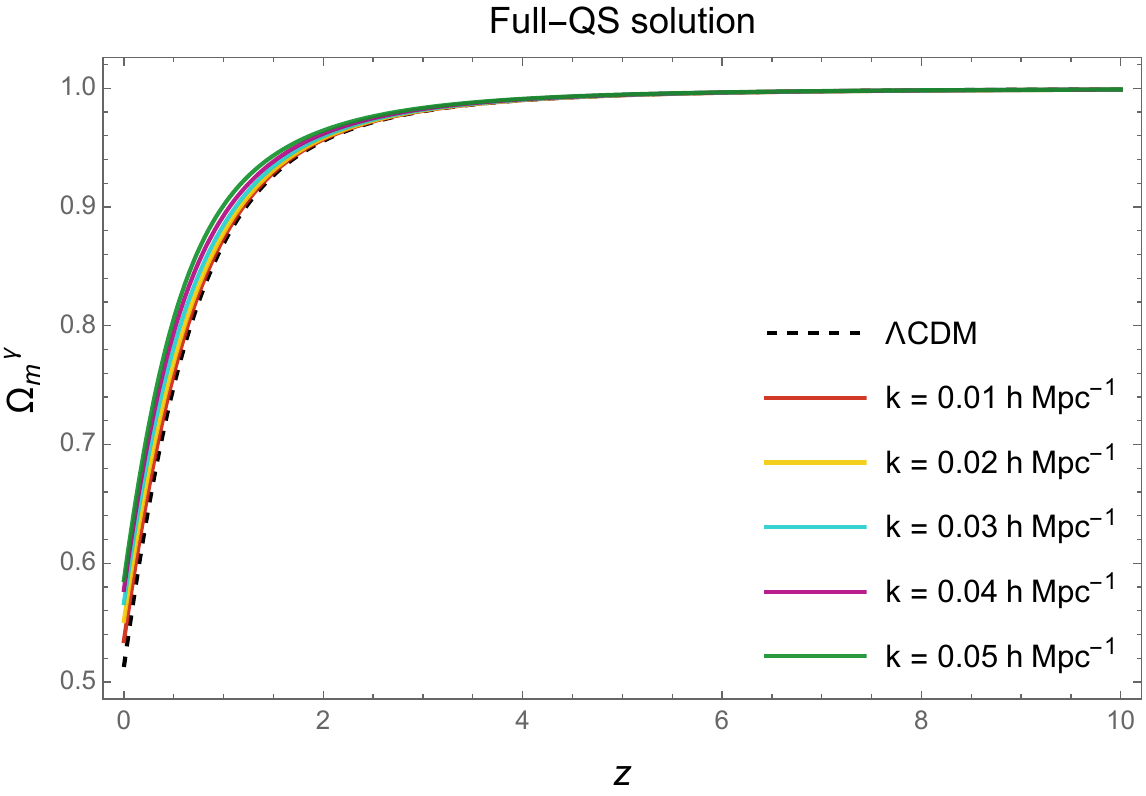}
     \caption{}
        \label{fig:growthfuncQSsmallk2}
    \end{subfigure}
    \begin{subfigure}[b]{0.49\linewidth}
    \includegraphics[width=\linewidth]{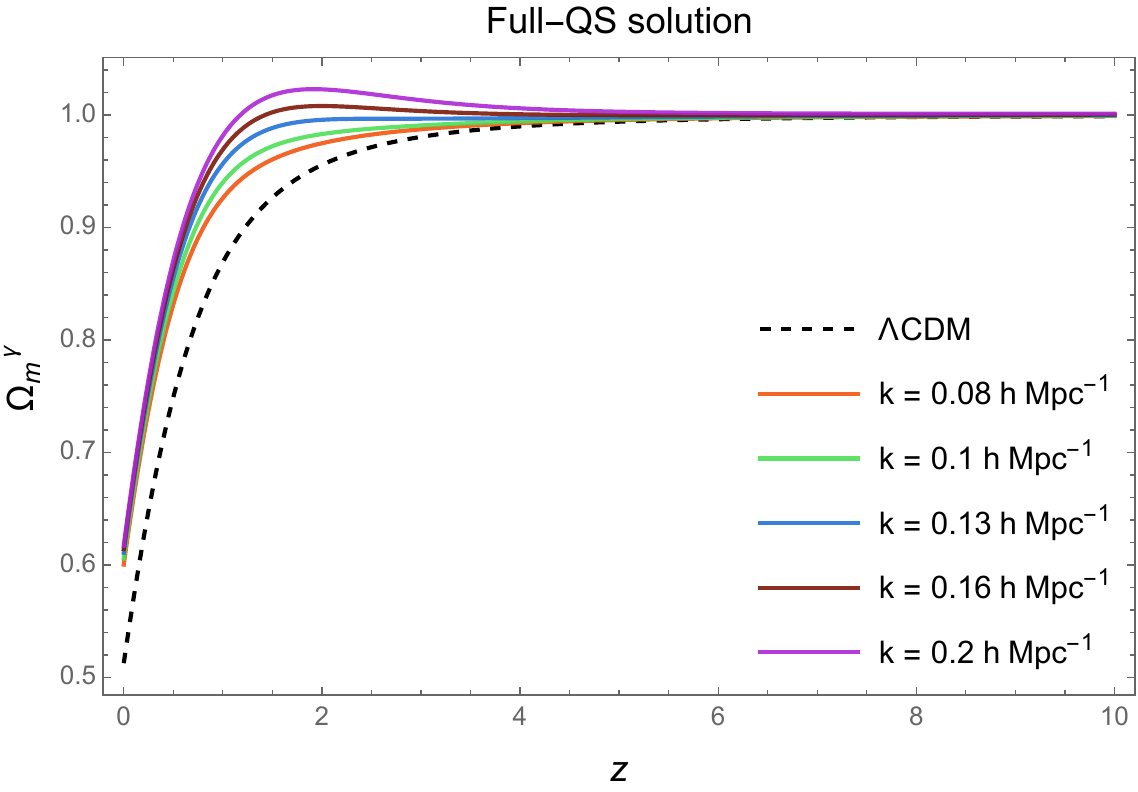}
     \caption{}
        \label{fig:growthfuncQSlargek2}
    \end{subfigure}
    \begin{subfigure}[b]{0.49\linewidth}
    \includegraphics[width=\linewidth]{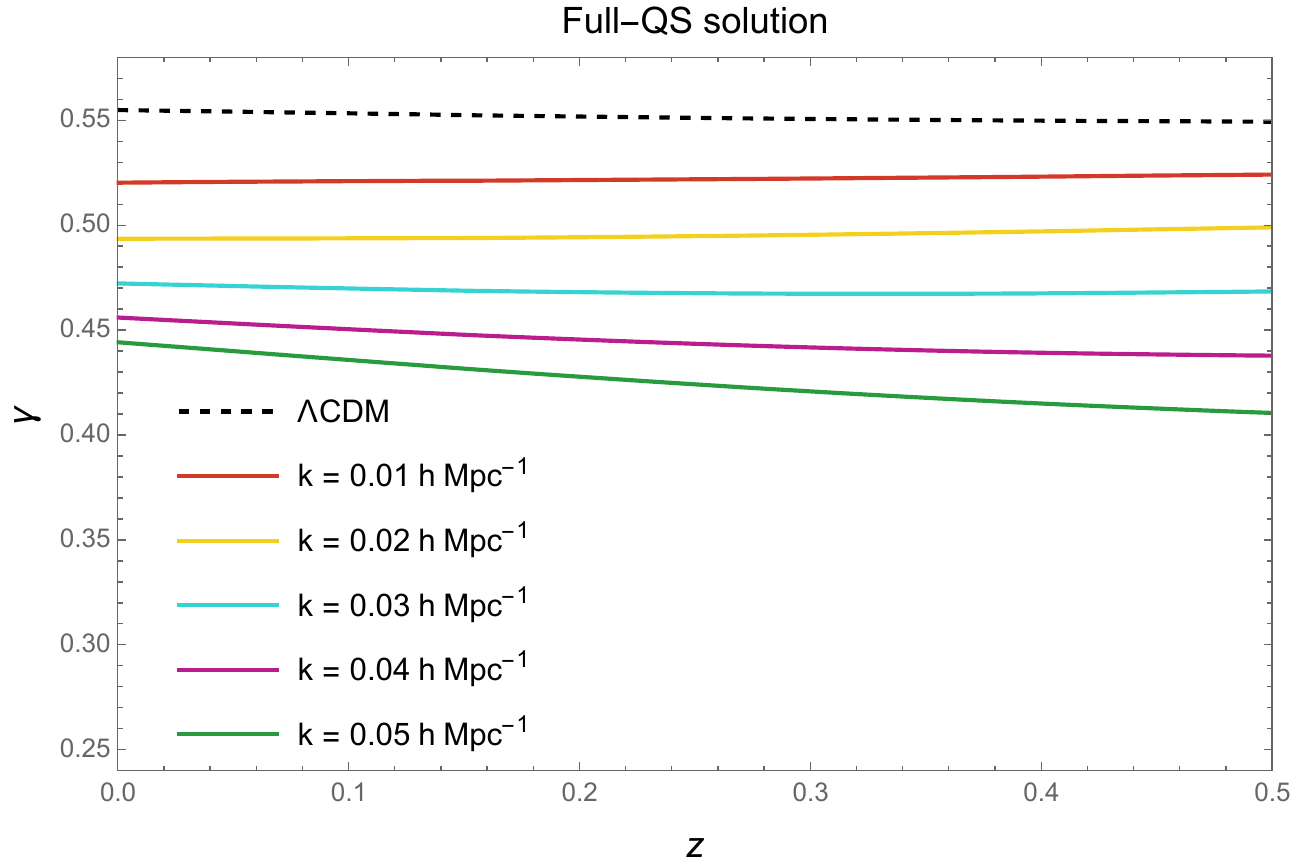}
     \caption{}
        \label{fig:growthindexQSsmallk}
    \end{subfigure}
    \begin{subfigure}[b]{0.49\linewidth}
    \includegraphics[width=\linewidth]{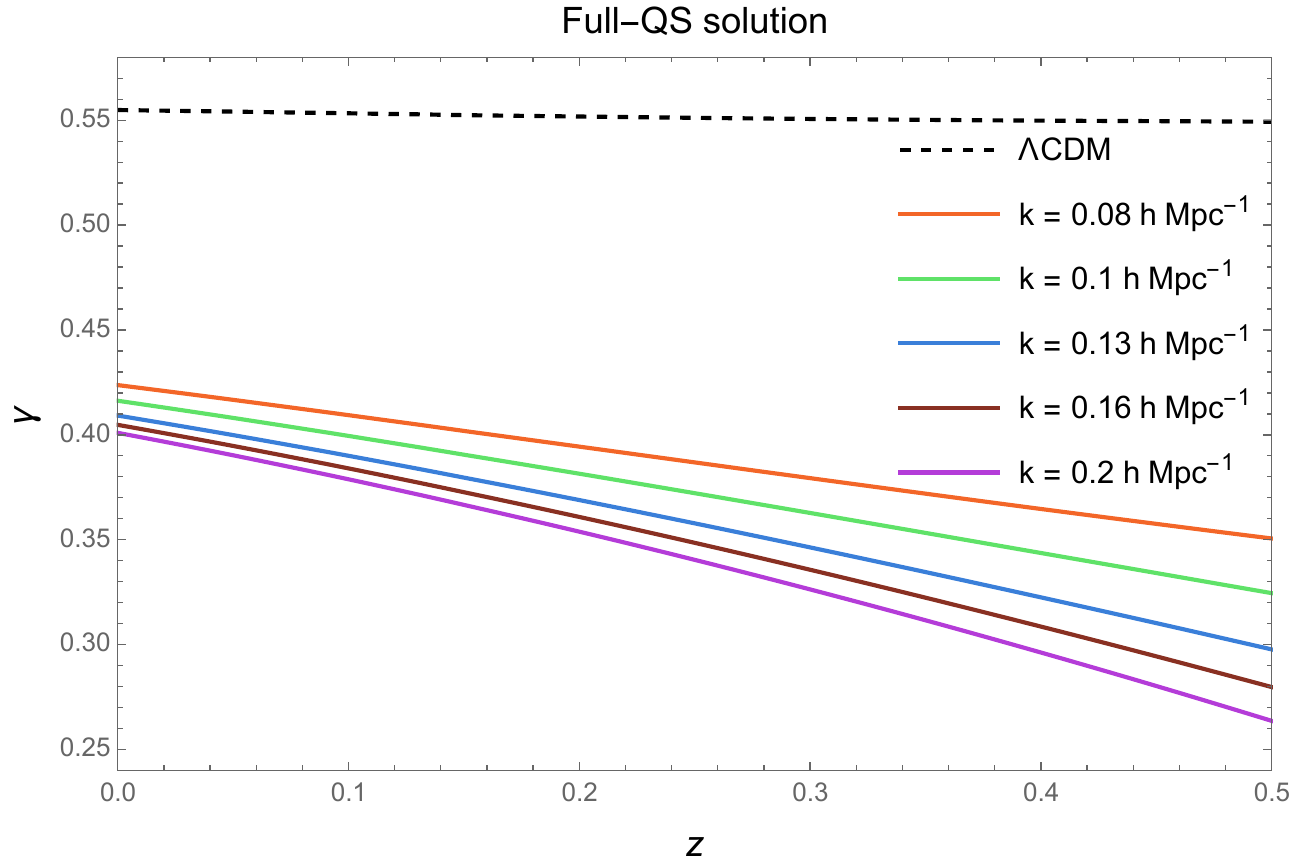}
     \caption{}
        \label{fig:growthindexQSlargek}
    \end{subfigure}
    \caption{(a-d) Growth function $S$ and (e-f) growth index $\gamma$ for the full quasi-static perturbation method. Left column is large scale, right column is small scale.}
    \label{fig:growthQS}
\end{figure}

The growth rate function $S$ defined in Eq.\eqref{eq:growthfunction} and the growth index parameter $\gamma$ defined in Eq.\eqref{eq:growthindexparam} associated with the full covariant, semi QS and full QS perturbation method are shown in Figs \ref{fig:growthfull}, \ref{fig:growthsemi}, and \ref{fig:growthQS} respectively. Here the scales are split for each method to show general large vs small scale trends. All methods show similar results for smaller scales (right columns), with the scale dependence of $S$ being only minor at $z=0$ and becoming more pronounced at around  $z\sim 1.5$. From Fig \ref{fig:transitionredshift}, it is apparent that all of these scales have been in the $f(R)$ regime since $z \simeq 1$ and have thus had time to feel the effect of the fifth force, leading to significant deviation from $\Lambda$CDM. For larger scales (left columns), the trend is inverted with scale-dependence being much larger at $z=0$. From Fig \ref{fig:growthQS}, we see that the growth function and growth index for the full QS method differ significantly from the other methods and, in the case of the growth function, only marginally from $\Lambda$CDM. For the growth index, while the values of $\gamma$ are not exactly those found for $\Lambda$CDM, the suppression of $\gamma_0$ is far weaker and the growth is far slower than in the methods. This again emphasises the inability to distinguish $f(R)$ from $\Lambda$CDM at these scales using the full QS method. 

\begin{figure}
    \centering
     \begin{subfigure}[b]{0.5\linewidth}
         \includegraphics[width=\linewidth]{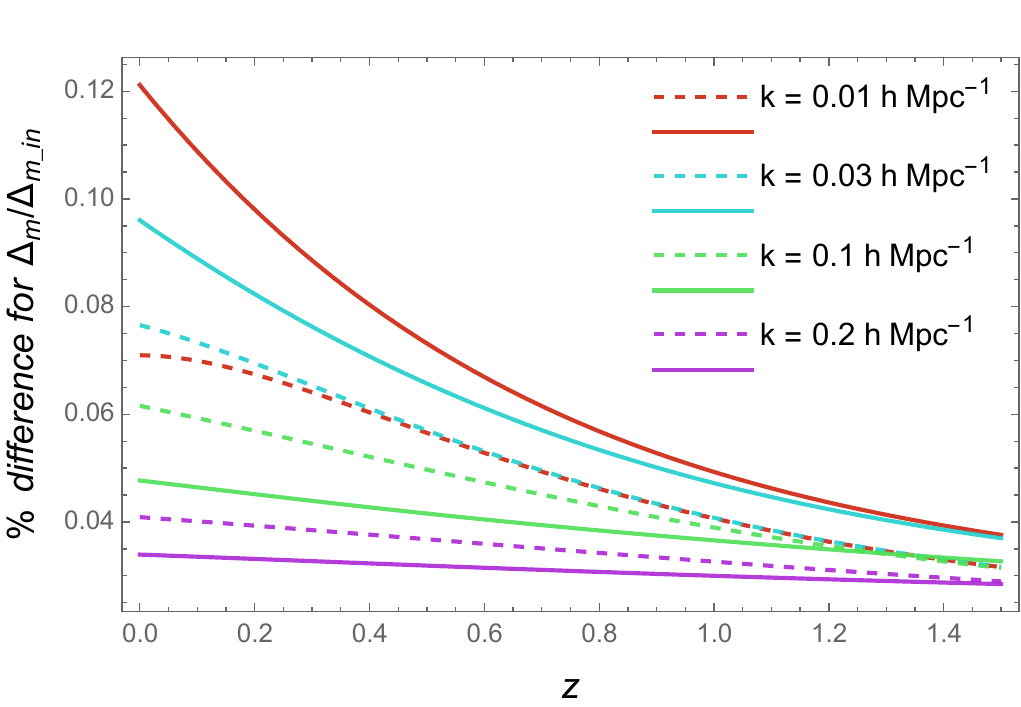}
         \caption{}
         \label{fig:percentcomparisonDelta}
    \end{subfigure}
    \begin{subfigure}[b]{0.5\linewidth}
         \includegraphics[width=\linewidth]{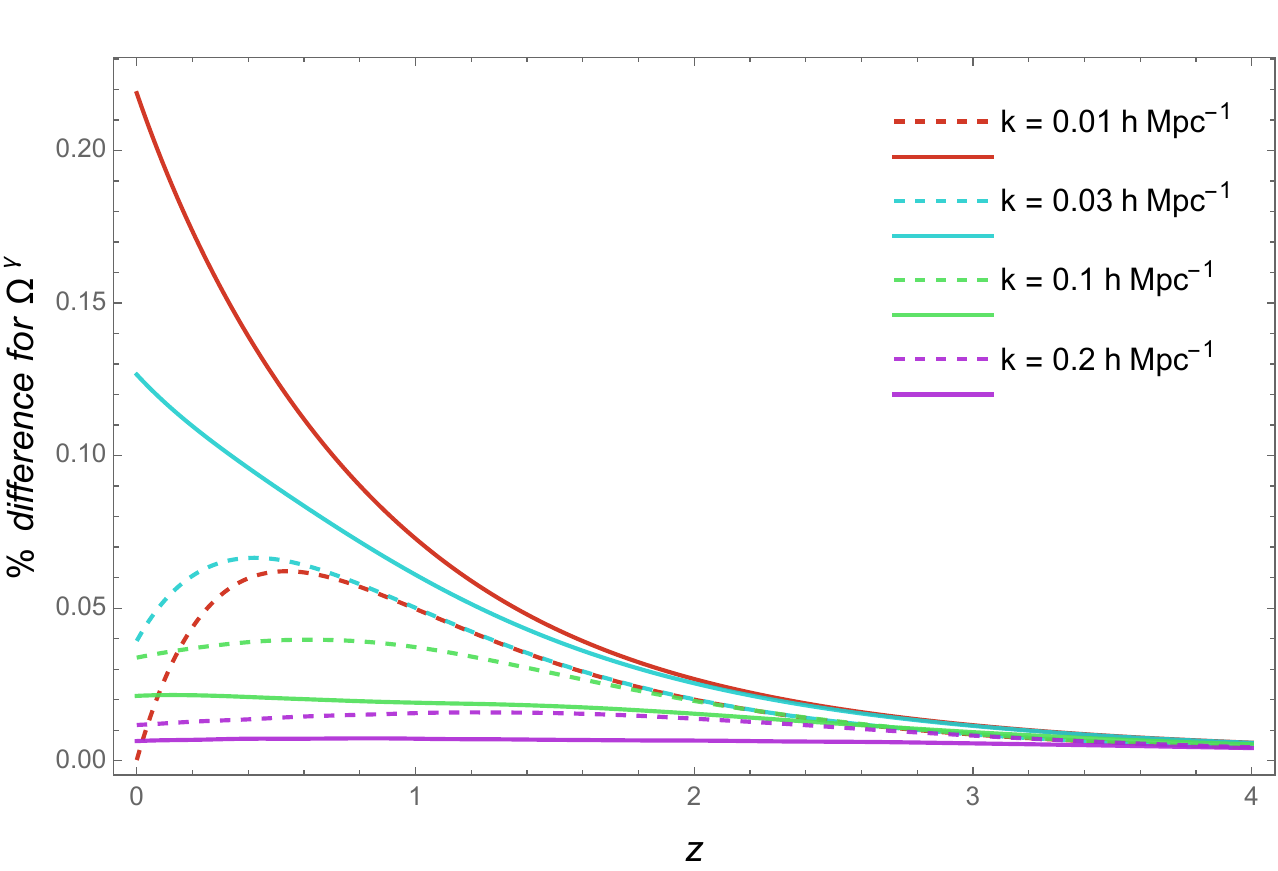}
         \caption{}
         \label{fig:percentcomparisonS}
    \end{subfigure}
    \begin{subfigure}[b]{0.5\linewidth}
         \includegraphics[width=\linewidth]{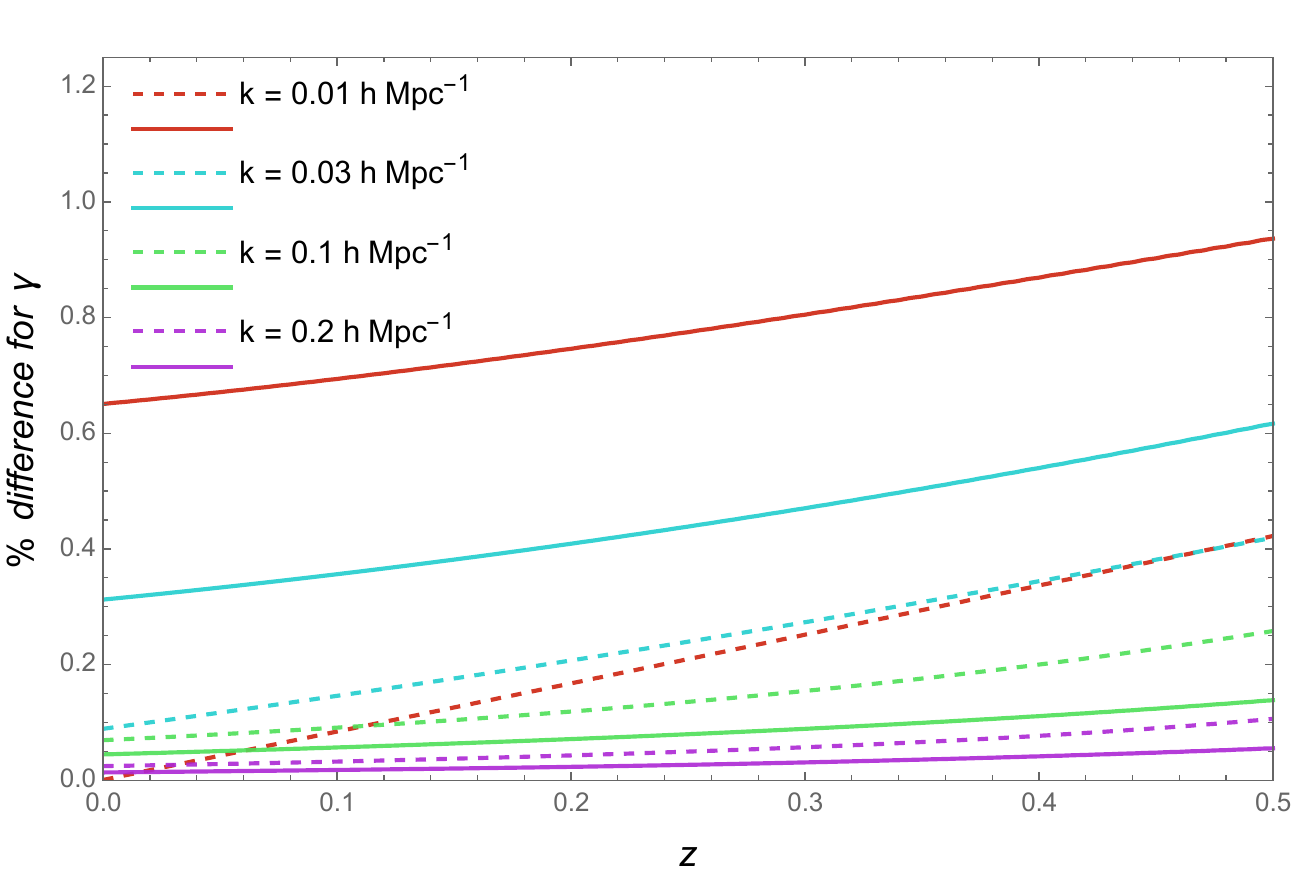}
         \caption{}
         \label{fig:percentcomparisonindex}
    \end{subfigure}
    \caption{Percentage difference between exact solutions and semi QS (dashed) and full QS (solid) solutions for (a) the normalised perturbation $\Delta_m/\Delta_{m}|_{in}$, (b) the growth function $S \equiv \Omega_m^\gamma$ and (c) the growth index $\gamma$.}
    \label{fig:percentcomparison}
\end{figure}

The differences between methods are examined in a more quantitative manner in Fig. \ref{fig:percentcomparison}, where the percentage differences between the exact solution and both the semi QS and full QS solutions are plotted for both large and small scales. Two general trends that are observed are the following
\begin{itemize}
    \item The difference between the solutions obtained with QS approximations (semi or full) and the exact solutions are larger for larger scales and smaller for smaller scales. 
    \item For smaller scales, solutions with full QS approximation appear closer to the exact solutions than solutions with semi QS approximation. The opposite is true for larger scales. 
\end{itemize}
 This scale-dependent behavior is evidently an important factor to take into account when applying the QS approximation in $f(R)$ gravity. 

Despite differences across methods at different scales, the results are in general in agreement with various model-specific or constrained $f(R)$ structure analyses \cite{HUTERER201523, Narikawa2010, Mirzatuny_2019, motohashi2011f, POLARSKI2008439, gannouji_growth, Tsujikawa2010chapt}. They display clear scale dependence, which becomes more pronounced at smaller scales, along with larger growth rate and significant deviation from $\Lambda$CDM for the growth index $\gamma$ on small redshifts. 

\begin{figure}
    \centering
    \begin{subfigure}[b]{0.49\linewidth}
         \includegraphics[width=\linewidth]{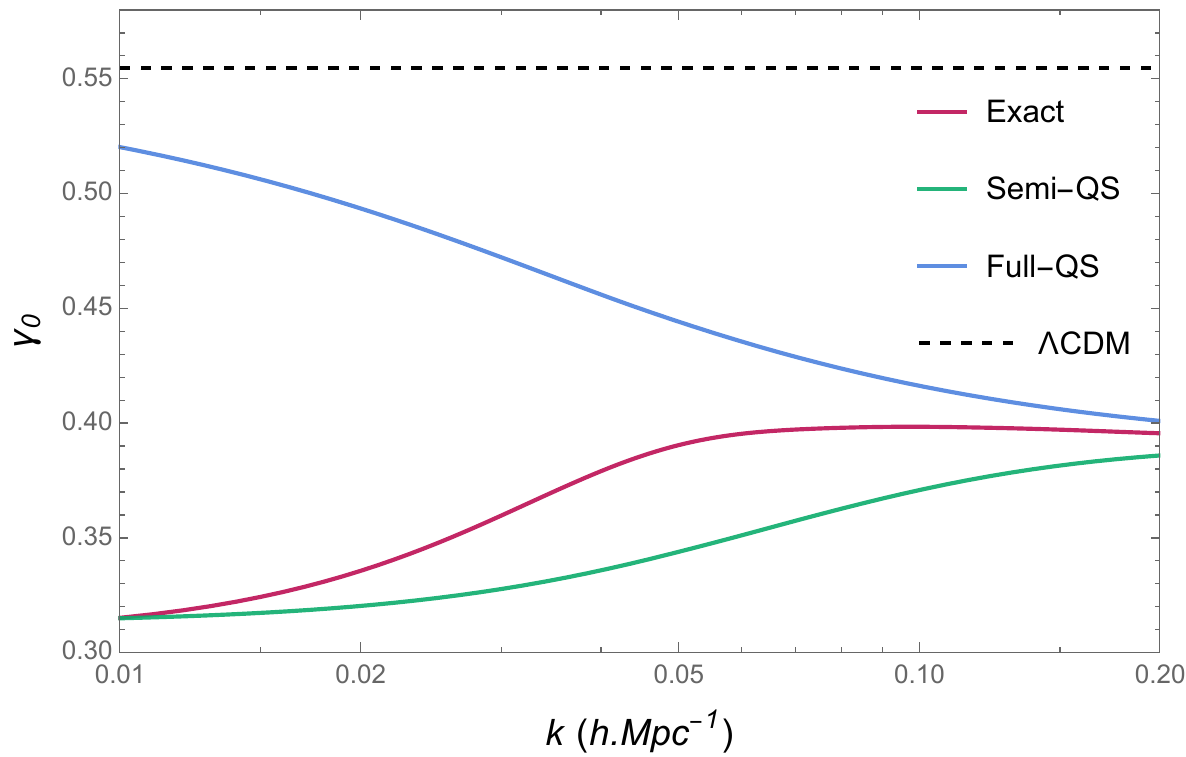}
         \caption{}
         \label{fig:gammacomparison1}
    \end{subfigure}
    \begin{subfigure}[b]{0.49\linewidth}
         \includegraphics[width=\linewidth]{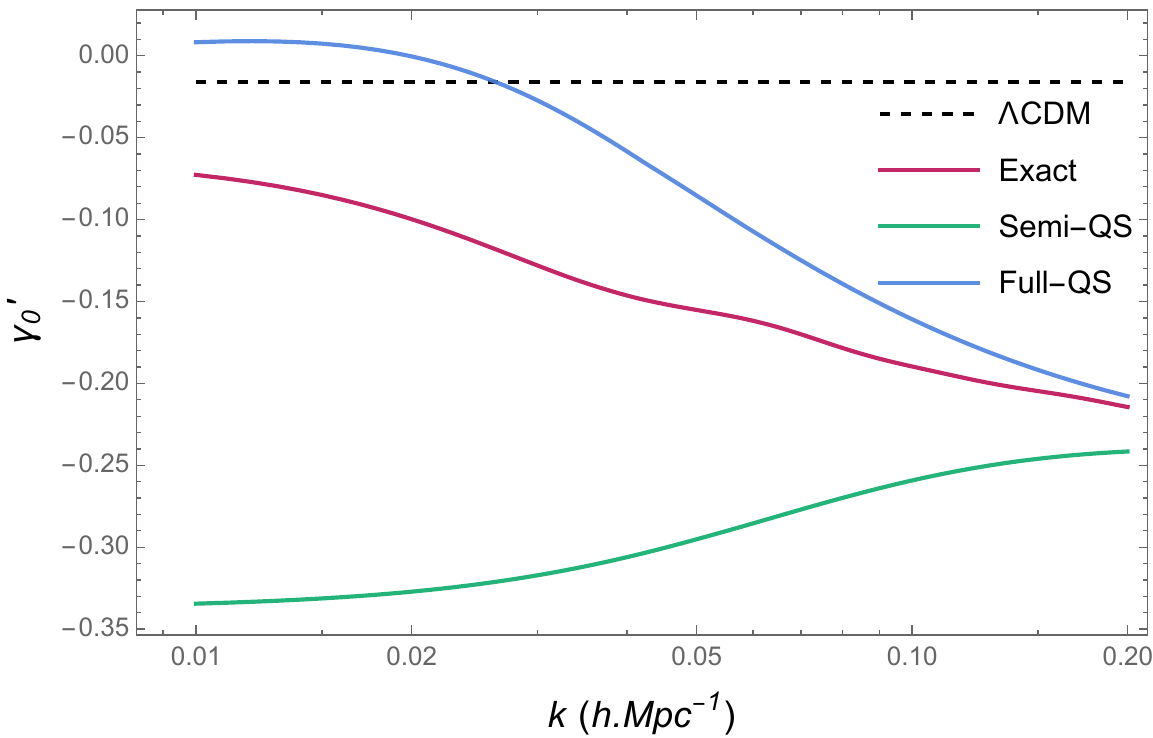}
         \caption{}
         \label{fig:gammacomparison2}
    \end{subfigure}
    \caption{Comparison of scale dependence for (a) $\gamma_0$ and (b) $\gamma_0'$ for each perturbation method. }
    \label{fig:gammacomparison}
\end{figure}

Further emphasising the difference from $\Lambda$CDM, there is clear time-dependence for $\gamma$ displayed on even small redshifts. $f(R)$ models are known to admit a growth index of this form, often approximated as 
\begin{equation}
    \gamma \simeq \gamma_0 + \gamma_0' \,z,
\end{equation}
for low redshifts. Both $\gamma_0$ and $\gamma_0'$ can be inferred from a number of observables, including rich clusters, redshift space distortions and tomographic weak lensing (see \cite{Reid2015, Planck2020, Amendola2008, Thomas2009, Belloso2011, Pouri2014}), however current constraints are not tight enough to definitively rule out a number of DE and modified gravity models. As a means of direct comparison with previous $f(R)$ model-dependent analysis, the scale dependence of $\gamma_0$ and $\gamma_0'$ for each method is shown in Fig \ref{fig:gammacomparison}. 


 All three methods come close to converging for both $\gamma_0$ and $\gamma_0'$ for $k \gtrsim 0.1$, as these are scales where both of the QS approximations are relevant. Again, we see in Fig \ref{fig:gammacomparison1} that only applying the semi QS approximation drives the value of $\gamma_0$ further from the values of both the full QS and exact solutions for small scales, while for the largest scales this approximation for $\gamma_0$ matches the exact solution. The full QS solutions at large scales show a value for $\gamma_0$ much closer to that of $\Lambda$CDM, validating the notion that this approximation is insufficient for distinguishing between $f(R)$ and $\Lambda$CDM at these scales. However, this is markedly different when looking at $\gamma_0'$ in Fig \ref{fig:gammacomparison2} where, on all scales, the semi QS method shows more significant deviation from the exact solution than the full QS method. 
 
 Confining our view to only the exact solutions for simplicity, we see that these model-independent results are in agreement with previous model-dependent $f(R)$ analyses, giving $ 0.3 \lesssim\gamma_0\lesssim 0.4$ and $-0.2 \lesssim\gamma_0'\lesssim -0.07$ \cite{POLARSKI2008439, gannouji_growth, Tsujikawa2010chapt, Tsujikawa:2009ku, Narikawa2010}. Regardless of marginal differences between methods, there is a dispersion across scales of at least $\sim 0.1$ for both $\gamma_0$ and $\gamma_0'$. This would be a strong indication of $f(R)$ modified gravity if detected in sufficiently sensitive observations.  

\clearpage
\newpage

\section{Discussion and Conclusion}

In this paper, we extend our model-independent, cosmography-based approach to studying $f(R)$ cosmologies which mimic $\Lambda$CDM \cite{Chakraborty:2021jku} to include analysis of the growth factor. To the best of our knowledge, this is the first time this type of analysis has been performed in a model-independent manner, i.e., without a-priori specifying the functional form of $f(R)$. This is achieved by recasting the terms involving $f(R)$ and its derivatives in terms of the cosmographic parameters ($q,\, j,\, s,\,l$), which appear in the $f(R)$ 1+3 covariant perturbation equations. We begin by reviewing the background model-independent analysis and emphasising the power of cosmography to describe the evolution of the cosmological expansion history. We then reformulate the exact linear perturbation equations and employ the separate quasi-static assumptions, leading to the semi and full quasi-static approximation equations. We explore the cosmographic condition $j=1$ and use this to specify a cosmic evolution that mimics the $\Lambda$CDM evolution at the background level. Then, we consider the dynamics of matter perturbation using the 1+3 covariant gauge-invariant formalism. Finally, we examine the evolution of the growth rate function and the growth index parameter for different scales at the levels of exact, semi and full quasi-static approximations and compare them with the model-specific $f(R)$ trends. We also show the dispersion across scales of the growth index parameter, which is a characteristic of modified gravity theories.

From these results we found that our approach (both with and without quasi-static approximations) is capable of capturing in a totally general way features that have earlier been observed in various earlier model-specific or constrained $f(R)$ structure analyses \cite{POLARSKI2008439,gannouji_growth,growthrateconstraints}. The density contrast $\Delta_m$, growth function $\Omega_m^\gamma$ and growth index $\gamma$ display clear scale-dependence,  which becomes more pronounced on smaller scales for redshift $z \gtrsim 0.5$. Additionally, $\gamma$ exhibits strong time-dependence and significantly smaller values, representing a substantial deviation from the $\Lambda$CDM model. A major achievement of our approach is the ability to obtain information about the present day values of the growth index and its derivative, $\gamma_0$ and $\gamma_0'$, for $\Lambda$CDM-mimicking $f(R)$ models without needing to specify any particular $f(R)$ theory. 
As explicitly reconstructed forms of $f(R)$ which mimic $\Lambda$CDM tend to be hypergeometric functions \cite{Dunsby:2010wg, He2013}, calculating these quantities and their dispersion would have been extremely complex without our approach. 

Finally, by comparing the exact, semi and full quasi-static methods, we were able to determine the shortcomings of the full QS method on large scales, where the perturbations have not spent significant time in the $f(R)$ regime and are thus trend closer to $\Lambda$CDM. For this reason, we caution against using the full quasi-static approximation for scales $k \lesssim 0.08$ h.Mpc$^{-1}$. For smaller scales, the differences between these methods are marginal. Regardless of the method, these results provide strong support and confirmation of many $f(R)$ structure growth studies, while avoiding both the need to narrow the analysis by choosing a form of $f(R)$ and the great complexity that results from the reconstruction programme. Even with the consideration of the exact perturbation equations (i.e. without any approximation), our approach is relatively uncomplicated,
computationally simple and allows us to determine predictions that can be tested with accurate data from future weak lensing surveys. The approach can also be easily implemented for other modifications of gravity and dynamical dark energy models, as well as background expansion histories which differ from $\Lambda$CDM.

\section{Acknowledgements}
KM thanks the University of Cape Town for financial assistance. SC acknowledges funding support from the NSRF via the Program Management Unit for
Human Resources and Institutional Development, Research and Innovation (Thailand) [grant
number B13F670063].
JW thanks the National Astrophysics and Space Science Programme and the University of Cape Town for their financial support. PKSD thanks the First Rand Bank for financial support. 

\clearpage
\newpage

\bibliographystyle{mnras}
\bibliography{apssamp}

\begin{thebibliography}{}
\makeatletter
\relax
\def\mn@urlcharsother{\let\do\@makeother \do\$\do\&\do\#\do\^\do\_\do\%\do\~}
\def\mn@doi{\begingroup\mn@urlcharsother \@ifnextchar [ {\mn@doi@}
  {\mn@doi@[]}}
\def\mn@doi@[#1]#2{\def\@tempa{#1}\ifx\@tempa\@empty \href
  {http://dx.doi.org/#2} {doi:#2}\else \href {http://dx.doi.org/#2} {#1}\fi
  \endgroup}
\def\mn@eprint#1#2{\mn@eprint@#1:#2::\@nil}
\def\mn@eprint@arXiv#1{\href {http://arxiv.org/abs/#1} {{\tt arXiv:#1}}}
\def\mn@eprint@dblp#1{\href {http://dblp.uni-trier.de/rec/bibtex/#1.xml}
  {dblp:#1}}
\def\mn@eprint@#1:#2:#3:#4\@nil{\def\@tempa {#1}\def\@tempb {#2}\def\@tempc
  {#3}\ifx \@tempc \@empty \let \@tempc \@tempb \let \@tempb \@tempa \fi \ifx
  \@tempb \@empty \def\@tempb {arXiv}\fi \@ifundefined
  {mn@eprint@\@tempb}{\@tempb:\@tempc}{\expandafter \expandafter \csname
  mn@eprint@\@tempb\endcsname \expandafter{\@tempc}}}

\bibitem[\protect\citeauthoryear{Abdelwahab, Goswami  \& Dunsby}{Abdelwahab
  et~al.}{2012}]{Abdelwahab:2011dk}
Abdelwahab M.,  Goswami R.,   Dunsby P. K.~S.,  2012, \mn@doi [Phys. Rev. D]
  {10.1103/PhysRevD.85.083511}, 85, 083511

\bibitem[\protect\citeauthoryear{Abebe, de~la Cruz-Dombriz  \& Dunsby}{Abebe
  et~al.}{2013}]{Abebe:2013zua}
Abebe A.,  de~la Cruz-Dombriz A.,   Dunsby P. K.~S.,  2013, \mn@doi [Phys. Rev.
  D] {10.1103/PhysRevD.88.044050}, 88, 044050

\bibitem[\protect\citeauthoryear{Adler, Casey  \& Jacob}{Adler
  et~al.}{1995}]{Adler1995}
Adler R.~J.,  Casey B.,   Jacob O.~C.,  1995, \mn@doi [American Journal of
  Physics] {10.1119/1.17850}, 63, 620

\bibitem[\protect\citeauthoryear{Amendola, Kunz  \& Sapone}{Amendola
  et~al.}{2008}]{Amendola2008}
Amendola L.,  Kunz M.,   Sapone D.,  2008, \mn@doi [Journal of Cosmology and
  Astroparticle Physics] {10.1088/1475-7516/2008/04/013}, 2008, 013

\bibitem[\protect\citeauthoryear{Amirhashchi \& Amirhashchi}{Amirhashchi \&
  Amirhashchi}{2020}]{Amirhashchi:2018vmy}
Amirhashchi H.,  Amirhashchi S.,  2020, \mn@doi [Gen. Rel. Grav.]
  {10.1007/s10714-020-2664-5}, 52, 13

\bibitem[\protect\citeauthoryear{Ananda, Carloni  \& Dunsby}{Ananda
  et~al.}{2009}]{Ananda_2009}
Ananda K.~N.,  Carloni S.,   Dunsby P. K.~S.,  2009, \mn@doi [Classical and
  Quantum Gravity] {10.1088/0264-9381/26/23/235018}, 26, 235018

\bibitem[\protect\citeauthoryear{Bean, Bernat, Pogosian, Silvestri  \&
  Trodden}{Bean et~al.}{2007}]{Bean_2007}
Bean R.,  Bernat D.,  Pogosian L.,  Silvestri A.,   Trodden M.,  2007, \mn@doi
  [Physical Review D] {10.1103/physrevd.75.064020}, 75

\bibitem[\protect\citeauthoryear{Belloso, García-Bellido  \& Sapone}{Belloso
  et~al.}{2011}]{Belloso2011}
Belloso A.~B.,  García-Bellido J.,   Sapone D.,  2011, \mn@doi [Journal of
  Cosmology and Astroparticle Physics] {10.1088/1475-7516/2011/10/010}, 2011,
  010

\bibitem[\protect\citeauthoryear{Blumenthal, Faber, Primack  \&
  Rees}{Blumenthal et~al.}{1984}]{Blumenthal:1984bp}
Blumenthal G.~R.,  Faber S.~M.,  Primack J.~R.,   Rees M.~J.,  1984, \mn@doi
  [Nature] {10.1038/311517a0}, 311, 517

\bibitem[\protect\citeauthoryear{Bolotin, Cherkaskiy, Ivashtenko, Konchatnyi
  \& Zazunov}{Bolotin et~al.}{2018}]{Bolotin:2018xtq}
Bolotin Y.~L.,  Cherkaskiy V.~A.,  Ivashtenko O.~Y.,  Konchatnyi M.~I.,
  Zazunov L.~G.,  2018, arXiv e-prints

\bibitem[\protect\citeauthoryear{Bruni, Ellis  \& Dunsby}{Bruni
  et~al.}{1992}]{MBruni_1992}
Bruni M.,  Ellis G. F.~R.,   Dunsby P. K.~S.,  1992, \mn@doi [Classical and
  Quantum Gravity] {10.1088/0264-9381/9/4/010}, 9, 921

\bibitem[\protect\citeauthoryear{Capozziello \& Francaviglia}{Capozziello \&
  Francaviglia}{2008}]{Capozziello:2007ec}
Capozziello S.,  Francaviglia M.,  2008, \mn@doi [Gen. Rel. Grav.]
  {10.1007/s10714-007-0551-y}, 40, 357

\bibitem[\protect\citeauthoryear{Capozziello, Cardone  \& Salzano}{Capozziello
  et~al.}{2008}]{Capozziello:2008qc}
Capozziello S.,  Cardone V.~F.,   Salzano V.,  2008, \mn@doi [Phys. Rev. D]
  {10.1103/PhysRevD.78.063504}, 78, 063504

\bibitem[\protect\citeauthoryear{Carloni}{Carloni}{2015}]{Carloni_2015}
Carloni S.,  2015, \mn@doi [Journal of Cosmology and Astroparticle Physics]
  {10.1088/1475-7516/2015/09/013}, 2015, 013

\bibitem[\protect\citeauthoryear{Carloni \& Dunsby}{Carloni \&
  Dunsby}{2007}]{Carloni:2006mr}
Carloni S.,  Dunsby P. K.~S.,  2007, \mn@doi [J. Phys. A]
  {10.1088/1751-8113/40/25/S40}, 40, 6919

\bibitem[\protect\citeauthoryear{Carloni, Dunsby, Capozziello  \&
  Troisi}{Carloni et~al.}{2005}]{Carloni:2004kp}
Carloni S.,  Dunsby P. K.~S.,  Capozziello S.,   Troisi A.,  2005, \mn@doi
  [Class. Quant. Grav.] {10.1088/0264-9381/22/22/011}, 22, 4839

\bibitem[\protect\citeauthoryear{Carloni, Dunsby  \& Troisi}{Carloni
  et~al.}{2008}]{Carloni_PRD77}
Carloni S.,  Dunsby P. K.~S.,   Troisi A.,  2008, \mn@doi [Phys. Rev. D]
  {10.1103/PhysRevD.77.024024}, 77, 024024

\bibitem[\protect\citeauthoryear{Carloni, Troisi  \& Dunsby}{Carloni
  et~al.}{2009}]{Carloni:2007br}
Carloni S.,  Troisi A.,   Dunsby P. K.~S.,  2009, \mn@doi [Gen. Rel. Grav.]
  {10.1007/s10714-008-0747-9}, 41, 1757

\bibitem[\protect\citeauthoryear{Carloni, Goswami  \& Dunsby}{Carloni
  et~al.}{2012}]{Carloni:2010ph}
Carloni S.,  Goswami R.,   Dunsby P. K.~S.,  2012, \mn@doi [Class. Quant.
  Grav.] {10.1088/0264-9381/29/13/135012}, 29, 135012

\bibitem[\protect\citeauthoryear{Carroll, Duvvuri, Trodden  \& Turner}{Carroll
  et~al.}{2004}]{Carroll:2003wy}
Carroll S.~M.,  Duvvuri V.,  Trodden M.,   Turner M.~S.,  2004, \mn@doi [Phys.
  Rev. D] {10.1103/PhysRevD.70.043528}, 70, 043528

\bibitem[\protect\citeauthoryear{Chakraborty, MacDevette  \&
  Dunsby}{Chakraborty et~al.}{2021}]{Chakraborty:2021jku}
Chakraborty S.,  MacDevette K.,   Dunsby P.,  2021, \mn@doi [Phys. Rev. D]
  {10.1103/PhysRevD.103.124040}, 103, 124040

\bibitem[\protect\citeauthoryear{Chakraborty, Gregoris  \& Mishra}{Chakraborty
  et~al.}{2023}]{Chakraborty:2022evc}
Chakraborty S.,  Gregoris D.,   Mishra B.,  2023, \mn@doi [Phys. Lett. B]
  {10.1016/j.physletb.2023.137962}, 842, 137962

\bibitem[\protect\citeauthoryear{De~Felice \& Tsujikawa}{De~Felice \&
  Tsujikawa}{2010}]{DeFelice:2010aj}
De~Felice A.,  Tsujikawa S.,  2010, \mn@doi [Living Rev. Rel.]
  {10.12942/lrr-2010-3}, 13, 3

\bibitem[\protect\citeauthoryear{De~la Cruz-Dombriz, Dobado  \& Maroto}{De~la
  Cruz-Dombriz et~al.}{2008}]{de_la_Cruz_Dombriz_2008}
De~la Cruz-Dombriz A.,  Dobado A.,   Maroto A.~L.,  2008, \mn@doi [Physical
  Review D] {10.1103/physrevd.77.123515}, 77

\bibitem[\protect\citeauthoryear{Dicke}{Dicke}{1961}]{DICKE1961}
Dicke R.~H.,  1961, \mn@doi [Nature] {10.1038/192440a0}, 192, 440

\bibitem[\protect\citeauthoryear{Dunajski \& Gibbons}{Dunajski \&
  Gibbons}{2008}]{Dunajski:2008tg}
Dunajski M.,  Gibbons G.,  2008, \mn@doi [Class. Quant. Grav.]
  {10.1088/0264-9381/25/23/235012}, 25, 235012

\bibitem[\protect\citeauthoryear{Dunsby \& Luongo}{Dunsby \&
  Luongo}{2016}]{Dunsby:2015ers}
Dunsby P. K.~S.,  Luongo O.,  2016, \mn@doi [Int. J. Geom. Meth. Mod. Phys.]
  {10.1142/S0219887816300026}, 13, 1630002

\bibitem[\protect\citeauthoryear{Dunsby, Bruni  \& Ellis}{Dunsby
  et~al.}{1992}]{dunsby1992covariant}
Dunsby P.~K.,  Bruni M.,   Ellis G.~F.,  1992, Astrophysical Journal, Part 1
  (ISSN 0004-637X), vol. 395, no. 1, p. 54-74., 395, 54

\bibitem[\protect\citeauthoryear{Dunsby, Elizalde, Goswami, Odintsov  \&
  Gomez}{Dunsby et~al.}{2010}]{Dunsby:2010wg}
Dunsby P. K.~S.,  Elizalde E.,  Goswami R.,  Odintsov S.,   Gomez D.~S.,  2010,
  \mn@doi [Phys. Rev. D] {10.1103/PhysRevD.82.023519}, 82, 023519

\bibitem[\protect\citeauthoryear{Ellis \& Bruni}{Ellis \&
  Bruni}{1989}]{Ellis:1989jt}
Ellis G. F.~R.,  Bruni M.,  1989, \mn@doi [Phys. Rev. D]
  {10.1103/PhysRevD.40.1804}, 40, 1804

\bibitem[\protect\citeauthoryear{Fonseca, Viljoen  \& Maartens}{Fonseca
  et~al.}{2019}]{growthrateconstraints}
Fonseca J.,  Viljoen J.-A.,   Maartens R.,  2019, \mn@doi [Journal of Cosmology
  and Astroparticle Physics] {10.1088/1475-7516/2019/12/028}, 2019, 028

\bibitem[\protect\citeauthoryear{Gannouji \& Polarski}{Gannouji \&
  Polarski}{2008}]{Gannouji_2008DE}
Gannouji R.,  Polarski D.,  2008, \mn@doi [Journal of Cosmology and
  Astroparticle Physics] {10.1088/1475-7516/2008/05/018}, 2008, 018

\bibitem[\protect\citeauthoryear{Gannouji, Moraes  \& Polarski}{Gannouji
  et~al.}{2009}]{gannouji_growth}
Gannouji R.,  Moraes B.,   Polarski D.,  2009, \mn@doi [Journal of Cosmology
  and Astroparticle Physics] {10.1088/1475-7516/2009/02/034}, 2009, 034

\bibitem[\protect\citeauthoryear{He \& Wang}{He \& Wang}{2013}]{He2013}
He J.-h.,  Wang B.,  2013, \mn@doi [Phys. Rev. D] {10.1103/PhysRevD.87.023508},
  87, 023508

\bibitem[\protect\citeauthoryear{Hojjati, Pogosian, Silvestri  \&
  Talbot}{Hojjati et~al.}{2012}]{Hojjati2012}
Hojjati A.,  Pogosian L.,  Silvestri A.,   Talbot S.,  2012, \mn@doi [Phys.
  Rev. D] {10.1103/PhysRevD.86.123503}, 86, 123503

\bibitem[\protect\citeauthoryear{Huterer et~al.,}{Huterer
  et~al.}{2015}]{HUTERER201523}
Huterer D.,  et~al., 2015, \mn@doi [Astroparticle Physics]
  {https://doi.org/10.1016/j.astropartphys.2014.07.004}, 63, 23

\bibitem[\protect\citeauthoryear{Kinney}{Kinney}{2002}]{Kinney:2002qn}
Kinney W.~H.,  2002, \mn@doi [Phys. Rev. D] {10.1103/PhysRevD.66.083508}, 66,
  083508

\bibitem[\protect\citeauthoryear{Liddle}{Liddle}{2003}]{Liddle:2003py}
Liddle A.~R.,  2003, \mn@doi [Phys. Rev. D] {10.1103/PhysRevD.68.103504}, 68,
  103504

\bibitem[\protect\citeauthoryear{{Lightman} \& {Schechter}}{{Lightman} \&
  {Schechter}}{1990}]{Lightman1990}
{Lightman} A.~P.,  {Schechter} P.~L.,  1990, \mn@doi [\apjs] {10.1086/191521},
  \href {https://ui.adsabs.harvard.edu/abs/1990ApJS...74..831L} {74, 831}

\bibitem[\protect\citeauthoryear{Mehrabi \& Rezaei}{Mehrabi \&
  Rezaei}{2021}]{Mehrabi:2021cob}
Mehrabi A.,  Rezaei M.,  2021, \mn@doi [Astrophys. J.]
  {10.3847/1538-4357/ac2fff}, 923, 274

\bibitem[\protect\citeauthoryear{Mirzatuny \& Pierpaoli}{Mirzatuny \&
  Pierpaoli}{2019}]{Mirzatuny_2019}
Mirzatuny N.,  Pierpaoli E.,  2019, \mn@doi [Journal of Cosmology and
  Astroparticle Physics] {10.1088/1475-7516/2019/09/066}, 2019, 066

\bibitem[\protect\citeauthoryear{Motohashi, Starobinsky  \& Yokoyama}{Motohashi
  et~al.}{2011}]{motohashi2011f}
Motohashi H.,  Starobinsky A.~A.,   Yokoyama J.,  2011, International Journal
  of Modern Physics D, 20, 1347

\bibitem[\protect\citeauthoryear{Mukherjee \& Banerjee}{Mukherjee \&
  Banerjee}{2016}]{Mukherjee:2016trt}
Mukherjee A.,  Banerjee N.,  2016, \mn@doi [Phys. Rev. D]
  {10.1103/PhysRevD.93.043002}, 93, 043002

\bibitem[\protect\citeauthoryear{Mukherjee \& Banerjee}{Mukherjee \&
  Banerjee}{2017}]{Mukherjee:2016shl}
Mukherjee A.,  Banerjee N.,  2017, \mn@doi [Class. Quant. Grav.]
  {10.1088/1361-6382/aa54c8}, 34, 035016

\bibitem[\protect\citeauthoryear{Mukherjee \& Banerjee}{Mukherjee \&
  Banerjee}{2021}]{Mukherjee:2020ytg}
Mukherjee P.,  Banerjee N.,  2021, \mn@doi [Eur. Phys. J. C]
  {10.1140/epjc/s10052-021-08830-5}, 81, 36

\bibitem[\protect\citeauthoryear{Narikawa \& Yamamoto}{Narikawa \&
  Yamamoto}{2009}]{Narikawa:2009lca}
Narikawa T.,  Yamamoto K.,  2009, in {19th Workshop on General Relativity and
  Gravitation in Japan}.

\bibitem[\protect\citeauthoryear{Narikawa \& Yamamoto}{Narikawa \&
  Yamamoto}{2010}]{Narikawa2010}
Narikawa T.,  Yamamoto K.,  2010, \mn@doi [Phys. Rev. D]
  {10.1103/PhysRevD.81.043528}, 81, 043528

\bibitem[\protect\citeauthoryear{Nojiri, Odintsov  \& Saez-Gomez}{Nojiri
  et~al.}{2009}]{Nojiri:2009kx}
Nojiri S.,  Odintsov S.~D.,   Saez-Gomez D.,  2009, \mn@doi [Phys. Lett. B]
  {10.1016/j.physletb.2009.09.045}, 681, 74

\bibitem[\protect\citeauthoryear{Noller, von Braun-Bates  \& Ferreira}{Noller
  et~al.}{2014}]{Noller2014}
Noller J.,  von Braun-Bates F.,   Ferreira P.~G.,  2014, \mn@doi [Phys. Rev. D]
  {10.1103/PhysRevD.89.023521}, 89, 023521

\bibitem[\protect\citeauthoryear{{Ostriker} \& {Steinhardt}}{{Ostriker} \&
  {Steinhardt}}{1995}]{Ostriker:1995rn}
{Ostriker} J.~P.,  {Steinhardt} P.~J.,  1995, \mn@doi [arXiv e-prints]
  {10.48550/arXiv.astro-ph/9505066}, \href
  {https://ui.adsabs.harvard.edu/abs/1995astro.ph..5066O} {pp
  astro--ph/9505066}

\bibitem[\protect\citeauthoryear{{Peebles}}{{Peebles}}{1980}]{Peebles1980}
{Peebles} P.~J.~E.,  1980, {The large-scale structure of the universe}

\bibitem[\protect\citeauthoryear{{Planck Collaboration} et~al.,}{{Planck
  Collaboration} et~al.}{2020}]{Planck2020}
{Planck Collaboration} et~al., 2020, \mn@doi [\aap]
  {10.1051/0004-6361/201833910}, \href
  {https://ui.adsabs.harvard.edu/abs/2020A&A...641A...6P} {641, A6}

\bibitem[\protect\citeauthoryear{Pogosian \& Silvestri}{Pogosian \&
  Silvestri}{2008}]{Pogosian2008}
Pogosian L.,  Silvestri A.,  2008, \mn@doi [Phys. Rev. D]
  {10.1103/PhysRevD.77.023503}, 77, 023503

\bibitem[\protect\citeauthoryear{Polarski \& Gannouji}{Polarski \&
  Gannouji}{2008}]{POLARSKI2008439}
Polarski D.,  Gannouji R.,  2008, \mn@doi [Physics Letters B]
  {https://doi.org/10.1016/j.physletb.2008.01.032}, 660, 439

\bibitem[\protect\citeauthoryear{Pouri, Basilakos  \& Plionis}{Pouri
  et~al.}{2014}]{Pouri2014}
Pouri A.,  Basilakos S.,   Plionis M.,  2014, \mn@doi [Journal of Cosmology and
  Astroparticle Physics] {10.1088/1475-7516/2014/08/042}, 2014, 042

\bibitem[\protect\citeauthoryear{Reid et~al.,}{Reid et~al.}{2015}]{Reid2015}
Reid B.,  et~al., 2015, \mn@doi [Monthly Notices of the Royal Astronomical
  Society] {10.1093/mnras/stv2382}, 455, 1553

\bibitem[\protect\citeauthoryear{Sahni, Saini, Starobinsky  \& Alam}{Sahni
  et~al.}{2003}]{Sahni:2002fz}
Sahni V.,  Saini T.~D.,  Starobinsky A.~A.,   Alam U.,  2003, \mn@doi [JETP
  Lett.] {10.1134/1.1574831}, 77, 201

\bibitem[\protect\citeauthoryear{Sawicki \& Bellini}{Sawicki \&
  Bellini}{2015}]{Sawicki2015}
Sawicki I.,  Bellini E.,  2015, \mn@doi [Phys. Rev. D]
  {10.1103/PhysRevD.92.084061}, 92, 084061

\bibitem[\protect\citeauthoryear{Silvestri, Pogosian  \& Buniy}{Silvestri
  et~al.}{2013}]{Silvestri2013}
Silvestri A.,  Pogosian L.,   Buniy R.~V.,  2013, \mn@doi [Phys. Rev. D]
  {10.1103/PhysRevD.87.104015}, 87, 104015

\bibitem[\protect\citeauthoryear{Sotiriou \& Faraoni}{Sotiriou \&
  Faraoni}{2010}]{Sotiriou:2008rp}
Sotiriou T.~P.,  Faraoni V.,  2010, \mn@doi [Rev. Mod. Phys.]
  {10.1103/RevModPhys.82.451}, 82, 451

\bibitem[\protect\citeauthoryear{Spalinski}{Spalinski}{2007}]{Spalinski:2007ef}
Spalinski M.,  2007, \mn@doi [JCAP] {10.1088/1475-7516/2007/08/016}, 08, 016

\bibitem[\protect\citeauthoryear{Steinhardt}{Steinhardt}{1998}]{Steinhardt+1998+123+146}
Steinhardt P.~J.,  1998, 7 COSMOLOGICAL CHALLENGES FOR THE 21ST CENTURY.
Princeton University Press, Princeton, pp 123--146,
  \mn@doi{doi:10.1515/9780691227498-008}

\bibitem[\protect\citeauthoryear{Stelle}{Stelle}{1978}]{Stelle:1977ry}
Stelle K.~S.,  1978, \mn@doi [Gen. Rel. Grav.] {10.1007/BF00760427}, 9, 353

\bibitem[\protect\citeauthoryear{Tada \& Terada}{Tada \& Terada}{2024}]{tada}
Tada Y.,  Terada T.,  2024, \mn@doi [Phys. Rev. D]
  {10.1103/PhysRevD.109.L121305}, 109, L121305

\bibitem[\protect\citeauthoryear{Thomas, Abdalla  \& Weller}{Thomas
  et~al.}{2009}]{Thomas2009}
Thomas S.~A.,  Abdalla F.~B.,   Weller J.,  2009, \mn@doi [Monthly Notices of
  the Royal Astronomical Society] {10.1111/j.1365-2966.2009.14568.x}, 395, 197

\bibitem[\protect\citeauthoryear{Tsujikawa}{Tsujikawa}{2007}]{tsuj2007}
Tsujikawa S.,  2007, \mn@doi [Phys. Rev. D] {10.1103/PhysRevD.76.023514}, 76,
  023514

\bibitem[\protect\citeauthoryear{Tsujikawa}{Tsujikawa}{2010}]{Tsujikawa2010chapt}
Tsujikawa S.,  2010, Modified Gravity Models of Dark Energy.
Springer Berlin Heidelberg, Berlin, Heidelberg, pp 99--145,
  \mn@doi{10.1007/978-3-642-10598-2_3}

\bibitem[\protect\citeauthoryear{Tsujikawa, Gannouji, Moraes  \&
  Polarski}{Tsujikawa et~al.}{2009}]{Tsujikawa:2009ku}
Tsujikawa S.,  Gannouji R.,  Moraes B.,   Polarski D.,  2009, \mn@doi [Phys.
  Rev. D] {10.1103/PhysRevD.80.084044}, 80, 084044

\bibitem[\protect\citeauthoryear{Zhai, Zhang, Zhang, Liu  \& Zhang}{Zhai
  et~al.}{2013}]{Zhai:2013fxa}
Zhai Z.-X.,  Zhang M.-J.,  Zhang Z.-S.,  Liu X.-M.,   Zhang T.-J.,  2013,
  \mn@doi [Phys. Lett. B] {10.1016/j.physletb.2013.10.020}, 727, 8

\makeatother
\end{thebibliography}

\label{lastpage}
\end{document}